\documentclass[12pt]{article}
\ifdefined\pdfminorversion\pdfminorversion=7\fi
\usepackage[top=1.25in, bottom=1.25in, left=1.25in, right=1.25in]{geometry}
\usepackage{setspace}
\usepackage{graphicx}
\graphicspath{ {./pix/} }
\usepackage[page]{appendix}
\usepackage{amsfonts}
\usepackage{amsmath}
\usepackage{amsthm}
\usepackage{amssymb}
\usepackage{epsfig}
\usepackage{url}
\usepackage{lscape}
\usepackage[colorlinks=true, allcolors=blue]{hyperref}
\usepackage{nameref}
\usepackage{bm}
\usepackage{xcolor,algorithmicx,algorithm}
\usepackage{bbm}
\usepackage{booktabs}
\usepackage{adjustbox}
\usepackage{mathtools}
\usepackage{siunitx}
\usepackage{tikz-cd}
\tikzcdset{ampersand replacement=\&}
\usetikzlibrary{calc}
\usetikzlibrary{arrows}
\usetikzlibrary{positioning}
\usetikzlibrary{fit}
\usetikzlibrary{shapes.geometric}
\usepackage[font=small]{subcaption}
\usepackage{rotating}
\usepackage{dcolumn}
\usepackage{float}
\usepackage{natbib}

\DeclareMathOperator{\spn}{span}
\DeclareMathOperator{\Ima}{Im}
\DeclareMathOperator{\diag}{diag}

\DeclareMathOperator{\tr}{tr}
\DeclareMathOperator{\rank}{rank}
\DeclareMathOperator{\length}{len}
\DeclareMathOperator{\size}{size}

\DeclareMathOperator{\var}{Var}
\DeclareMathOperator{\cov}{Cov}
\DeclareMathOperator{\corr}{Corr}

\DeclareMathOperator*{\esssup}{ess\,sup}

\newcommand{\E}{{\mathbb E}}

\newcommand{\Pa}{{\mathbb P}}

\newcommand{\R}{{\mathbb R}}

\newcommand{\N}{{\mathbb N}}

\newcommand{\Ecal}{{\mathcal E}}
\newcommand{\Fcal}{{\mathcal F}}

\newcommand{\Scal}{{\mathcal S}}

\newcommand{\Vcal}{{\mathcal V}}

\renewcommand{\labelenumi}{\rm{(\roman{enumi})}}
\renewcommand{\theenumi}{\rm{(\roman{enumi})}}

\newcommand{\id}{{\rm id}}

\newtheorem{proposition}{Proposition}[section]
\newtheorem{lemma}[proposition]{Lemma}
\newtheorem{theorem}[proposition]{Theorem}

\newtheorem{corollary}[proposition]{Corollary}
\newtheorem{remark}[proposition]{Remark}

\usepackage{thmtools}
\declaretheorem[sibling=proposition]{example}
\renewcommand\thmcontinues[1]{continued}
\makeatletter
\def\thmt@patch@grabnum#1#2\thmt@patch@nil{\def\thmt@patch@refnum{#1}}
\def\KV@continues@continues#1{\thmt@thmuse@iskvtrue
  \@namedef{thmt@unusedkey@continues}{}%
  \expandafter\ifx\csname r@#1\endcsname\relax
    \def\thmt@patch@refnum{??}
  \else
    \expandafter\expandafter\expandafter\thmt@patch@grabnum
      \csname r@#1\endcsname\relax\thmt@patch@nil
  \fi
  \thmt@suspendcounter{\thmt@envname}{\thmt@patch@refnum}%
  \g@addto@macro\thmt@newoptarg{{, }%
    \thmcontinues{#1}%
    \@iden}%
}
\makeatother

\author{Damir\ Filipovic\footnote{
\textsc{EPFL and SFI},
 \texttt{damir.filipovic@epfl.ch}}
 \and
 Paul\ Schneider\footnote{
 \textsc{Universit\`a della Svizzera italiana and SFI},
 \texttt{paul.schneider@usi.ch}
 }}

\newcommand{\Twin}{T_{\mathrm{w}}}  

\begin{document}

\title{Fundamental Properties of Linear Factor Models\footnote{Thanks to Mikhail Chernov, Alain Fortin, Lukas Gonon, Olivier Scaillet for helpful comments. Part of this paper was written while the first author was a guest of the FinsureTech Hub, Department of Mathematics, ETH Zurich. We used Claude (Anthropic), under our direction, for language editing and drafting assistance, for consistency checks of cross-references and of the numerical statements against the tables and figures, and for edits to the estimation and plotting code in the replication package.}
}

\date{8 September 2026}

\begin{titlepage}
\maketitle

\begin{abstract}

We characterize the loading matrices that admit a conditional linear factor representation for excess returns in which the factors are traded, residual risk is unpriced, and the loadings are the betas of the factors. The characterization is a joint restriction on the loadings, the risk premia, and the second moment matrix of returns. It requires no rank assumptions, and exactly one traded factor representation attains it for given loadings: the moment-weighted factor portfolios. On Fama--French panels, characteristic loadings differ from the regression loadings of their factors by 19 to 48 percent of their size, at every estimation window.

\vspace{2ex}

\noindent\textbf{Keywords:} asset pricing; conditional factor models; factor representations; moment-weighted factor portfolios; mean--variance efficiency; stochastic discount factor; factor spanning; loading discrepancy

\vspace{2ex}

\noindent\textbf{JEL classification:} G11, G12, C52
\end{abstract}

\end{titlepage}

\setcounter{page}{2}

\section{Introduction}\label{sec:introduction}
Conditional linear factor models are the workhorse of empirical asset pricing in both academia and industry. In a typical application the excess returns of a cross-section of $n_t$ assets are represented as
\begin{equation}\label{eq:linfacmodelPhi}
\bm x_{t+1} = \bm\Phi_t \bm f_{t+1} + \bm\epsilon_{t+1},
\end{equation}
where the $n_t\times m_t$ loading matrix $\bm\Phi_t$ is built from firm characteristics and macroeconomic variables observable at time $t$, and $\bm f_{t+1}$ is a vector of $m_t$ common factors recovered period by period as the cross-sectional regression coefficients of returns on the loadings.
This paper asks a simple question about this practice: \emph{do the loadings $\bm\Phi_t$ correspond to the betas of the factors the representation reports?}\footnote{Throughout, beta means the slope matrix of the conditional regression of returns on a constant and the factors, Section~\ref{sec:alphatests}. Property~\ref{it:prop1} below concerns the regression without the constant, whose coefficient matrix we call the regression loadings. For traded factors under the law of one price the two coincide exactly when the alpha vanishes.}
Since the loadings are prespecified, it would be an unlikely coincidence if they did. This paper works out when they do, in terms of the second moment matrix of returns, and develops the population properties of linear factor models for finite cross-sections.

Our analysis speaks to many existing methods at once. Instrumented principal components \citep{kellypruittsu19}, projected principal components \citep{fanliaowang16}, regressed principal components \citep{chenroussanovwang23}, heuristic characteristics-based factor portfolios \citep{kozaknagel23}, and risk-premium principal components \citep{lettaupelger20} all recover their factors as cross-sectional regression coefficients in the tradition of \citet{famamacbeth73}, and differ only in how they choose the span of the loadings. Regressed principal components add a fixed compression step after the regression and test pricing through an estimated intercept. Both steps fall within the same framework. One condition therefore answers the question for all of them at once. It also identifies where the answer is yes: in the conditional model class of \citet{kozaknagel23}, in which risk premia are linear in the characteristics and covariances inherit the factor structure of the same characteristics, the correspondence holds automatically, and we carry this class through the paper as a running example. Outside of it, the correspondence is a property the data may or may not grant, and Section~\ref{sec:empirical} finds that in US portfolio returns it does not hold, on every panel and at every window length we examine: the measured distance between the loadings and the regression loadings of the factors averages 19 to 48 percent of the size of the loadings and does not close as the window lengthens.

Our first and main contribution is an analysis of four properties that empirical work attributes to a factor representation \eqref{eq:linfacmodelPhi}, and of which combinations of them are attainable:
\begin{enumerate}
\renewcommand{\labelenumi}{\rm{(P\arabic{enumi})}}
\renewcommand{\theenumi}{\rm{(P\arabic{enumi})}}
\item\label{it:prop1} the loadings are the regression loadings of the factors,
\item\label{it:prop2} residual risk is unpriced,
\item\label{it:prop3} the factors span the maximal Sharpe ratio attainable from the assets,
\item\label{it:prop4} the residuals are idiosyncratic: uncorrelated with the factors, with a second moment matrix of full rank.
\end{enumerate}
The opening question combines properties~\ref{it:prop1} and~\ref{it:prop2}: for traded factors under the law of one price, the loadings are the betas of the factors when they are the regression loadings of the factors, property~\ref{it:prop1}, and residual risk is unpriced, property~\ref{it:prop2}, since then the alpha of the intercept regression vanishes, Proposition~\ref{prop:tradeq}. Property~\ref{it:prop1} pins the factor representation down. Whenever traded factors that carry a prespecified loading matrix as their own regression loadings exist at all, they are unique and available in closed form from the loadings and the second moments of returns, Theorem~\ref{thm:rigid}. We call them moment-weighted factor portfolios. They come from the same cross-sectional regression that empirical work already runs, with a weighting supplied by the second moments in place of equal weighting: once the loadings are named, the factors are named with them. There is no extraction step and no rotation to resolve, in contrast to latent-factor methods, which recover loadings and factors jointly and only up to rotation \citep{lettaupelger20}. Nothing about the loadings needs to change, only the weighting of the cross-sectional regression that recovers the factors from them. We also show how to compute the moment-weighted portfolios and identify the specifications under which current practice already produces them.

Property~\ref{it:prop2} brings in asset pricing and restricts the loadings rather than the factors: residual risk is unpriced exactly when the risk premia are a combination of the loadings, with nothing left over for the residuals. We call loading matrices that can carry properties \ref{it:prop1} and \ref{it:prop2} admissible, and factor representations that have them coherent. Admissibility asks of the span of the loadings only that it be large enough to reach the risk premia and small enough to stay within what returns can span, and it needs no rank or invertibility assumption, which matters because large panels place estimated moment matrices at or beyond the singular boundary. Under the law of one price, the only pricing assumption of the paper, coherence delivers property~\ref{it:prop3} as well: a coherent factor representation spans the maximal Sharpe ratio, so a mean--variance investor loses nothing by trading a handful of factor portfolios in place of the full cross-section. Coherence also clarifies two fixtures of empirical practice. For factor portfolios under the law of one price, the zero-alpha restriction of \citet{gib_ros_sha_89} and the pricing restriction are equivalent, Proposition~\ref{prop:tradeq}, so alpha-based tests test what they are stated to test. And that characteristics are covariances, the operational form of the characteristics-versus-covariances debate of \citet{danieltitman97}, which \citet{kellypruittsu19} impose as a specification and test through an anomaly intercept, is a property coherent factor representations deliver by themselves, for traded factors. The Appendix treats the general case, where factors need not be traded.

\citet{kozaknagel23} is the paper closest to ours. They ask when characteristics-based factor portfolios span the stochastic discount factor. We extend their analysis in three directions. Our statements allow the moment and loading matrices to be singular where theirs assume invertibility, which is what fails with redundant test assets and estimation windows shorter than the cross-section. The factors of a coherent factor representation are shown to be unique and given in closed form, where they establish sufficient conditions for particular representations. Under their Assumption~1 and the law of one price, their spanning result coincides with our Corollary~\ref{cor:OLScoh}, as Section~\ref{sec:leadingreps} spells out. And the two conditions entering their analysis, factors uncorrelated with residuals and residual risk unpriced, are derived and characterized here rather than assumed.

Our second contribution makes the discrepancy between the reported loadings and the regression loadings of the reported factors measurable. Corollary~\ref{cor:cohgap} identifies the smallest such discrepancy that any traded factor representation can achieve. This floor depends only on the loading matrix and the second moments of returns, the moment-weighted portfolios attain it, and whatever an implementation pays above the floor measures its misspecification, in every sample and at essentially no computational cost. A second number, the share of the risk premia that the loadings fail to reach, measures the distance from the pricing requirement in the same way.

Our third contribution is an impossibility result with a reassuring reading. For traded factors with a nonzero factor component, requiring the residuals to be uncorrelated with the factors forces the residual second moment matrix to be singular, Proposition~\ref{prop:rankdef}. A factor structure with full-rank residuals uncorrelated with the factors, property~\ref{it:prop4}, therefore cannot be carried by traded factors in any finite cross-section, and once residual risk is unpriced, factors and residuals must be correlated: the factor--residual correlations documented for characteristics-based portfolios are implied by the theory rather than anomalous. The result also places the finite cross-section next to the arbitrage pricing tradition. \citet{connor84} derives exact factor pricing, our unpriced-residual condition, from diversification and market clearing for one particular loading matrix, sharpening the approximate pricing of \citet{ross76}. We characterize all loading matrices that can carry the condition, such that his conclusion is our hypothesis. And since the forced singularity concerns at most as many directions as there are factors, it vanishes as the cross-section grows, so the full-rank idealization of large economies is the correct limit and our characterization its exact finite cross-section counterpart in population.

A word on scope. This is a paper about the population, not about estimation. All statements take the conditional moments as given and describe what factor representations built from them can and cannot achieve, so no sampling error enters anywhere in the theory. For the same reason we do not study shrinkage or regularization outside of the empirical section, where sample moments make it a practical necessity. Wherever a matrix fails to be invertible the theory works with its pseudoinverse, which is the mathematical limit of ridge regularization as the penalty vanishes. The population analysis therefore already sits at the idealized endpoint that any regularized estimator approaches, and carrying an explicit penalty through the theory would add notation without adding content.

The remainder of the paper is organized as follows. Section~\ref{sec:CLFM} introduces the setting and records the factor representations of the literature in a common notation. Section~\ref{sec:OFM} takes the four properties in turn and establishes what each requires. Section~\ref{sec:verdict} draws the implications for current practice. Section~\ref{sec:empirical} contains the empirical evidence, and Section~\ref{sec:conclusion} concludes. Appendices~\ref{app:matrix} to~\ref{app:CAP} collect the matrix, probabilistic, and asset pricing preliminaries. Appendix~\ref{app:optimalfactors} extends the theory to general, possibly non-traded, factors and shows that the restriction to traded factors drives none of the conclusions on properties \ref{it:prop1} and \ref{it:prop2}. Appendix~\ref{app:proofs} contains the proofs, and Appendix~\ref{app:existence} answers the existence question for statistical factor representations.

\paragraph{Related literature.}
Our framework is the conditional asset pricing setting of \citet{han_ric_87}: conditional information sets, conditional projections, and conditional mean--variance efficiency, constructed in Appendices~\ref{app:probsetup} and~\ref{app:CAP}. Their question meets ours at one point. They ask what conditioning information does to the testable content of a pricing model, and in particular what is lost when a conditional model is confronted with unconditional moments. For the object this paper is about, nothing is lost: under the law of one price, the portfolio attaining the maximal conditional Sharpe ratio attains the maximal unconditional one as well, by Proposition~\ref{prop:span}~\ref{it:span4} and Remark~\ref{rem:SharpeCU}, so factor spanning can be evaluated without conditioning information even though it is a conditional property. This equivalence justifies the design of Section~\ref{sec:empirical}, in which weights formed from trailing conditional moments are evaluated on their realized unconditional Sharpe ratios.

Factor structures with full-rank residual covariance matrices are the standard idealization of large economies, going back to \citet{chamberlainrothschild83}, as in the performance measurement framework of \citet{ConnorKorajckzyk1986} and the large-dimensional inferential theory of \citet{bai03}. Our results are exact statements for any finite cross-section, without rank assumptions, and Section~\ref{sec:largeeconomies} spells out the mapping between the large-economy settings: the residual singularity and the factor--residual covariance that our results imply in a finite cross-section vanish as the cross-section grows, at the rates given there, so the two views are consistent.\footnote{A second strand, the literature on short panels, keeps the cross-section large and the time dimension fixed, where the sample second moment matrix is rank deficient by construction \citep{raponirobottizaffaroni19,fortin2025eigenvalue,fortingagliardiniscaillet24}. A singular second moment matrix reduces an $n_t$-asset economy to one of its own rank, so that setting is the sample image of the population case studied here.}

On the inference side, \citet{gagliardiniossolascaillet16} develop two-pass inference on time-varying risk premia in large unbalanced panels of individual stocks, with betas instrumented by observable characteristics and common instruments, and \citet{zaffaroni25} develops inference on the pricing ability of conditional models through local principal components. These papers are complementary to ours: they provide asymptotic inference as the cross-section and the time series grow, and we provide the exact population structure that conditional specifications must respect period by period. The instrumented loadings of \citet{gagliardiniossolascaillet16} are a special case of our loading matrix, which may be an arbitrary function of the information available at time $t$. Because our results restrict the joint specification of loadings, risk premia, and second moments within a period, and do not restrict how the mapping from characteristics to loadings or premia varies across periods, they are also compatible with the post-publication decay of characteristic premia documented by \citet{mcleanpontiff16}.

A broader literature builds return representations from characteristics through flexible functional forms and machine learning, as in the semiparametric characteristic-based factors of \citet{connorhagmannlinton12}, the nonparametric characteristic model of \citet{freybergerneuhierlweber20}, the autoencoder representations of \citet{gukellyxiu20}, the structural deep-learning factors of \citet{fankeliaoneuhierl22}, and the deep factor models of \citet{kellykuznetsovmalamudzhang26}. In our notation, added flexibility is a larger loading matrix, so these specifications remain within the scope of our results, and the loading discrepancy of Section~\ref{sec:diagnostics} can be computed for any of them. Finally, on the econometrics side, Lemma~\ref{lem:GLSinv} generalizes the GLS equality conditions of \citet[Theorem~2]{lu_sch_12} to singular matrices, and the residual singularity of Section~\ref{sec:rankdeficiency}, which makes the population target of residual covariance estimation rank deficient under traded factors, relates our results to the covariance estimation literature surveyed in \citet{ledoitwolf20}.

\section{Factor Representations}\label{sec:CLFM}
This section fixes notation, states the standing assumption that factors are portfolios, defines the two objects the paper compares: the loadings a researcher specifies and the loadings a regression on the factors produces. Subsequently, it records what the empirical literature builds, and closes with the law of one price and the pricing benchmarks of Section~\ref{sec:pricingbenchmarks}. We make heavy use of elementary notions of linear algebra, in particular image, kernel,  direct sums, and the Moore--Penrose pseudoinverse $\bm A^+$: the matrix that inverts $\bm A$ on its image and vanishes off it, reviewed in Appendix~\ref{app:matrix}.

\subsection{Notation}\label{sec:notation}
Time is discrete. We fix a generic period $[t,t+1]$ and denote by $\Fcal_t$ the information set at time $t$. Let $\bm x_{t+1}$ be the vector of excess returns of $n_t$ assets over the period, where the dimension $n_t$ is $\Fcal_t$-measurable, that is, known at $t$. We impose the standard regularity condition that all random vectors under consideration have finite conditional second moments, written $\bm y_{t+1}\in L^2_t$.\footnote{A formal construction of the conditional $L^2$ spaces and their projection properties is given in Appendix~\ref{app:probsetup}.}

For any $\bm y_{t+1}\in L^2_t$ we write $\bm\mu_{\bm y,t}\coloneqq\E_t[\bm y_{t+1}]$, $\bm\Sigma_{\bm y,t}\coloneqq\cov_t[\bm y_{t+1}]$, and $\bm Q_{\bm y,t}\coloneqq\E_t[\bm y_{t+1}\bm y_{t+1}^\top]=\bm\Sigma_{\bm y,t}+\bm\mu_{\bm y,t}\bm\mu_{\bm y,t}^\top$ for its conditional mean vector, covariance matrix, and second moment matrix. We show in Remark~\ref{rem:support} in Appendix~\ref{app:probsetup} that $\Ima\bm Q_{\bm y,t}$ is the smallest subspace containing the support of the conditional distribution of $\bm y_{t+1}$. To lighten notation, we suppress the subscript $\bm y=\bm x$ for the returns and write $\bm\mu_t$, $\bm\Sigma_t$, $\bm Q_t$, henceforth omit the qualifier ``conditional'', and refer to excess returns simply as returns.

The returns $\bm x_{t+1}$ are the primary object of the paper. Everything else is a way of representing them. A \emph{factor representation} is a pair $(\bm\Phi_t,\bm f_{t+1})$ consisting of an $\Fcal_t$-measurable $n_t\times m_t$ loading matrix $\bm\Phi_t$ and $m_t$ factors $\bm f_{t+1}\in L^2_t$, understood as the decomposition \eqref{eq:linfacmodelPhi},
in which the residuals are implicit, $\bm\epsilon_{t+1}\coloneqq\bm x_{t+1}-\bm\Phi_t\bm f_{t+1}$. The dimension $m_t$ is exogenous and $\Fcal_t$-measurable. In applications, $\bm\Phi_t$ is typically a function of observable firm characteristics and macroeconomic variables at time $t$. Every pair $(\bm\Phi_t,\bm f_{t+1})$ is a factor representation of $\bm x_{t+1}$ in this sense: the equation defines the residuals rather than restricting the pair. Both $\E_t[\bm\epsilon_{t+1}]$ and the conditional covariance between $\bm f_{t+1}$ and $\bm\epsilon_{t+1}$ are then consequences of the joint conditional distribution of $(\bm x_{t+1},\bm f_{t+1})$ and of the choice of $\bm\Phi_t$. In Section~\ref{sec:OFM} below we ask what may be required of them.

Every statement below holds for arbitrary $m_t$ and $n_t$ and for loading matrices of any rank. What matters throughout is $\rank\bm\Phi_t$, the effective number of factors, against $\rank\bm Q_t$. The column count $m_t$ plays no role. The typical case in characteristics-based applications is $m_t\ll n_t$, a few characteristics against a large cross-section, and the case $m_t>n_t$ typical for overparameterized machine learning models is covered as well, see Lemma~\ref{lem:zeroloss} and Proposition~\ref{prop:minPhi} in Appendix~\ref{app:optimalfactors}.

The second moment matrix aggregates the supports of covariances and risk premia, $\Ima\bm Q_t=\Ima\bm\Sigma_t+\Ima\bm\mu_t$. This follows from $\bm Q_t=\bm\Sigma_t+\bm\mu_t\bm\mu_t^\top$ and the image identity $\Ima(\bm A+\bm B)=\Ima\bm A+\Ima\bm B$ for symmetric positive semidefinite matrices stated in Appendix~\ref{app:matrix}. In particular, $\bm\mu_t\in\Ima\bm Q_t$ always holds, and $\Ima\bm Q_t\supseteq\Ima\bm\Sigma_t$, with equality if and only if the law of one price holds. The \emph{law of one price} is the existence of a stochastic discount factor for the assets, and Proposition~\ref{prop:LOPmain} shows that it holds if and only if
\begin{equation}\label{eq:LoP}
\bm\mu_t\in\Ima\bm\Sigma_t.
\end{equation}
We use \eqref{eq:LoP} as the working form of the law of one price throughout.

Representation \eqref{eq:linfacmodelPhi} carries no intercept, unlike the more familiar formulation of \citet{gib_ros_sha_89}. We prefer it because it leaves the behaviour of $\E_t[\bm\epsilon_{t+1}]$ to be examined rather than normalized away. Section~\ref{sec:maintheorem} relates the two formulations.

\subsection{Factors as portfolios}\label{sec:tradednontraded}
Two readings of \eqref{eq:linfacmodelPhi} coexist in the factor pricing literature. On the statistical reading, the right-hand side is a data-generating process in which general, possibly latent and non-traded, risk factors $\bm f_{t+1}$, together with idiosyncratic shocks $\bm\epsilon_{t+1}$, generate asset returns conditional on $\bm\Phi_t$. The loading matrix captures the conditional exposure of assets to common sources of risk, and Appendix~\ref{app:existence} shows that such representations exist under mild conditions on the prescribed factor and residual moments.

The second reading is economic, and it is the one this paper takes. Throughout the main text we assume that the factors are traded portfolios,
\begin{equation}\label{eq:factorportfolios}
\bm f_{t+1}=\bm W_t^\top\bm x_{t+1},
\end{equation}
for an $\Fcal_t$-measurable $n_t\times m_t$ portfolio weight matrix $\bm W_t$. We refer to such factors as factor portfolios, and to factors, loading matrices, and factor representations satisfying \eqref{eq:factorportfolios} as traded. In representation \eqref{eq:linfacmodelPhi}, we call $\bm\Phi_t\bm f_{t+1}$ the factor component of returns and, for factor portfolios, $\bm\Pi_t\coloneqq\bm\Phi_t\bm W_t^\top$ the factor component operator, as it maps returns to their factor component. Rewriting \eqref{eq:linfacmodelPhi} with \eqref{eq:factorportfolios}, the residuals can be expressed as $\bm\epsilon_{t+1}=(\bm I_{n_t}-\bm\Pi_t)\bm x_{t+1}$.

The restriction \eqref{eq:factorportfolios} is a choice of subject, rather than a technical convenience. It is where the factor representations of the empirical literature live, and it turns every statement of Section~\ref{sec:OFM} into a matrix identity in $(\bm\Phi_t,\bm W_t,\bm Q_t)$.
It also costs little, as in the appendices we consider general, possibly untraded factors, for which the theory applies as well with few exceptions that we are explicit about.

\subsection{Regression loadings and weighted least squares}\label{sec:weightedconstructions}

Empirical work builds the factors from a given loading matrix by weighted least squares, with specifications differing only in the weighting matrix. This section introduces two objects in this context: the loadings that a regression on given factors produces, and the factors that a weighting matrix produces from given loadings. Appendices~\ref{app:TSproblem} and~\ref{app:CSproblem} supply the corresponding optimization problems that the empirical literature usually starts from.

Given any factors $\bm f_{t+1}\in L^2_t$, the returns have a canonical set of loadings on them, namely the loadings of the time-series regression of $\bm x_{t+1}$ on $\bm f_{t+1}$. We write
\begin{equation}\label{eq:TSreg}
\bm\Phi^{\bm f_{t+1}}_t \coloneqq \E_t[\bm x_{t+1}\bm f_{t+1}^\top]\bm Q_{\bm f,t}^+
\end{equation}
for the \emph{time-series regression loading matrix}, so that $\bm\Phi^{\bm f_{t+1}}_t\bm f_{t+1}=P_t[\bm x_{t+1}\mid\bm f_{t+1}]$ is the componentwise conditional least-squares projection of the returns on the factors, by Proposition~\ref{prop:Sty}~\ref{it:Sty3} in Appendix~\ref{app:probsetup}. Appendix~\ref{app:optimalfactors} shows that $\bm\Phi^{\bm f_{t+1}}_t$ minimizes the weighted panel loss over loading matrices for a given factor.

Let $\bm S_t$ be an $\Fcal_t$-measurable square weight matrix satisfying
\begin{equation}\label{eq:NoNDS}
\Ima\bm Q_t\cap\ker \bm S_t=\{\bm 0\},
\end{equation}
which holds in particular if $\Ima\bm Q_t\subseteq \Ima\bm S_t^\top$. The \emph{weighted least-squares (WLS) factor portfolios} of a loading matrix $\bm\Phi_t$ are
\begin{equation}\label{eq:SLSfact}
 \bm f^{\bm\Phi_t}_{t+1}\coloneqq (\bm S_t\bm\Phi_t)^+\bm S_t \bm x_{t+1},
 \end{equation}
that is, \eqref{eq:factorportfolios} with $\bm W_t^\top = (\bm S_t\bm\Phi_t)^+\bm S_t$. We write $\bm\Pi^{\bm\Phi_t}_t\coloneqq\bm\Phi_t(\bm S_t\bm\Phi_t)^+\bm S_t$ for the corresponding factor component operator, which depends on the weighting as well as on the loadings, and
\begin{equation}\label{eq:resPhidef}
\bm\epsilon^{\bm\Phi_t}_{t+1} \coloneqq \bm x_{t+1} - \bm\Phi_t\bm f^{\bm\Phi_t}_{t+1} = (\bm I_{n_t} - \bm\Pi^{\bm\Phi_t}_t)\bm x_{t+1}.
\end{equation}
 Three choices of $\bm S_t$ recur. The unweighted case $\bm S_t=\bm I_{n_t}$ gives the \emph{OLS factor portfolios} $\bm f^{\bm\Phi_t}_{t+1}=\bm\Phi_t^+\bm x_{t+1}$, the \emph{GLS weighting} is $\bm S_t=\bm\Sigma_t^{+1/2}$, for which \eqref{eq:NoNDS} is equivalent to the law of one price \eqref{eq:LoP},\footnote{Here $\ker\bm S_t=\ker\bm\Sigma_t=(\Ima\bm\Sigma_t)^\perp$. If $\bm\mu_t\in\Ima\bm\Sigma_t$ then $\Ima\bm Q_t=\Ima\bm\Sigma_t$ meets $\ker\bm\Sigma_t$ trivially. If not, the component of $\bm\mu_t$ orthogonal to $\Ima\bm\Sigma_t$ is a nonzero element of $\Ima\bm Q_t\cap\ker\bm\Sigma_t$.}
 and the \emph{moment weighting} is $\bm S_t=\bm Q_t^{+1/2}$.
The next subsection records which members of this family the empirical literature uses.

\subsection{Factor representations in the literature}\label{sec:representations}
The leading characteristics-based and latent-factor representations of the empirical literature all build on the unweighted member of \eqref{eq:SLSfact}, $\bm S_t=\bm I_{n_t}$, and differ in how they choose $\Ima\bm\Phi_t$. Regressed PCA runs it at its sieve matrix, with a fixed compression after. We record each here as a pair $(\bm\Phi_t,\bm f_{t+1})$ in the sense of \eqref{eq:linfacmodelPhi}, and defer their discussion to Section~\ref{sec:leadingreps}. Throughout, these are statements about the population analogue of each estimating equation, not about the sampling properties of any estimator.

\emph{Instrumented principal components.} In the notation of \citet{kellypruittsu19}, loadings are instrumented as $\bm\beta_t=\bm Z_t\bm\Gamma_\beta$ with $\bm Z_t$ a matrix of characteristics, and their first-order condition for the factor realizations, their equation~(6), is $\bm f_{t+1}=(\bm\Gamma_\beta^\top\bm Z_t^\top\bm Z_t\bm\Gamma_\beta)^{-1}\bm\Gamma_\beta^\top\bm Z_t^\top\bm x_{t+1}$, which is \eqref{eq:SLSfact} for $\bm S_t=\bm I_{n_t}$ and $\bm\Phi_t=\bm Z_t\bm\Gamma_\beta$. The paper itself describes this choice as a period-by-period cross-sectional regression of returns on the instrumented loadings. The factor representation is the pair $\bigl(\bm Z_t\bm\Gamma_\beta,\ (\bm Z_t\bm\Gamma_\beta)^+\bm x_{t+1}\bigr)$.

\emph{Projected principal components.} \citet{fanliaowang16} write loadings as functions of observed covariates, expand those functions in a sieve basis $\bm\Phi(\bm X)$, and apply principal components to the projected panel $\bm P\bm Y$, where $\bm Y$ is the $n\times T$ panel of returns and $\bm P=\bm\Phi(\bm X)(\bm\Phi(\bm X)^\top\bm\Phi(\bm X))^{-1}\bm\Phi(\bm X)^\top$ is their equation~(2.6). Their loading estimate is $\widehat{\bm G}(\bm X)=T^{-1}\bm P\bm Y\widehat{\bm F}$, their equation~(2.7), and the factor extracted in a single period is the cross-sectional least-squares coefficient of that period's returns on $\widehat{\bm G}(\bm X)$, so that the factor representation is $\bigl(\widehat{\bm G}(\bm X),\ \widehat{\bm G}(\bm X)^+\bm x_{t+1}\bigr)$.\footnote{Equation numbers are those of the published version. The factor is not written in this form in \citet{fanliaowang16}. It follows from (2.6), (2.7) and the sample normalization $T^{-1}\widehat{\bm F}^\top\widehat{\bm F}=\bm I$ implicit in their definition of $\widehat{\bm F}$ as the scaled eigenvectors of $\bm Y^\top\bm P\bm Y$, which is the sample counterpart of their identification condition~(2.3). Two features distinguish this representation from the others: the covariates are time invariant, so the loading matrix carries no $t$, and the panel is not centered, so the object implicitly analyzed is a second moment matrix.}

\emph{Regressed PCA.} \citet{chenroussanovwang23} estimate a semiparametric conditional model in which the intercept and loading functions are sieve expansions of the characteristics. Writing $\bm S_t$ for the resulting $n_t\times J$ sieve loading matrix, their first step runs the period-by-period unweighted cross-sectional regression of returns on $\bm S_t$, so the Fama--MacBeth managed portfolios it produces, $\bm S_t^+\bm x_{t+1}$, are the OLS factor portfolios \eqref{eq:SLSfact} of the sieve loading matrix. Their second step extracts a time-invariant $J\times K$ matrix $\widehat{\bm B}$ with orthonormal columns by principal component analysis of the demeaned time series of these managed portfolios, together with an intercept, and reports the factor representation $\bigl(\bm S_t\widehat{\bm B},\ \widehat{\bm B}^\top\bm S_t^+\bm x_{t+1}\bigr)$. The loading span is thus confined to the sieve span of the characteristics, with the $K$-dimensional selection inside it read off the time-series second moments of the managed-portfolio panel, rather than $\bm Q_t$. Nonlinearity of the loading functions transfers, in our notation, into a larger loading matrix.

\emph{Heuristic characteristics-based factor portfolios.} \citet{kozaknagel23} study factor portfolios with weight matrix $\bm W_t=\bm X_t\bm S_t$ for a characteristics matrix $\bm X_t$ and a nonsingular $\bm S_t$, so that $\Ima\bm W_t=\Ima\bm X_t$ and the representation is again the unweighted one up to the reparameterization $\bm S_t$. The pair is $\bigl(\bm X_t,\ \bm X_t^+\bm x_{t+1}\bigr)$ up to that reparameterization, which moves the individual factors but not their span.

\emph{Risk-premium PCA.} \citet{lettaupelger20} estimate latent loadings $\widehat{\bm\Lambda}$ as the leading eigenvectors of $T^{-1}\bm X^\top\bm X+\gamma\bar{\bm x}\bar{\bm x}^\top$, where $\bm X$ is the raw, uncentered panel of returns, $\bar{\bm x}$ is the vector of sample means and $\gamma$ weights the mean. Their factors are $\widehat{\bm F}=\bm X\widehat{\bm\Lambda}(\widehat{\bm\Lambda}^\top\widehat{\bm\Lambda})^{-1}$, which is once more the OLS factor representation, $\bm f_{t+1}=\widehat{\bm\Lambda}^+\bm x_{t+1}$. Because the first term is the uncentered second moment matrix, the weighted matrix is $\widehat{\bm\Sigma}+(1+\gamma)\bar{\bm x}\bar{\bm x}^\top$ in terms of the sample covariance matrix $\widehat{\bm\Sigma}=T^{-1}\bm X^\top\bm X-\bar{\bm x}\bar{\bm x}^\top$, so that $\gamma$ interpolates through the sample second moment matrix: at $\gamma=-1$ it is $\widehat{\bm\Sigma}$ and the procedure is conventional PCA, the endpoint \citet{lettaupelger20} name explicitly, and at $\gamma=0$ it is exactly the sample analogue of $\bm Q_t$.\footnote{Their representation is unconditional, with one loading matrix estimated from the whole panel, so statements about it concern the unconditional second moment matrix rather than $\bm Q_t$.}

The five differ in the choice of $\Ima\bm\Phi_t$. Four confine it to a span fixed by firm characteristics, and are in this sense \emph{prespecified} without reference to the population moments of returns.
For the heuristic portfolios the loading matrix is prespecified outright, for instrumented principal components it is the instrument span $\Ima\bm Z_t$ that is, for projected principal components the sieve span $\Ima\bm\Phi(\bm X)$, and for regressed PCA the sieve span of the characteristics, with the loading matrix in each case then selected inside that span from the panel. Risk-premium PCA does not prespecify a loading matrix at all, but reads its column space off the second moments of the panel. Section~\ref{sec:rulesout} shows that this single difference decides whether the loadings each representation reports are the regression loadings of the factors it reports.
The smallest model in which that difference is visible is the market model, and it can be worked out by hand.

\begin{example}[The Market Model with a Characteristic]\label{ex:market}
Let returns be generated by a market model,
\[
\bm x_{t+1}=\bm\beta_t r^{\rm m}_{t+1}+\bm u_{t+1},
\]
with market excess return $r^{\rm m}_{t+1}$ with conditional mean $\mu_{{\rm m},t}>0$ and variance $\sigma^2_{{\rm m},t}$, market betas $\bm\beta_t\neq\bm 0$, and idiosyncratic shocks, with a scalar $\sigma_t>0$, $\E_t[\bm u_{t+1}]=\bm 0$ and $\cov_t[\bm u_{t+1}]=\sigma_t^2\bm I_{n_t}$, uncorrelated with the market. Then
\begin{equation}\label{eq:marketmoments}
\bm\mu_t=\bm\beta_t\mu_{{\rm m},t},\qquad
\bm Q_t=q_t\,\bm\beta_t\bm\beta_t^\top+\sigma_t^2\bm I_{n_t},\qquad
q_t\coloneqq\sigma^2_{{\rm m},t}+\mu^2_{{\rm m},t}.
\end{equation}
Suppose the betas are not observed and a single characteristic is reported as the loadings in their place, $\bm\Phi_t=\bm\varphi_t\neq\bm 0$, with the OLS factor portfolio $f_{t+1}=\bm\varphi_t^+\bm x_{t+1}$ of \eqref{eq:SLSfact} for $\bm S_t=\bm I_{n_t}$. From \eqref{eq:marketmoments},
\[
\bm Q_t\bm\varphi_t=q_t(\bm\beta_t^\top\bm\varphi_t)\,\bm\beta_t+\sigma_t^2\bm\varphi_t,
\]
so $\bm Q_t\bm\varphi_t$ remains a multiple of $\bm\varphi_t$, equivalently $\bm\varphi_t$ is an eigenvector of $\bm Q_t$, if and only if
\begin{equation}\label{eq:marketaccident}
\bm\beta_t^\top\bm\varphi_t=0
\qquad\text{or}\qquad
\bm\varphi_t\in\spn\{\bm\beta_t\}.
\end{equation}
Outside these two cases the regression loading vector \eqref{eq:TSreg} of the reported factor,
\[
\bm\Phi^{f_{t+1}}_t
=\frac{\|\bm\varphi_t\|_2^2\bigl(q_t(\bm\beta_t^\top\bm\varphi_t)\bm\beta_t+\sigma_t^2\bm\varphi_t\bigr)}
{q_t(\bm\beta_t^\top\bm\varphi_t)^2+\sigma_t^2\|\bm\varphi_t\|_2^2},
\]
is a combination of $\bm\varphi_t$ and $\bm\beta_t$ carrying nonzero weight on $\bm\beta_t$, rather than $\bm\varphi_t$ itself, and the distance between the two is
\begin{equation}\label{eq:marketgap}
\frac{\bigl\|\bm\Phi^{f_{t+1}}_t-\bm\varphi_t\bigr\|_2}{\|\bm\varphi_t\|_2}
=\frac{s_t\,|\rho_t|\sqrt{1-\rho_t^2}}{s_t\rho_t^2+1},
\qquad
\rho_t\coloneqq\frac{\bm\beta_t^\top\bm\varphi_t}{\|\bm\beta_t\|_2\|\bm\varphi_t\|_2},
\qquad
s_t\coloneqq\frac{q_t\|\bm\beta_t\|_2^2}{\sigma_t^2},
\end{equation}
in terms of the cosine $\rho_t$ between the characteristic and the betas and the strength $s_t$ of the market factor.
\end{example}

The two cases \eqref{eq:marketaccident} are the two uninteresting ones: a characteristic orthogonal to the betas carries no exposure to the only priced source of risk here, and a characteristic collinear with them is the market beta under another name. The case the empirical literature is concerned with, a characteristic correlated with the betas without being a function of them, which is what size, value and momentum are, is neither. The discrepancy \eqref{eq:marketgap} vanishes at $|\rho_t|\in\{0,1\}$, is strictly positive between, and increases in $s_t$: the better the market explains returns, the further a characteristic sits from the beta it is reported as. Section~\ref{sec:rulesout} shows that \eqref{eq:marketaccident} is an instance of the general criterion, and Section~\ref{sec:empirical} measures the same distance on data.

The second example is the one the paper carries throughout. It is a conditional specification widely used in the empirical literature, and it is built so that the coincidence which fails in Example~\ref{ex:market}, $\bm Q_t$-invariance of $\Ima\bm\Phi_t$, holds instead.

\begin{example}[Running Example]\label{ex:running}
The following specializes the conditional factor pricing setting of \citet{kozaknagel23}: the mean equation is their Assumption~1, that expected returns are linear in characteristics, and the covariance equation is their equation~(19) with the non-characteristic factor block set to zero and the isotropic variance allowed to depend on $t$. Given an $\Fcal_t$-measurable $n_t\times m_t$ loading matrix $\bm\Phi_t$ of asset characteristics, an $\Fcal_t$-measurable $m_t$-vector $\bm b_t$, a symmetric positive semidefinite $\Fcal_t$-measurable $m_t\times m_t$ matrix $\bm C_t$, and a scalar $\sigma_t>0$, let
\begin{equation}\label{eq:running}
\bm\mu_t=\bm\Phi_t\bm b_t,\qquad \bm\Sigma_t=\bm\Phi_t\bm C_t\bm\Phi_t^\top+\sigma_t^2\bm I_{n_t}.
\end{equation}
Risk premia are spanned by the loadings, factor risk enters through $\bm\Phi_t\bm C_t\bm\Phi_t^\top$, and the structural noise is isotropic with the full-rank covariance matrix $\sigma_t^2\bm I_{n_t}$. The second moment matrix $\bm Q_t=\bm\Phi_t(\bm C_t+\bm b_t\bm b_t^\top)\bm\Phi_t^\top+\sigma_t^2\bm I_{n_t}$ is invertible, as is $\bm\Sigma_t=\bm\Phi_t\bm C_t\bm\Phi_t^\top+\sigma_t^2\bm I_{n_t}$, and the law of one price \eqref{eq:LoP} holds for the latter reason.
\end{example}

The example is continued twice below, in Sections~\ref{sec:maintheorem} and~\ref{sec:impossible}, as ``Example~\ref{ex:running}, continued'', and returns in the guidelines of Section~\ref{sec:guidelines}.

\subsection{The law of one price and factor spanning}\label{sec:pricingbenchmarks}
This section collects the conditional asset pricing apparatus the rest of the paper uses, building on \citet{han_ric_87}. Appendix~\ref{app:CAP} contains the full statements and proofs. The law of one price, in its form \eqref{eq:LoP}, is the only pricing assumption this paper ever uses. It is strictly weaker than absence of arbitrage, as Example~\ref{ex:LOOPnoarb} in Appendix~\ref{app:CAP} shows.

We denote by $\Scal_t(\bm x_{t+1})$ the space of returns attainable through conditional portfolios formed from $\bm x_{t+1}$, and by $\Scal(\bm x_{t+1})\subseteq\Scal_t(\bm x_{t+1})$ its square-integrable subspace, see Proposition~\ref{prop:Sty} in Appendix~\ref{app:probsetup}. For $X\in\Scal_t(\bm x_{t+1})$ the conditional Sharpe ratio is $\text{SR}_t(X)\coloneqq\E_t[X]/\sqrt{\var_t[X]}$, with the convention $0/0=0$, and for $X\in\Scal(\bm x_{t+1})$ the unconditional Sharpe ratio is $\text{SR}(X)\coloneqq\E[X]/\sqrt{\var[X]}$. The maximal Sharpe ratios are\footnote{The essential supremum $\esssup$ takes the best over all portfolios separately in each state at time $t$, ignoring events of probability zero: the tightest ceiling that no attainable Sharpe ratio exceeds with positive probability.}
\[
\text{SR}_t^\star\coloneqq\esssup_{X\in\Scal_t(\bm x_{t+1})}\text{SR}_t(X)
\qquad\text{and}\qquad
\text{SR}^\star\coloneqq\sup_{X\in\Scal(\bm x_{t+1})}\text{SR}(X).
\]
A portfolio $X$ is conditionally mean--variance efficient (CMVE) if $\text{SR}_t(X)=\text{SR}_t^\star$, and unconditionally mean--variance efficient (UMVE) if $\text{SR}(X)=\text{SR}^\star$.

A central object is the conditional least-squares projection of the constant $1$ on the returns,
\begin{equation}\label{eq:defXstar}
X^\star_{t+1}\coloneqq P_t[1\mid\bm x_{t+1}]=\bm\mu_t^\top\bm Q_t^+\bm x_{t+1},
\end{equation}
where $Y\mapsto P_t[Y\mid\bm x_{t+1}]\coloneqq\E_t[Y\bm x_{t+1}^\top]\bm Q_t^+\bm x_{t+1}$ denotes the conditional least-squares projector onto $\Scal_t(\bm x_{t+1})$ given in Proposition~\ref{prop:Sty}. We emphasize that $X^\star_{t+1}$ is a $\bm Q_t$-object whose definition requires no invertibility and no pricing assumption. We further set $M^\star_{t+1}\coloneqq 1-X^\star_{t+1}$. The following proposition states the headline equivalences characterizing the law of one price. The full list, including an extended Sherman--Morrison formula for $\bm Q_t^+$, is Proposition~\ref{prop:LOP} in Appendix~\ref{app:CAP}.

\begin{proposition}[Law of One Price]\label{prop:LOPmain}
The following conditions are equivalent:
\begin{enumerate}
\item\label{it:LOPmain1} There exists a stochastic discount factor $M_{t+1}\in L^2_t$ satisfying $\E_t[M_{t+1}]>0$ and $\E_t[M_{t+1}\bm x_{t+1}]=\bm 0$.
\item\label{it:LOPmain2} Conditional Sharpe ratios are uniformly bounded, $\text{SR}_t^\star<\infty$.
\item\label{it:LOPmain3} Asset risk premia are spanned by return covariances, \eqref{eq:LoP}.
\item\label{it:LOPmain4} $M^\star_{t+1}$ is a stochastic discount factor.
\end{enumerate}
\end{proposition}

Under the law of one price \eqref{eq:LoP}, the extended Sherman--Morrison formula gives $\bm\mu_t^\top\bm Q_t^+=(1+\bm\mu_t^\top\bm\Sigma_t^+\bm\mu_t)^{-1}\bm\mu_t^\top\bm\Sigma_t^+$, so that, when $\bm\mu_t\neq\bm 0$, $X^\star_{t+1}$ is the unique CMVE portfolio up to a positive conditional scaling, with maximal conditional Sharpe ratio $\text{SR}_t^\star=\sqrt{\bm\mu_t^\top\bm\Sigma_t^+\bm\mu_t}$. It is also the unique UMVE portfolio up to a positive constant scaling, see Proposition~\ref{prop:Xstar} in Appendix~\ref{app:CAP}.

Given factor portfolios \eqref{eq:factorportfolios}, we denote by $\Scal_t(\bm f_{t+1})\subseteq\Scal_t(\bm x_{t+1})$ the subspace of returns attainable by conditional trading in the factor portfolios, and by $\Scal(\bm f_{t+1})\subseteq\Scal(\bm x_{t+1})$ its square-integrable subspace, with maximal Sharpe ratios $\text{SR}_{\bm f,t}^\star$ and $\text{SR}_{\bm f}^\star$ defined accordingly. By construction, $\text{SR}_{\bm f,t}^\star\le\text{SR}_t^\star$ and $\text{SR}_{\bm f}^\star\le\text{SR}^\star$. Following \citet{kozaknagel23}, we say that factor spanning holds if
\begin{equation}\label{eq:maxSR}
\text{SR}_{\bm f,t}^\star=\text{SR}_t^\star.
\end{equation}
The following proposition characterizes factor spanning.

\begin{proposition}[Characterization of Factor Spanning]\label{prop:span}
Assume the law of one price \eqref{eq:LoP} holds. Then the following conditions are equivalent:
\begin{enumerate}
\item\label{it:span1} Factor spanning \eqref{eq:maxSR} holds.
\item\label{it:span2} Asset risk premia are spanned by the return--factor covariance $\cov_t[\bm x_{t+1},\bm f_{t+1}]=\bm\Sigma_t\bm W_t$,
\begin{equation}\label{eq:spanNEWeq1}
\bm\mu_t\in\Ima(\bm\Sigma_t\bm W_t).
\end{equation}
\item\label{it:span3} Factors span the UMVE portfolio \eqref{eq:defXstar},
\begin{equation}\label{eq:spanNEWeqX}
P_t[1\mid\bm x_{t+1}]=P_t[1\mid\bm f_{t+1}].
\end{equation}
\item\label{it:span4} Factor spanning holds for unconditional Sharpe ratios, $\text{SR}_{\bm f}^\star=\text{SR}^\star$.
\end{enumerate}
\end{proposition}

The equivalence of \ref{it:span1} and \ref{it:span4} justifies the unconditional evaluation of conditional models, which we exploit in Section~\ref{sec:empirical}.

\section{Coherent Factor Representations}\label{sec:OFM}
This section takes the four properties of the introduction in turn: \ref{it:prop1} the loadings are the regression loadings of the factors, \ref{it:prop2} residual risk is unpriced, \ref{it:prop3} the factors span the maximal conditional Sharpe ratio, and \ref{it:prop4} the residuals are idiosyncratic, uncorrelated with the factors with a second moment matrix of full rank.
We refer to a factor representation with properties \ref{it:prop1} and \ref{it:prop2} as \emph{coherent}.

\subsection{The loadings are the regression loadings}\label{sec:rigidity}
To lead into property~\ref{it:prop1}, start with common empirical practice. A researcher specifies a matrix of firm characteristics as the loadings, forms factors from it by a cross-sectional regression, and reads betas off the loading matrix: this stock has a loading of $1.2$ on the value factor. The number is then used as a beta, in variance decompositions, in risk attributions, in the intercept of a pricing test. Yet it comes from a balance sheet, not from a regression of returns on the factors, and whether the two coincide is a question about the second moments of returns. Property~\ref{it:prop1} asks that they do.

Formally, the property that the loadings are the regression loadings requires
\begin{equation}\label{eq:coh1}
\bm\Phi_t=\bm\Phi_t^{\bm f_{t+1}},
\end{equation}
which implies that the factor component is the conditional least-squares projection of returns on the factors,
\begin{equation}\label{eq:CACproj}
\bm\Phi_t\bm f_{t+1}=P_t[\bm x_{t+1}\mid\bm f_{t+1}].
\end{equation}
The two statements are equivalent when the factors are linearly independent in $L^2_t$. In general \eqref{eq:coh1} is the stronger one, as it also fixes $\bm\Phi_t$ on the null directions of the factors (Lemma~\ref{lem:FOCf} in Appendix~\ref{app:optimalfactors}). It is a second-moment consistency requirement, as the prespecified loadings must be the regression loadings they claim to be. It is what \citet{kellypruittsu19} impose implicitly by reporting characteristics as loadings, and for a coherent factor representation Proposition~\ref{prop:tradeq} below shows that it is the precise sense in which characteristics are covariances. It also says, by Lemma~\ref{lem:TSregral0}, that the loadings minimize the weighted panel loss given the factors, simultaneously for every $\bm S_t$.

Attaining it requires the loadings to be consistent with the support of returns, which by Remark~\ref{rem:support} is contained in $\Ima\bm Q_t$. Denote by
\begin{equation}\label{eq:Phistarass}
\Ima\bm\Phi_t\subseteq\Ima\bm Q_t
\end{equation}
the \emph{support condition}. It is exactly what \eqref{eq:coh1} forces, since $\Ima\bm\Phi^{\bm f_{t+1}}_t\subseteq\Ima\bm Q_t$ always, and it is the hypothesis of the result below.

Property~\ref{it:prop1} is attainable in exactly one way. The theorem says that traded factors carrying $\bm\Phi_t$ as their own regression loadings exist \ref{it:rigid1}, presents them in closed form and shows they are unique \ref{it:rigid2}, and then describes them: as a projection \ref{it:rigid3}, and through their rank in \eqref{eq:rankinherit}.

\begin{theorem}[Existence and Uniqueness]\label{thm:rigid}
Let $\bm\Phi_t$ be an $\Fcal_t$-measurable $n_t\times m_t$ loading matrix. Traded factors with
\begin{equation}\label{eq:coh1restate}
\bm\Phi_t=\bm\Phi_t^{\bm f_{t+1}}
\end{equation}
exist if and only if the support condition \eqref{eq:Phistarass} holds. In that case, for traded factors $\bm f_{t+1}=\bm W_t^\top\bm x_{t+1}$ with $\Ima\bm W_t\subseteq\Ima\bm Q_t$, conditions \ref{it:rigid1} and \ref{it:rigid2} below are equivalent, and each of them implies \ref{it:rigid3}. If moreover $\rank\bm\Phi_t=m_t$, all three are equivalent.
\begin{enumerate}
\item\label{it:rigid1} The loadings are the regression loadings of the factors, \eqref{eq:coh1restate}.

\item\label{it:rigid2} The weight matrix is
\begin{equation}\label{eq:MWfactors}
\bm W_t^\top=(\bm\Phi_t^\top\bm Q_t^+\bm\Phi_t)^+\bm\Phi_t^\top\bm Q_t^+.
\end{equation}

\item\label{it:rigid3} The factor component operator $\bm\Pi_t=\bm\Phi_t\bm W_t^\top$ is the projection onto $\Ima\bm\Phi_t$ that is orthogonal for the semi-inner product
\begin{equation}\label{eq:Qplusinnercrit}
\langle\bm u,\bm v\rangle_{\bm Q^+}\coloneqq\bm u^\top\bm Q_t^+\bm v,
\end{equation}
an inner product on $\Ima\bm Q_t$ and degenerate off it.
\end{enumerate}
We call $\bm f_{t+1}=\bm W_t^\top\bm x_{t+1}$ with \eqref{eq:MWfactors} the \emph{moment-weighted factor portfolios}. They satisfy
\begin{equation}\label{eq:rankinherit}
\rank\bm Q_{\bm f,t}=\rank\bm W_t=\rank\bm\Phi_t,
\end{equation}
so the factors inherit the rank of the loading matrix. In particular they are linearly independent in $L^2_t$, equivalently $\bm Q_{\bm f,t}$ is nonsingular, if and only if $\rank\bm\Phi_t=m_t$. Moreover $\bm W_t^\top\bm\Phi_t$ is the orthogonal projection onto $\Ima\bm Q_{\bm f,t}$, so that $\bm W_t^\top\bm\Phi_t=\bm I_{m_t}$ if and only if $\rank\bm\Phi_t=m_t$.
\end{theorem}

Condition \ref{it:rigid3} constrains only the factor component operator $\bm\Pi_t$, and $\bm\Pi_t$ determines the weight matrix only when the loading matrix carries no redundant columns. For $\rank\bm\Phi_t<m_t$ a whole family of weight matrices produces the same projection, and condition \ref{it:rigid1} selects \eqref{eq:MWfactors} among them.

Part \ref{it:rigid3} is the most intuitive of the equivalent properties. The second moments of returns supply their own notion of distance and angle on the asset span, and in that geometry the factor component is nothing but the orthogonal projection of returns onto the span of the loadings. Uniqueness is then no accident, because the projection onto a given subspace is unique. Appendix~\ref{app:bijection} shows that weight matrices and regression loading matrices are two coordinate systems for one and the same object.

The weight matrix in \eqref{eq:MWfactors} coincides with the weighted least-squares representation \eqref{eq:SLSfact} for the weighting $\bm S_t=\bm Q_t^{+1/2}$,
\begin{equation}\label{eq:MWnosqrt}
(\bm Q_t^{+1/2}\bm\Phi_t)^+\bm Q_t^{+1/2}=(\bm\Phi_t^\top\bm Q_t^+\bm\Phi_t)^+\bm\Phi_t^\top\bm Q_t^+,
\end{equation}
an identity valid for every $\bm\Phi_t$ and $\bm Q_t$ without rank conditions, by $\bm A^+=(\bm A^\top\bm A)^+\bm A^\top$ applied to $\bm A=\bm Q_t^{+1/2}\bm\Phi_t$. The right-hand side involves no matrix square root and is the form we use. The economic point is that the moment weighting is not a modeling preference to be defended against alternatives. It is the unique weighting whose factor portfolios can carry a prespecified loading matrix as their own regression loadings, and Appendix~\ref{app:optimalfactors} shows what the alternatives deliver instead.

Parts~\ref{it:rigid3} and~\eqref{eq:rankinherit} discharge the claims that Remark~\ref{rem:onlyimage} makes for traded factors: the factor component operator, the factor component, the residuals and the support condition \eqref{eq:Phistarass} all depend on $\bm\Phi_t$ only through $\Ima\bm\Phi_t$, and the nonsingularity of $\bm Q_{\bm f,t}$ assumed there is exactly $\rank\bm\Phi_t=m_t$. The loading matrix fixes only the coordinates in which the factors are expressed.\footnote{Requirement \eqref{eq:coh1} cannot be weakened to factor--residual orthogonality, $\E_t[\bm\epsilon_{t+1}\bm f_{t+1}^\top]=\bm 0$, without destroying uniqueness. Under the support condition the traded solutions of $\E_t[\bm\epsilon_{t+1}\bm f_{t+1}^\top]=\bm 0$ form a family indexed by the symmetric $\{2\}$-inverses of $\bm\Phi_t^\top\bm Q_t^+\bm\Phi_t$, containing the null factors, and \eqref{eq:coh1} selects one member. See Remark~\ref{rem:twoinverse} in Appendix~\ref{app:proofs}.}

\begin{remark}[Only the Column Space Matters]\label{rem:onlyimage}
Coherence is a property of the pair, but it does not depend on the coordinates in which the factors are expressed. If $\bm Q_{\bm f,t}$ is nonsingular, then for any invertible $\Fcal_t$-measurable $m_t\times m_t$ matrix $\bm A_t$ the rotated pair $(\bm\Phi_t\bm A_t,\bm A_t^{-1}\bm f_{t+1})$ has the same factor component and the same residuals, and is coherent if and only if $(\bm\Phi_t,\bm f_{t+1})$ is, because $\bm\Phi_t^{\bm A_t^{-1}\bm f_{t+1}}=\bm\Phi_t^{\bm f_{t+1}}\bm A_t$. When $\bm Q_{\bm f,t}$ is singular the equality can fail, since the pseudoinverse in \eqref{eq:TSreg} does not transform covariantly, and \eqref{eq:coh1} then selects the minimum-norm regression loadings, in line with the pseudoinverse convention used throughout.

For traded factors the invariance is sharper, and Theorem~\ref{thm:rigid} establishes it, as the paragraph preceding this remark records. Every economic notion in this paper, pricing, spanning, Sharpe ratios, residual moments, alphas, is therefore a statement about a subspace of $\R^{n_t}$ carrying at most $\rank\bm Q_t$ dimensions, and the loading matrix is a chart for it.
\end{remark}

The moment-weighted factor portfolios have a classical antecedent. Given non-traded factors $\bm g_{t+1}$, the factor-mimicking portfolio has weight matrix $\bm W_t=\bm Q_t^+\E_t[\bm x_{t+1}\bm g_{t+1}^\top]$, as in \citet{hub_kan_sta_87} and \citet{leh_mod_88} and the maximum-correlation portfolios of \citet{bre_gib_lit_89}, see \citet{admatipfleiderer85} for the arbitrage pricing reading and \citet{bal_rob_08} for the associated pricing tests. In the notation of this paper (in particular Appendix~\ref{app:probsetup}), the mimicking portfolios are the conditional least-squares projection of the factors onto the asset span,
\begin{equation}\label{eq:mimicproj}
\bm W_t^\top\bm x_{t+1}=\E_t[\bm g_{t+1}\bm x_{t+1}^\top]\bm Q_t^+\bm x_{t+1}=P_t[\bm g_{t+1}\mid\bm x_{t+1}],
\end{equation}
componentwise, by \eqref{eq:Styeq2}. Taking $\bm g_{t+1}=1$ makes the efficient portfolio \eqref{eq:defXstar} the mimicking portfolio of the constant. Since $\E_t[\bm x_{t+1}\bm g_{t+1}^\top]=\bm\Phi_t^{\bm g_{t+1}}\bm Q_{\bm g,t}$, the weights see the factors only through their regression loadings and the scaling $\bm Q_{\bm g,t}$, and $\Ima\bm W_t=\bm Q_t^+\Ima\bm\Phi_t^{\bm g_{t+1}}$. Theorem~\ref{thm:rigid}~\ref{it:rigid2} identifies this span with the one moment weighting builds from $\bm\Phi_t^{\bm g_{t+1}}$, so both routes give the same portfolios and differ only in scale: mimicking inherits that of $\bm g_{t+1}$, whereas \eqref{eq:MWfactors} picks the scale that turns the prespecified loadings into the regression loadings. The two literatures ask different questions of the same object. Mimicking begins with a non-traded factor and asks what survives its replacement by a traded one. We begin with a loading matrix, with no factor to mimic, and ask which factors fit it. Theorem~\ref{thm:rigid} says exactly one does, so the mimicking portfolio, scaled this way, is not a convenient substitute but the only traded factor whose regression loadings are $\bm\Phi_t$, and Corollary~\ref{cor:premium} says when it prices as well.

Failing \eqref{eq:coh1} has three costs. First, reported loadings are not the regression loadings: variance decompositions and risk attributions are computed against a matrix that is not the loading matrix of any regression on the reported factors. Second, the factor component is not the projection it is taken to be: by \eqref{eq:CACproj}, the variance a factor representation attributes to its factors is then not the variance those factors explain, and by Lemma~\ref{lem:TSregral0} the reported loadings need not minimize the weighted panel loss for the reported factors, for any weighting, though when $\bm Q_{\bm f,t}$ is singular that minimizer is not unique and the reported loadings may still be one of them. Third, alpha tests detach from the reported loadings: a test in the tradition of \citet{gib_ros_sha_89} examines the intercept of the regression on the reported factors, whose loading matrix is $\bm\beta^{\bm f_{t+1}}_t$, defined in Section~\ref{sec:alphatests}. The test keeps its warrant as a test of whether those factors price, by Proposition~\ref{prop:tradeq}, but if $\bm\Phi_t\neq\bm\beta^{\bm f_{t+1}}_t$ a zero intercept certifies pricing against loadings the representation does not report, and the reported pair can leave $\E_t[\bm x_{t+1}-\bm\Phi_t\bm f_{t+1}]\neq\bm 0$ while the alpha vanishes. Section~\ref{sec:alphatests} shows that this cost reverses under \eqref{eq:coh1}.

\subsection{Residual risk is unpriced}\label{sec:maintheorem}
Property~\ref{it:prop1} concerns second moments alone and imposes no restriction on the compensation investors receive for bearing risk. The second introduces pricing. A factor model asserts that a small number of common sources of risk account for that compensation, and that the remainder of a stock's return, the component not spanned by its loadings, is risk for which investors are not compensated. Property~\ref{it:prop2} states this assertion formally. It restricts the loading matrix, and it does so independently of the first: the factors can transmit only those premia the loadings span, so residual risk is unpriced only if the risk premia lie in the column space of the loadings, and, for the factors of Theorem~\ref{thm:rigid}, exactly then.

The property is that residual risk carries no premium,
\begin{equation}\label{eq:coh2}
\E_t[\bm\epsilon_{t+1}]=\bm 0.
\end{equation}
It is a pricing hypothesis: the factors carry all risk premia. By Lemma~\ref{lem:FOC1} it holds if and only if asset risk premia are spanned by the factors, and it therefore forces the loading matrix to span them. Denote by
\begin{equation}\label{eq:arpspanning}
\bm\mu_t\in\Ima \bm\Phi_t
\end{equation}
the \emph{risk-premium condition}. A loading matrix satisfying both the support condition \eqref{eq:Phistarass} and the risk-premium condition \eqref{eq:arpspanning} is called \emph{admissible}. The two together read
\begin{equation}\label{eq:admissible}
\spn\{\bm\mu_t\}\subseteq\Ima\bm\Phi_t\subseteq\Ima\bm Q_t.
\end{equation}

Theorem~\ref{thm:rigid} settles \eqref{eq:coh1} without reference to pricing. What \eqref{eq:coh2} adds is a second, independent restriction on the loading matrix.

\begin{corollary}[Admissibility]\label{cor:premium}
Let $\bm\Phi_t$ satisfy the support condition \eqref{eq:Phistarass}, so that the factors \eqref{eq:MWfactors} are defined. The representation $(\bm\Phi_t,\bm f_{t+1})$ has unpriced residual risk, \eqref{eq:coh2}, if and only if $\bm\Phi_t$ is admissible, \eqref{eq:admissible}.
\end{corollary}

By Theorem~\ref{thm:rigid}\ref{it:rigid1}, Corollary~\ref{cor:premium} and Lemma~\ref{lem:FOC1}, a loading matrix admits a coherent factor representation if and only if it satisfies \eqref{eq:admissible}, the factors realizing it are \eqref{eq:MWfactors}, and under \eqref{eq:LoP} they are CMVE. Four aspects deserve emphasis.

\begin{figure}
\centering
\begin{tikzpicture}[x=1cm,y=1cm,scale=0.84, every node/.style={font=\small}]

\path (-7.9,-0.7) (7.9,8.8);
\draw[gray!55] (0,0) -- (-6.2,4) -- (0,8) -- (6.2,4) -- cycle;

\draw[gray!80, dashed] (0,0) -- (-3.2,3) -- (0,6) -- (3.2,3) -- cycle;

\fill[black!12] (0,1) -- (-2.0,3.5) -- (0,6) -- (2.0,3.5) -- cycle;
\draw[thick] (0,1) -- (-2.0,3.5) -- (0,6) -- (2.0,3.5) -- cycle;

\draw[->,gray!70] (-7.4,-0.4) -- (-7.4,8.5);
\node[gray!70, above, font=\scriptsize] at (-7.4,8.5) {$\rank$};
\foreach \y/\t in {0/0, 1/1, 3.2/{\rank\bm\Phi_t}, 6/{\rank\bm Q_t}, 8/{n_t}}
  {\draw[gray!70] (-7.55,\y) -- (-7.25,\y);
   \node[gray!70, left, font=\scriptsize] at (-7.6,\y) {$\t$};}

\node[circle,fill=black,inner sep=1.5pt] at (0,8) {};
\node[above, font=\scriptsize] at (0,8.12) {$\R^{n_t}$};
\node[circle,fill=black,inner sep=1.5pt] at (0,0) {};
\node[below, font=\scriptsize] at (0,-0.12) {$\{\bm 0\}$};

\node[circle,fill=black,inner sep=2pt] at (0,6) {};
\node[right, font=\scriptsize, fill=white, inner sep=1.5pt] at (0.22,6.42) {$\Ima\bm Q_t$: saturated, $\bm\Phi_t=\bm Q_t\bm Q_t^+$};
\node[circle,fill=black,inner sep=2pt] at (0,1) {};
\node[right, font=\scriptsize, fill=white, inner sep=1.5pt] at (0.22,0.62) {$\spn\{\bm\mu_t\}$: CMVE one-factor};

\node[circle,fill=black,inner sep=1.7pt] (adm)    at (-0.9,3.2) {};
\node[circle,draw,fill=white,inner sep=1.5pt] (nospan) at (-2.5,3.2) {};
\node[circle,draw,fill=white,inner sep=1.5pt] (nosupp) at (4.1,3.2) {};
\node[circle,fill=black,inner sep=1.3pt] (proj)   at (2.35,3.2) {};

\node[above, font=\scriptsize, align=center] at (-0.9,3.36) {$\Ima\bm\Phi_t$\\[-2pt]admissible};
\node[font=\scriptsize, align=center, fill=white, inner sep=1.5pt] at (-3.75,2.42)
  {$\bm\mu_t\notin\Ima\bm\Phi_t$\\[-2pt]\eqref{eq:arpspanning} fails};
\node[font=\scriptsize, align=center] at (4.45,2.10)
  {$\Ima\bm\Phi_t\not\subseteq\Ima\bm Q_t$\\[-2pt]\eqref{eq:Phistarass} fails};
\draw[gray!70, thin] (nospan) -- (-3.6,2.78);
\draw[gray!70, thin] (nosupp) -- (4.45,2.46);

\draw[->,>=stealth,dashed] (nosupp) -- (proj);
\node[font=\scriptsize, above, inner sep=1.5pt] at (3.25,3.28) {loading gap};
\node[font=\scriptsize, below, inner sep=2.5pt] at (2.35,3.10) {$\bm Q_t\bm Q_t^+\bm\Phi_t$};

\node[draw, rounded corners, font=\scriptsize, align=center, fill=white, inner sep=3pt]
  (rig) at (-4.65,0.95) {Theorem~\ref{thm:rigid}:\\[-2pt]one such $\bm\Pi_t$};
\draw[->,>=stealth] (rig) -- (adm);

\end{tikzpicture}
\caption{Loading matrices satisfying \eqref{eq:admissible}. Points are subspaces of $\R^{n_t}$ ordered by inclusion, with $\rank$ on the vertical axis and horizontal position schematic. The outer diamond is the lattice of all subspaces, narrowing to $\{\bm 0\}$ and $\R^{n_t}$ at its two vertices. The dashed diamond is the lattice of subspaces of $\Ima\bm Q_t$. The shaded region is the interval $\spn\{\bm\mu_t\}\subseteq\Ima\bm\Phi_t\subseteq\Ima\bm Q_t$, its two endpoints labelled. Four loading matrices are drawn at the same rank: one admissible, one failing the risk-premium condition \eqref{eq:arpspanning}, one failing the support condition \eqref{eq:Phistarass}, and the projection $\bm Q_t\bm Q_t^+\bm\Phi_t$ of the last. The dashed arrow joining these two is labelled by the loading gap of Corollary~\ref{cor:cohgap}. The projection restores the support condition but not, in general, the risk-premium condition, so the projected matrix may still sit outside the shaded interval. The box records Theorem~\ref{thm:rigid}: on each loading matrix whose column space lies in the interval, there is exactly one coherent traded factor representation.}
\label{fig:interval}
\end{figure}

First, condition \eqref{eq:admissible} depends on the loading matrix only through its column space, which it confines between $\spn\{\bm\mu_t\}$ and $\Ima\bm Q_t$, and such subspaces always exist. Within these bounds the rank of $\bm\Phi_t$ measures parsimony, and the paper asks whether a loading matrix chosen on economic grounds, typically firm characteristics, lies strictly inside rather than at an endpoint. Figure~\ref{fig:interval} illustrates the geometry. The support condition is close to a normalization but not free of content, and Corollary~\ref{cor:cohgap} quantifies the difference.

Second, \eqref{eq:admissible} disciplines specification. It restricts the conditional mean, the loading matrix and the second moment matrix jointly, so these cannot be specified independently. Its substance is that the loading matrix must span the risk premia while remaining consistent with the support of returns. When $\bm Q_t$ has full rank the support condition is vacuous and \eqref{eq:admissible} reduces to \eqref{eq:arpspanning}. In conditional applications $n_t$ is large and estimated second moment matrices are routinely of reduced rank, as in the empirical Section~\ref{sec:datadesign}.

Third, if \eqref{eq:admissible} fails, no choice of factors delivers coherence, and not merely no choice of traded factors, since the necessity argument in Appendix~\ref{app:proofs} applies to general factors. Condition \eqref{eq:admissible} is the exact dividing line, rather than a regularity condition.

Fourth, the law of one price is used in this section only where a statement concerns portfolios: the spanning payoff of Proposition~\ref{prop:span}, the tradability equivalence of Proposition~\ref{prop:tradeq}, and the exclusion of degenerate traded factors below. Characterization \eqref{eq:admissible} and uncorrelatedness hold without it. Example~\ref{ex:sccounter} shows what fails when it is dropped.

\begin{example}[continues=ex:running]\label{ex:runningmain}
In Example~\ref{ex:running}, $\bm\mu_t=\bm\Phi_t\bm b_t\in\Ima\bm\Phi_t$, and the support condition \eqref{eq:Phistarass} holds, as $\bm Q_t$ is invertible and thus $\Ima\bm Q_t=\R^{n_t}$. The loading matrix is therefore admissible, for any arbitrary $\bm\Phi_t$. The interval \eqref{eq:admissible} is the full lattice above $\spn\{\bm\Phi_t\bm b_t\}$, so the specification places no restriction on the loadings beyond those it imposes by construction. Under the law of one price \eqref{eq:LoP}, which holds here as $\bm\Sigma_t$ is invertible, the coherent traded factor representation is CMVE.
\end{example}

\subsubsection{Coherent factor representations are not scarce}\label{sec:trivialconstructions}
Coherent traded factor representations always exist: each endpoint of the interval \eqref{eq:admissible} carries one, and a third representation lies between them, but none of the three turns out to be empirically relevant. By Remark~\ref{rem:onlyimage} a loading matrix contributes only its column space, and at either endpoint that column space is dictated by the moments of returns rather than chosen by the researcher, so the representation asserts nothing about firm characteristics and no observation could contradict it. Each endpoint also forfeits one of the two purposes a factor model serves.

\emph{Minimal rank: the CMVE one-factor representation.} For $\bm\mu_t\neq\bm 0$, the single traded factor $f_{t+1}=X^\star_{t+1}$ of \eqref{eq:defXstar}, with weights $\bm W_t=\bm Q_t^+\bm\mu_t$ and loading vector $\bm\Phi_t=\bm\mu_t/(\bm\mu_t^\top\bm Q_t^+\bm\mu_t)$, is coherent: \eqref{eq:coh1} holds with no pricing assumption, by Proposition~\ref{prop:Xstar}~\ref{it:Xstar6}, the residual has conditional mean zero, so \eqref{eq:coh2} holds, and under the law of one price \eqref{eq:LoP} the factor spans the maximal Sharpe ratio by Proposition~\ref{prop:span}. Its column space, however, is fixed at the lower endpoint $\spn\{\bm\mu_t\}$ by the risk premia themselves, so the representation can be written down only by someone who already knows the premia it is meant to explain, and it captures only the variance that pricing forces.

\emph{Maximal rank: the saturated representation.} At the other end, $\bm\Phi_t=\bm Q_t\bm Q_t^+$ with $\bm f_{t+1}=\bm x_{t+1}$ satisfies \eqref{eq:coh1}, \eqref{eq:coh2}, \eqref{eq:factorportfolios} and \eqref{eq:maxSR}, since $\bm\Phi_t\bm f_{t+1}=\bm x_{t+1}$ almost surely by Remark~\ref{rem:support} and the residual vanishes. The factors are the returns themselves, so the representation states only that returns equal returns: what is lost is parsimony. The familiar choice $\bm\Phi_t=\bm I_{n_t}$ requires $\bm Q_t$ to be invertible, and otherwise violates \eqref{eq:coh1} by assigning exposures in directions in which returns do not move.

\emph{Between the two: the constant factor.} Whenever $\bm\mu_t=\bm\Phi_t\bm b_t$, the constant factor $\bm f_{t+1}=\bm b_t$ prices correctly, with residual $\bm x_{t+1}-\bm\mu_t$, and leaves the loading matrix free. But \eqref{eq:coh1} fails in general once $m_t\ge2$, since the regression loading matrix is $\bm\Phi_t\bm b_t\bm b_t^\top/(\bm b_t^\top\bm b_t)$. The factor is not traded and cannot be, since an $\Fcal_t$-measurable nonzero traded excess return would have an unbounded conditional Sharpe ratio, excluded by Proposition~\ref{prop:LOPmain}~\ref{it:LOPmain2}, and it explains no variance at all, $\bm\Sigma_{\bm\epsilon,t}=\bm\Sigma_t$. Appendix~\ref{app:intercept} takes up such degenerate covariance directions for non-traded factors.

Coherence is therefore not scarce. What is scarce is coherence at a loading matrix of small rank whose column space the researcher chooses.

\subsubsection{Alpha tests and betas}\label{sec:alphatests}
The third cost of Section~\ref{sec:rigidity} reverses once \eqref{eq:coh1} holds: a researcher who has secured property~\ref{it:prop1} gets property~\ref{it:prop2} from a regression already being run. Empirical practice tests pricing not through \eqref{eq:coh2} but through a zero-alpha restriction, formulated in a second type of factor representation that carries an explicit intercept,
\begin{equation}\label{eq:linfacmodelN}
\bm x_{t+1} = \bm\alpha_t + \bm\beta_t \bm f_{t+1} + \bm\eta_{t+1},
\end{equation}
with $\Fcal_t$-measurable intercept $\bm\alpha_t$ and slope matrix $\bm\beta_t$, and with the residuals normalized to $\E_t[\bm\eta_{t+1}]=\bm 0$, in the tradition of \citet{gib_ros_sha_89}. This is the alternative to \eqref{eq:linfacmodelPhi} anticipated in Section~\ref{sec:notation}: where \eqref{eq:linfacmodelPhi} leaves $\E_t[\bm\epsilon_{t+1}]$ to be examined, \eqref{eq:linfacmodelN} normalizes it into the intercept. The canonical coefficients are those of the conditional regression of returns on a constant and the factors, $\bm\alpha_t=\bm\alpha^{\bm f_{t+1}}_t$ and $\bm\beta_t=\bm\beta^{\bm f_{t+1}}_t$, recorded in Appendix~\ref{app:intercept}, and we reserve the term \emph{beta} for the slope matrix $\bm\beta^{\bm f_{t+1}}_t$ of this regression. The zero-alpha restriction $\bm\alpha^{\bm f_{t+1}}_t=\bm 0$ is strictly stronger than \eqref{eq:coh2} in general. For traded factors under the law of one price \eqref{eq:LoP} the two coincide.

\begin{proposition}[Tradability Equivalence]\label{prop:tradeq}
Assume the law of one price \eqref{eq:LoP} holds.
\begin{enumerate}
\item\label{it:tradeq2} The zero-alpha restriction and the pricing restriction are equivalent,
\begin{equation}\label{eq:tradeq}
\bm\alpha^{\bm f_{t+1}}_t=\bm 0 \quad\Longleftrightarrow\quad \E_t[\bm\epsilon^{\bm f_{t+1}}_{t+1}]=\bm 0,
\end{equation}
and in either case the residuals of the two regressions coincide, $\bm\eta^{\bm f_{t+1}}_{t+1}=\bm\epsilon^{\bm f_{t+1}}_{t+1}$.

\item\label{it:tradeq1} Factor risk premia are spanned by the factor covariance matrix, $\bm\mu_{\bm f,t}\in\Ima\bm\Sigma_{\bm f,t}$, and the betas and the regression loadings differ by the intercept alone,
\begin{equation}\label{eq:betaPhi}
\bm\Phi^{\bm f_{t+1}}_t=\bm\beta^{\bm f_{t+1}}_t+\bm\alpha^{\bm f_{t+1}}_t\bm\mu_{\bm f,t}^\top\bm Q_{\bm f,t}^+.
\end{equation}
In particular $\bm\beta^{\bm f_{t+1}}_t=\bm\Phi^{\bm f_{t+1}}_t$ if and only if $\bm\alpha^{\bm f_{t+1}}_t\bm\mu_{\bm f,t}^\top=\bm 0$, and hence whenever either restriction in part~\ref{it:tradeq2} holds.

\end{enumerate}
\end{proposition}

With Lemma~\ref{lem:alphazero}, under the law of one price \eqref{eq:LoP} a factor representation satisfying \eqref{eq:coh1} is coherent if and only if its alpha vanishes, and then $\bm\beta^{\bm f_{t+1}}_t=\bm\Phi^{\bm f_{t+1}}_t=\bm\Phi_t$.
Part~\ref{it:tradeq1} is what allows loadings to be called betas at all. Requirement \eqref{eq:coh1} constrains the regression loading matrix, an uncentered second-moment object. For traded factors under the law of one price \eqref{eq:LoP}, that object coincides with the beta as soon as the alpha vanishes, by \eqref{eq:betaPhi}. A coherent factor representation therefore carries its prespecified loadings as genuine betas, and that characteristics are covariances becomes a derived property rather than a specification choice. Appendix~\ref{app:optimalfactors} shows by example that neither hypothesis can be dropped.

\subsection{The factors span the efficient portfolio}\label{sec:spanning}
Properties \ref{it:prop1} and \ref{it:prop2} concern second moments and pricing, and neither mentions portfolios. The economic question is whether an investor holding the factors does as well as one holding the assets. If the factor portfolios attain the Sharpe ratio of the full cross-section, they span the investment opportunity set. This is property~\ref{it:prop3}, and it turns out to follow from properties \ref{it:prop1} and \ref{it:prop2}.

The link is factor--residual uncorrelatedness,
\begin{equation}\label{eq:UNCORR}
\cov_t[\bm\epsilon_{t+1},\bm f_{t+1}]=\bm 0,
\end{equation}
which is usually assumed. A coherent factor representation implies it,
$\cov_t[\bm\epsilon_{t+1},\bm f_{t+1}]=\E_t[\bm\epsilon_{t+1}\bm f_{t+1}^\top]-\E_t[\bm\epsilon_{t+1}]\bm\mu_{\bm f,t}^\top=\bm 0$,
since the first term vanishes by \eqref{eq:coh1} and \eqref{eq:L2tortho} and the second by \eqref{eq:coh2}. Uncorrelatedness links the mean restriction \eqref{eq:coh2} to the Sharpe ratio restriction \eqref{eq:maxSR}, and this is how a pricing property acquires a portfolio consequence.

\begin{proposition}[Spanning and Unpriced Residual Risk]\label{prop:uncspan}
Assume the law of one price \eqref{eq:LoP} holds, and let the factors and residuals of the factor representation \eqref{eq:linfacmodelPhi} be uncorrelated, \eqref{eq:UNCORR}. Then factor spanning \eqref{eq:maxSR} holds if and only if residual risk is unpriced, \eqref{eq:coh2}.
\end{proposition}

Remark~\ref{rem:propspan} in Appendix~\ref{app:proofs} isolates the role of uncorrelatedness. The implication from unpriced residual risk to spanning uses only uncorrelatedness of the factor component, and the converse fails without it. Beyond the law of one price, a coherent factor representation assumes neither hypothesis, since uncorrelatedness is automatic and unpriced residual risk is \eqref{eq:coh2} itself. The first implication gives:

\begin{quote}
Under the law of one price \eqref{eq:LoP}, a coherent traded factor representation attains the maximal conditional Sharpe ratio, \eqref{eq:maxSR}.
\end{quote}

A mean--variance investor loses nothing by trading the $m_t$ factor portfolios instead of the $n_t$ assets, whatever $n_t$ is and however singular the moment matrices are. The $n_t$-dimensional investment problem collapses to an $m_t$-dimensional one at no cost in Sharpe ratio, and \eqref{eq:admissible} is the exact condition under which it does. A factor representation that needed spanning as a further hypothesis would have loadings and a pricing claim doing no work.

Two qualifications. First, the law of one price is needed, since it makes the maximal Sharpe ratio finite, by Proposition~\ref{prop:LOPmain}~\ref{it:LOPmain2}. Without it, conditional Sharpe ratios are unbounded and the alpha equivalence of Proposition~\ref{prop:tradeq} fails, as Example~\ref{ex:sccounter} in Appendix~\ref{app:optimalfactors} shows. Second, the implication runs one way. A factor representation can attain the maximal Sharpe ratio while its loadings are not the regression loadings of its factors, by the second half of Example~\ref{ex:spanning2} in Appendix~\ref{app:proofs}, such that spanning is a consequence, not a characterization.

\subsection{The residuals are (not) idiosyncratic}\label{sec:impossible}
Sections~\ref{sec:rigidity} and~\ref{sec:maintheorem} show that properties \ref{it:prop1} and \ref{it:prop2} are attainable, and Section~\ref{sec:spanning} that property~\ref{it:prop3} follows from them. Property~\ref{it:prop4} is not attainable at all, and its impossibility explains the factor--residual correlations that characteristic-based factor portfolios display in data.

Property~\ref{it:prop4} asks that the residuals be idiosyncratic: uncorrelated with the factors, with a second moment matrix of full rank. A strict factor structure requires more, a residual covariance matrix that is diagonal and nonsingular, and an approximate one a residual spectrum bounded as the cross-section grows. Full rank is the part a finite cross-section can decide. For traded factors it is unavailable. Proposition~\ref{prop:rankdef} shows that factor--residual orthogonality \eqref{eq:ORTHO} alone forces the residual second moment matrix, and with it the residual covariance matrix, to be singular in at least $\rank\bm\Pi_t$ directions, so a factor representation with property~\ref{it:prop1} and a nonzero loading matrix cannot have property~\ref{it:prop4}. Read the other way, a factor representation whose residual covariance matrix has full rank must violate \eqref{eq:ORTHO}, and once residual risk is unpriced it must violate uncorrelatedness \eqref{eq:UNCORR} as well. The correlations observed for characteristic-based factor portfolios are therefore implied by the theory rather than evidence against it. This section proves both halves and quantifies them.

\subsubsection{Residual rank deficiency}\label{sec:rankdeficiency}
Property~\ref{it:prop1}, \eqref{eq:coh1}, has a structural implication for residual moment matrices that holds for every traded factor representation, whether or not it is coherent. It rests on factor--residual orthogonality alone,
\begin{equation}\label{eq:ORTHO}
\E_t[\bm\epsilon_{t+1} \bm f_{t+1}^\top]=\bm 0.
\end{equation}
Under \eqref{eq:ORTHO} the second moment matrix decomposes into a factor part and a residual part,
\begin{equation}\label{eq:Qdecomp}
\bm Q_t=\bm\Phi_t\bm Q_{\bm f,t}\bm\Phi_t^\top+\bm Q_{\bm\epsilon,t},
\end{equation}
since $\E_t[\bm x_{t+1}\bm f_{t+1}^\top]=\bm\Phi_t\bm Q_{\bm f,t}$ and $\E_t[\bm x_{t+1}\bm\epsilon_{t+1}^\top]=\bm Q_{\bm\epsilon,t}$ then. By Remark~\ref{rem:centering} the covariance counterpart, $\bm\Sigma_t=\bm\Phi_t\bm\Sigma_{\bm f,t}\bm\Phi_t^\top+\bm\Sigma_{\bm\epsilon,t}$, holds under uncorrelatedness \eqref{eq:UNCORR} in the same way. The decomposition is the form in which the factor structure is usually written. Here, we discuss its relation to the residuals.

Orthogonality is strictly weaker than \eqref{eq:coh1}: it pins down the factor component but not the loading matrix, and it is too weak to determine the factors. Lemma~\ref{lem:FOCf} in Appendix~\ref{app:optimalfactors} records the two equivalent readings we use, that the residuals are the regression residuals of the given factors, $\bm\epsilon_{t+1}=\bm\epsilon^{\bm f_{t+1}}_{t+1}$, and that the projected loadings $\bm\Phi_t\bm Q_{\bm f,t}\bm Q_{\bm f,t}^+$ are the regression loadings. The latter is \eqref{eq:coh1} itself precisely when $\Ima\bm\Phi_t^\top\subseteq\Ima\bm Q_{\bm f,t}$. Orthogonality is nonetheless enough for the moment matrix decomposition and for the rank result below.

\begin{proposition}[Residual Rank Deficiency]\label{prop:rankdef}
Consider a factor representation \eqref{eq:linfacmodelPhi} with a non-degenerate factor component, $\bm\Pi_t\bm Q_t\neq\bm 0$. Assume that either factor--residual orthogonality \eqref{eq:ORTHO} holds or the factor component operator is a projection,
\begin{equation}\label{eq:FCopProj}
 \bm\Pi_t^2 =\bm\Pi_t. 
\end{equation}
Then the residual second moment matrix satisfies the matrix orthogonality
\begin{equation}\label{eq:matrixortho}
\bm\Pi_t\bm Q_{\bm\epsilon,t}=\bm 0,
\end{equation}
and thus
\begin{equation}\label{eq:RDambient}
\rank\bm\Sigma_{\bm\epsilon,t}\le\rank\bm Q_{\bm\epsilon,t}\le n_t-\rank\bm\Pi_t<n_t.
\end{equation}
\end{proposition}


A coherent traded factor representation satisfies both hypotheses of Proposition~\ref{prop:rankdef}, by Theorem~\ref{thm:rigid}~\ref{it:rigid3}, with $\rank\bm\Pi_t=\rank\bm\Phi_t$. This is the formal content of the impossibility announced at the head of this section: a fourth condition requiring idiosyncratic, or merely full-rank, residuals is not merely restrictive but inconsistent with \eqref{eq:coh1} and \eqref{eq:coh2}. The WLS factors of \eqref{eq:SLSfact} satisfy \eqref{eq:FCopProj} for every weighting, as $\bm\Pi_t$ is a projection by Lemma~\ref{lem:Pidefchar}, which yields the following corollary.

\begin{corollary}[Rank Deficiency under Weighted Least Squares]\label{cor:rankdefWLS}
WLS residuals $\bm\epsilon_{t+1}=\bm\epsilon^{\bm\Phi_t}_{t+1}$ satisfy \eqref{eq:matrixortho}, and thus
\[
\rank\bm\Sigma_{\bm\epsilon,t}\le\rank\bm Q_{\bm\epsilon,t}\le n_t-\rank(\bm S_t\bm\Phi_t).
\]
\end{corollary}

Lemma~\ref{lem:rankdefsupp} in Appendix~\ref{app:proofs} complements these bounds. It places the rank loss inside the support of returns, $\rank\bm Q_{\bm\epsilon,t}<\rank\bm Q_t$, which is the informative statement when $\bm Q_t$ is singular, and for the WLS factors it holds automatically under the support condition \eqref{eq:Phistarass}.

Proposition~\ref{prop:rankdef} implies in particular that there exists no factor representation \eqref{eq:linfacmodelPhi} with factor portfolios and a nonzero factor component whose residuals are orthogonal to the factors and have a full-rank second moment or covariance matrix. The covariance-based analogue, with uncorrelatedness \eqref{eq:UNCORR} in place of orthogonality \eqref{eq:ORTHO}, holds as well: it is this proposition read on the centered panel, by Remark~\ref{rem:centering} in Appendix~\ref{app:optimalfactors}. This has several ramifications. First, a strict factor structure in the sense of \citet{ross76}, with uncorrelated factors and residuals and a diagonal residual covariance matrix, can obtain only with non-traded factors. Second, in any finite cross-section with traded factors, rank deficiency of the residual covariance matrix is unavoidable under factor--residual uncorrelatedness. This provides a structural complement to the standard arbitrage pricing motivation for passing to infinite cross-sections in \citet{chamberlainrothschild83} and \citet{connor84}. Third, for the shrinkage of factor structured covariance matrices reviewed in \citet[][Section~5]{ledoitwolf20}, which estimates the residual covariance matrix of a factor model by nonlinear shrinkage and thereby lifts its spectrum away from zero, the population object being estimated is singular under traded factors with factor--residual uncorrelatedness: a full-rank estimate may well be preferred on statistical grounds, but its target is rank deficient.

Read in the contrapositive: if $\bm\Sigma_{\bm\epsilon,t}$ has full rank, for instance if it is diagonal or equal to $\sigma_t^2\bm I_{n_t}$, then \eqref{eq:RDambient} fails, so factor--residual orthogonality \eqref{eq:ORTHO} fails, and once residual risk is unpriced, uncorrelatedness \eqref{eq:UNCORR} fails as well. The nonzero factor--residual correlations of characteristic-based OLS factor portfolios, as in the OLS factor representation of \citet{kozaknagel23}, are therefore not evidence against imposing \eqref{eq:ORTHO}. Within each window they are the cost of a loading span that is not $\hat{\bm Q}_t$-invariant, Corollary~\ref{cor:OLScoh}, and Proposition~\ref{prop:rankdef} forecloses the repair, since no traded factor representation can remove them while retaining a full-rank residual covariance matrix. The large-economy discussion below quantifies their magnitude.

\begin{example}[continues=ex:running]\label{ex:runningRD}
In Example~\ref{ex:running}, $\bm Q_t=\bm\Phi_t(\bm C_t+\bm b_t\bm b_t^\top)\bm\Phi_t^\top+\sigma_t^2\bm I_{n_t}$ is invertible. Consider the OLS factor portfolios $\bm f_{t+1}=\bm\Phi_t^+\bm x_{t+1}$, given by \eqref{eq:SLSfact} for $\bm S_t=\bm I_{n_t}$, with residuals $\bm\epsilon_{t+1}=(\bm I_{n_t}-\bm\Phi_t\bm\Phi_t^+)\bm x_{t+1}$. Using $(\bm I_{n_t}-\bm\Phi_t\bm\Phi_t^+)\bm\Phi_t=\bm 0$ and $(\bm I_{n_t}-\bm\Phi_t\bm\Phi_t^+)\bm\Phi_t^{+\top}=\bm 0$, we obtain $\E_t[\bm\epsilon_{t+1}\bm f_{t+1}^\top]=\bm 0$ and $\E_t[\bm\epsilon_{t+1}]=\bm 0$ exactly, so that \eqref{eq:ORTHO}, \eqref{eq:coh2} and \eqref{eq:UNCORR} all hold, and the residual moment matrices
\[
\bm Q_{\bm\epsilon,t}=\bm\Sigma_{\bm\epsilon,t}=\sigma_t^2(\bm I_{n_t}-\bm\Phi_t\bm\Phi_t^+)
\]
have rank $n_t-\rank\bm\Phi_t$.
The rank deficiency \eqref{eq:RDambient} thus obtains with equality, and Corollary~\ref{cor:rankdefWLS} applies with $\bm S_t=\bm I_{n_t}$. The full-rank matrix $\sigma_t^2\bm I_{n_t}$ is the covariance matrix of the structural noise in the specification of $\bm\Sigma_t$, and not of the residuals of any traded OLS factor representation, whose second moment matrix is singular in exactly $\rank\bm\Phi_t$ directions. The example is also an instance of Corollary~\ref{cor:OLScoh}: here $\bm Q_t=\bm\Phi_t(\bm C_t+\bm b_t\bm b_t^\top)\bm\Phi_t^\top+\sigma_t^2\bm I_{n_t}$ maps $\Ima\bm\Phi_t$ into itself, so $\Ima\bm\Phi_t$ is $\bm Q_t$-invariant, the OLS factor representation is coherent, and its factors are equal to the moment-weighted ones by Theorem~\ref{thm:rigid}. Section~\ref{sec:largeeconomies} returns to this identity and quantifies the deficient fraction in large cross-sections.
\end{example}

\subsubsection{Large economies}\label{sec:largeeconomies}
Proposition~\ref{prop:rankdef} is an exact statement about a finite cross-section. We now relate it to the large-economy factor models of the arbitrage pricing literature, where a full-rank, in the strict factor structure of \citet{ross76} even diagonal, residual covariance matrix is the standard idealization \citep{chamberlainrothschild83,connor84,ConnorKorajckzyk1986}. The arguments are elementary consequences of the preceding results, and no new formal statements are made.

Consider any factor representation \eqref{eq:linfacmodelPhi}, traded or not. Since $\bm\epsilon_{t+1}\in\Ima\bm Q_{\bm\epsilon,t}$, returns lie in $\Ima\bm\Phi_t+\Ima\bm Q_{\bm\epsilon,t}$ almost surely, and $\Ima\bm Q_t$ is the smallest subspace containing them almost surely (Remark~\ref{rem:support}), so
\begin{equation}\label{eq:rankbound}
\rank\bm Q_{\bm\epsilon,t}\ge\rank\bm Q_t-m_t.
\end{equation}
The same argument applied to $\bm x_{t+1}-\bm\mu_t$ gives $\rank\bm\Sigma_{\bm\epsilon,t}\ge\rank\bm\Sigma_t-m_t$. Whenever \eqref{eq:RD} holds for the factor portfolios \eqref{eq:factorportfolios}, as under Proposition~\ref{prop:rankdef}, it combines with \eqref{eq:rankbound} to localize the residual rank,
\[
\rank\bm Q_t-m_t\le\rank\bm Q_{\bm\epsilon,t}\le\rank\bm Q_t-1.
\]
The residual second moment matrix is thus degenerate in at least one and at most $m_t$ directions of $\Ima\bm Q_t$, a fraction of at most $m_t/n_t$ that vanishes as $n_t\to\infty$ for fixed $m_t$.

The factor--residual covariance implied by a full-rank residual covariance matrix vanishes at rate $1/n_t$ under standard large-economy conditions. Suppose that
\begin{equation}\label{eq:approxfs}
\bm\Sigma_t=\bm\Phi_t\bm C_t\bm\Phi_t^\top+\bm D_t,
\end{equation}
for some positive semidefinite $m_t\times m_t$ matrix $\bm C_t$ and residual covariance matrix $\bm D_t$ with eigenvalues bounded uniformly in $n_t$ \citep{chamberlainrothschild83}. Take again the OLS factor portfolios of Example~\ref{ex:running}, continued. As $(\bm I_{n_t}-\bm\Phi_t\bm\Phi_t^+)\bm\Phi_t=\bm 0$, only $\bm D_t$ contributes to the factor--residual covariance,
\begin{equation}\label{eq:covOLSlargeecon}
\cov_t[\bm f_{t+1},\bm\epsilon_{t+1}]=\bm\Phi_t^+\bm D_t(\bm I_{n_t}-\bm\Phi_t\bm\Phi_t^+).
\end{equation}
Under regularity conditions on the loadings and on $\bm D_t$, every entry of \eqref{eq:covOLSlargeecon} is of order $O(1/n_t)$, as is every entry of $\E_t[\bm\epsilon_{t+1}\bm f_{t+1}^\top]$ when $\bm\mu_t\in\Ima\bm\Phi_t$.\footnote{Assume that the loadings are uniformly bounded and pervasive, $\lambda_{\min}(\bm\Phi_t^\top\bm\Phi_t)\ge c\,n_t$ for some $c>0$, and that $\max_i\sum_j|(\bm D_t)_{ij}|$ is bounded uniformly in $n_t$, a standard strengthening of the bounded-eigenvalue condition \citep{bai03,fanliaowang16}. Every entry of $\bm\Phi_t^+$ is then of order $O(1/n_t)$ and the absolute column sums of $\bm D_t(\bm I_{n_t}-\bm\Phi_t\bm\Phi_t^+)$ are bounded uniformly in $n_t$. In spectral norm, $\|\cov_t[\bm f_{t+1},\bm\epsilon_{t+1}]\|\le\lambda_{\min}(\bm\Phi_t^\top\bm\Phi_t)^{-1/2}\|\bm D_t\|=O(n_t^{-1/2})$. If in addition $\bm C_t$ is nonsingular and the residual variances are bounded away from zero, $\min_i(\bm D_t)_{ii}\ge c'$ for some $c'>0$, then $\cov_t[\bm f_{t+1}]=\bm C_t+\bm\Phi_t^+\bm D_t\bm\Phi_t^{+\top}=\bm C_t+O(1/n_t)$ stays bounded away from singularity and the factor--residual correlations are of order $O(1/n_t)$ as well. Without these last conditions only the covariance statement survives.} For $\bm D_t=\sigma_t^2\bm I_{n_t}$, as in Example~\ref{ex:running}, the covariance \eqref{eq:covOLSlargeecon} vanishes identically and Example~\ref{ex:runningRD} applies unchanged: the structural noise and the traded-factor residuals differ in exactly $\rank\bm\Phi_t$ directions in every finite cross-section, however large $n_t$ is.

The same limit softens Corollary~\ref{cor:OLScoh}. Under \eqref{eq:approxfs} with $\bm D_t=\sigma_t^2\bm I_{n_t}$ and risk premia spanned by the loadings, $\bm\mu_t\in\Ima\bm\Phi_t$, the subspace $\Ima\bm\Phi_t$ is exactly $\bm Q_t$-invariant and the OLS factor representation is coherent, as Example~\ref{ex:runningRD} records. The risk-premium condition is doing real work in both claims, because \eqref{eq:approxfs} restricts $\bm\Sigma_t$ alone while $\bm Q_t=\bm\Sigma_t+\bm\mu_t\bm\mu_t^\top$, so invariance needs $(\bm I_{n_t}-\bm\Phi_t\bm\Phi_t^+)\bm\mu_t\bm\mu_t^\top\bm\Phi_t\bm\Phi_t^+=\bm 0$. It is required for invariance up to the degenerate case $\bm\Phi_t^\top\bm\mu_t=\bm 0$, and for coherence outright. Under it, and for general $\bm D_t$, the invariance defect $(\bm I_{n_t}-\bm\Phi_t\bm\Phi_t^+)\bm Q_t\bm\Phi_t\bm\Phi_t^+=(\bm I_{n_t}-\bm\Phi_t\bm\Phi_t^+)\bm D_t\bm\Phi_t\bm\Phi_t^+$ is governed by $\bm D_t$ alone and is of the same order as \eqref{eq:covOLSlargeecon}. Failure of \eqref{eq:coh1} for the OLS factor representation is therefore a finite-$n_t$ phenomenon of the same size as the factor--residual correlation it produces, invisible in the limit and visible in the panels of Section~\ref{sec:empirical}.

The full-rank specifications of \citet{ross76}, \citet{chamberlainrothschild83}, \citet{ConnorKorajckzyk1986}, \citet{bai03} and \citet{fanliaowang16} are the large-$n_t$ idealization of the economies studied here, and our results are their exact finite-$n_t$ counterpart. The finite-$n_t$ case is also the empirically relevant one. At conventional panel sizes the sample second moment matrix is near singular, see Section~\ref{sec:datadesign}, and the pseudoinverse calculus used throughout keeps all statements exact without rank assumptions.
\section{Implications for Current Practice}\label{sec:verdict}
The factor representations of Section~\ref{sec:representations} are all built by unweighted least squares, \eqref{eq:SLSfact} with $\bm S_t=\bm I_{n_t}$. Regressed PCA runs it at its sieve matrix, with a fixed compression after. A single criterion from Section~\ref{sec:rigidity} therefore applies to all of them.

\subsection{Consistency with the reported loadings}\label{sec:rulesout}
The criterion is Theorem~\ref{thm:rigid}~\ref{it:rigid3}. Condition \eqref{eq:coh1} requires the factor component operator to be the $\bm Q_t^+$-orthogonal projection onto $\Ima\bm\Phi_t$, while unweighted least squares delivers the Euclidean projection. The following corollary states when the two coincide.

\begin{corollary}[OLS Factor Portfolios]\label{cor:OLScoh}
Assume the support condition \eqref{eq:Phistarass} and let $\bm f_{t+1}=\bm\Phi_t^+\bm x_{t+1}$ be the OLS factor portfolios. Then $\bm\Phi_t=\bm\Phi_t^{\bm f_{t+1}}$ if and only if $\Ima\bm\Phi_t$ is an invariant subspace of $\bm Q_t$,
\begin{equation}\label{eq:Qinvariant}
\bm Q_t\,\Ima\bm\Phi_t\subseteq\Ima\bm\Phi_t,
\end{equation}
equivalently, if and only if $\Ima\bm\Phi_t$ is spanned by eigenvectors of $\bm Q_t$.
\end{corollary}

The proof runs through Theorem~\ref{thm:rigid}. OLS delivers the Euclidean orthogonal projection $\bm\Phi_t\bm\Phi_t^+$ onto $\Ima\bm\Phi_t$, \eqref{eq:coh1} delivers the $\bm Q_t^+$-orthogonal projection onto the same subspace, and two orthogonal projections onto a common subspace coincide if and only if that subspace is invariant under the matrix relating the two inner products. The necessary half needs neither existence nor a closed form, and it carries the negative result.

Condition \eqref{eq:Qinvariant} asks that $\Ima\bm\Phi_t$ be spanned by some eigenvectors of $\bm Q_t$, not necessarily leading ones. For $\bm Q_t=\diag(3,2,1)$ and $\bm\Phi_t=\bm e_3$ the OLS factor portfolio has property~\ref{it:prop1} while $\Ima\bm\Phi_t$ is the trailing eigenspace. The optimality of Proposition~\ref{prop:minPhi} in Appendix~\ref{app:optimalfactors} selects principal subspaces, and therefore a strict subset of the loading matrices on which unweighted least squares has property~\ref{it:prop1}.

Corollary~\ref{cor:OLScoh} is the exact sense in which characteristic-based OLS factor portfolios fail property~\ref{it:prop1}. Their loadings are the regression loadings of their factors precisely on the $\bm Q_t$-invariant loading matrices, and a matrix of firm characteristics is invariant under the conditional second moment matrix of returns only by accident. Example~\ref{ex:market} is the smallest model carrying the statement, where \eqref{eq:Qinvariant} reduces to \eqref{eq:marketaccident} and is satisfied only by a characteristic orthogonal to the betas or proportional to them.

\subsection{Discussion of factor representations in the literature}\label{sec:leadingreps}

Applied to the five factor representations of Section~\ref{sec:representations}, Corollary~\ref{cor:OLScoh} gives an answer that turns only on how each chooses $\Ima\bm\Phi_t$. For instrumented principal components the reported and the regression loadings agree exactly when the instrument span $\Ima(\bm Z_t\bm\Gamma_\beta)$ is $\bm Q_t$-invariant, and for projected principal components the same holds for the sieve span $\Ima\widehat{\bm G}(\bm X)$. Regressed PCA fails on a second margin, because its factors are compressed from the sieve-level OLS portfolios rather than formed from its reported loading matrix. Matching the OLS factor representation of those loadings needs $\Ima\widehat{\bm B}$ invariant under $\bm S_t^\top\bm S_t$, and consistency then needs $\bm Q_t$-invariance of the compressed span $\Ima(\bm S_t\widehat{\bm B})$ as well. Neither condition holds other than by accident. The intercept the method estimates, and finds significant, is in our terms the failure of the risk-premium condition \eqref{eq:arpspanning} on the sieve span, the risk-premium gap of Section~\ref{sec:diagnostics}. For the heuristic portfolios of \citet{kozaknagel23} the correspondence is exact, and the authors establish it themselves.\footnote{Their Proposition~2 characterizes attainment of the maximal conditional Sharpe ratio through the decomposition $\bm\Sigma_t=\bm X_t\bm\Psi_t\bm X_t^\top+\bm U_t\bm\Omega_t\bm U_t^\top$ with $\bm U_t^\top\bm X_t=\bm 0$, their equations~(17)--(18), which is $\bm\Sigma_t$-invariance of $\Ima\bm X_t$ in the form $\bm\Sigma_t\bm X_t=\bm X_t\bm B_t$, by conditions (F$'$) and (B) of \citet[Theorem~1]{lu_sch_12}, which hold under positive definiteness. Under their Assumption~1, that expected returns are linear in characteristics, which is our risk-premium condition \eqref{eq:arpspanning}, and under the law of one price \eqref{eq:LoP}, we have $\Ima\bm Q_t=\Ima\bm\Sigma_t$ and $\bm\mu_t\in\Ima\bm X_t$, so their condition agrees with \eqref{eq:Qinvariant} and their Proposition~2 is Corollary~\ref{cor:OLScoh} in that case.}

Risk-premium PCA is the one of the five that satisfies \eqref{eq:coh1} by design rather than by accident. At $\gamma=0$ the eigendecomposed matrix is the sample analogue of $\bm Q_t$, so $\Ima\widehat{\bm\Lambda}$ is one of its principal subspaces and therefore invariant, which gives \eqref{eq:coh1} by Corollary~\ref{cor:OLScoh} and makes the OLS factor portfolios the moment-weighted ones by Corollary~\ref{cor:Phistar}. Neither identity survives at any other $\gamma$, apart from degenerate panels such as those in which $\bar{\bm x}$ is an eigenvector of $\widehat{\bm\Sigma}$.\footnote{The reading of $\gamma$ as an interpolation through the second moment matrix is ours. \citet{lettaupelger20} name the $\gamma=-1$ endpoint and not the $\gamma=0$ one.} The property is bought as at the endpoints of Section~\ref{sec:trivialconstructions}, by forfeiting the choice of column space, and it cannot be had together with \eqref{eq:coh2}, because a principal subspace of the second moment matrix need not contain the mean. Raising $\gamma$ shrinks the risk-premium gap but moves $\Ima\widehat{\bm\Lambda}$ off the invariant subspaces and reopens the loading discrepancy of Section~\ref{sec:diagnostics}.

The pattern is uniform. Each of the five fixes a loading matrix and forms factors by unweighted least squares, with the one compression noted above, so by Corollary~\ref{cor:OLScoh} each is consistent with its own loadings exactly on the $\bm Q_t$-invariant column spaces. Four fix that column space from firm characteristics, on grounds external to $\bm Q_t$, which is precisely when invariance holds only by accident. The fifth reads the column space off $\bm Q_t$ and gives up the choice. The answers hold at any cross-section size, but their magnitudes do not. By Section~\ref{sec:largeeconomies}, when the loadings span a pervasive factor structure and the risk premia, the invariance defect of the OLS factor representation is $O(1/n_t)$. On a large individual-stock panel a small measured discrepancy is then evidence that the characteristics are close to covariances and a large one that they are not, while the magnitudes of Section~\ref{sec:empirical} are magnitudes for portfolio panels of the size we study.

\subsection{The loading gap}\label{sec:cost}
A factor representation that fails \eqref{eq:coh1} reports a loading matrix that is not the loading matrix of a regression on the factors it reports. Section~\ref{sec:rigidity} records what this costs. The variance decompositions and risk attributions such a representation reports belong to no regression on its own factors, the factor component is not the projection it is taken to be, and a zero alpha certifies pricing against $\bm\beta^{\bm f_{t+1}}_t$ rather than against the reported $\bm\Phi_t$. The smallest distance to the regression loadings that any traded factor representation can achieve has a closed form, the loading gap, and it is measurable in any sample.

\begin{corollary}[Loading Gap]\label{cor:cohgap}
For any $\Fcal_t$-measurable $n_t\times m_t$ loading matrix $\bm\Phi_t$ and any unitarily invariant matrix norm,
\begin{equation}\label{eq:cohgap}
\min_{\bm f_{t+1}\ \text{traded}}\bigl\|\bm\Phi_t^{\bm f_{t+1}}-\bm\Phi_t\bigr\|
=\bigl\|(\bm I_{n_t}-\bm Q_t\bm Q_t^+)\bm\Phi_t\bigr\|,
\end{equation}
attained at the moment-weighted factor portfolios of $\bm Q_t\bm Q_t^+\bm\Phi_t$. The minimum is zero if and only if the support condition \eqref{eq:Phistarass} holds.
\end{corollary}

The gap \eqref{eq:cohgap} is the distance from $\bm\Phi_t$ to the loading matrices a traded factor can carry. By Theorem~\ref{thm:rigidgen} in Appendix~\ref{app:proofs} those are exactly the members of
\begin{equation}\label{eq:Vcal}
\Vcal_t\coloneqq\{\bm A\in\R^{n_t\times m_t}:\Ima\bm A\subseteq\Ima\bm Q_t\},
\end{equation}
a linear subspace, so the nearest point is the orthogonal projection $\bm Q_t\bm Q_t^+\bm\Phi_t$ and the distance is \eqref{eq:cohgap}. The gap vanishes identically when $\bm Q_t$ is invertible, so it is a purely rank-deficient phenomenon, and its sample analogue measures the smallest discrepancy from \eqref{eq:coh1} that any factor representation carrying the given loadings must pay in a finite panel.


\subsection{Computing the factors}\label{sec:computation}
Theorem~\ref{thm:rigid} gives the traded factors of an admissible loading matrix in closed form, but the formula \eqref{eq:MWfactors} involves the pseudoinverse of the $n_t\times n_t$ second moment matrix. In applications $n_t$ is large and $\bm Q_t$ is estimated, so this is the step one would like to avoid. This subsection shows that under the structured specifications used in empirical work the pseudoinverse of $\bm Q_t$ can be replaced by that of the residual part alone, and identifies two cases in which the moment weighting coincides with a simpler one: with GLS, which uses $\bm\Sigma_t$ in place of $\bm Q_t$, and with OLS, which uses no covariance information at all. One invariance drives all three statements: the weight matrix \eqref{eq:MWfactors} is unchanged by stripping a factor structure aligned with $\bm\Phi_t$ from the weighting matrix, Lemma~\ref{lem:GLSinv} in Appendix~\ref{app:proofs}, which generalizes the GLS equality conditions of \citet[Theorem~2]{lu_sch_12} to singular matrices.

Fix an $\Fcal_t$-measurable $n_t\times m_t$ loading matrix $\bm\Phi_t$. We call a pair $(\bm C_t,\bm D_t)$ of $\Fcal_t$-measurable symmetric positive semidefinite matrices, of dimensions $m_t\times m_t$ and $n_t\times n_t$, a \emph{regular decomposition} of $\bm Q_t$ with respect to $\bm\Phi_t$ if
\begin{equation}\label{eq:regulardec}
\bm Q_t=\bm\Phi_t\bm C_t\bm\Phi_t^\top+\bm D_t
\end{equation}
and
\begin{equation}\label{eq:regularim}
\Ima\bm\Phi_t\subseteq\Ima\bm D_t.
\end{equation}
This is the shape of the approximate factor structures of Section~\ref{sec:largeeconomies} and of the running example \eqref{eq:running}: a factor part driven by the loading matrix and a residual part. Regular decompositions of $\bm Q_t$ with respect to $\bm\Phi_t$ exist if and only if the support condition \eqref{eq:Phistarass} holds, see Lemma~\ref{lem:regdec} in Appendix~\ref{app:proofs}, which also collects the proofs for this subsection.

\begin{proposition}[Computing under a Regular Decomposition]\label{prop:COMO}
Let $(\bm C_t,\bm D_t)$ be any regular decomposition of $\bm Q_t$ with respect to $\bm\Phi_t$, which exists if and only if the support condition \eqref{eq:Phistarass} holds. Then the moment-weighted factor portfolios \eqref{eq:MWfactors} of Theorem~\ref{thm:rigid} are given by \eqref{eq:factorportfolios} with weight matrix
\begin{equation}\label{eq:MWD}
\bm W_t^\top=(\bm\Phi_t^\top\bm Q_t^+\bm\Phi_t)^+\bm\Phi_t^\top\bm Q_t^+=(\bm\Phi_t^\top\bm D_t^+\bm\Phi_t)^+\bm\Phi_t^\top\bm D_t^+.
\end{equation}
\end{proposition}
This is the invariance applied to $\bm Q_t$ itself. Computing the moment-weighted factor portfolios under a regular decomposition thus requires no square root of $\bm Q_t$ and no pseudoinverse of $\bm Q_t$, only the pseudoinverse of the residual part, a saving whenever that part is specified in structured form, as in the cases below. Closed-form expressions for the moments of the implied factor representation \eqref{eq:linfacmodelPhi} are given in Lemma~\ref{lem:closedmoments} in Appendix~\ref{app:proofs}.

Combining Lemma~\ref{lem:regdec} with \eqref{eq:admissible}, a loading matrix $\bm\Phi_t$ satisfies \eqref{eq:admissible} if and only if there exist an $\Fcal_t$-measurable $m_t$-vector $\bm b_t$ and a regular decomposition $(\bm C_t,\bm D_t)$ of $\bm Q_t$ with respect to $\bm\Phi_t$ such that $\bm\mu_t=\bm\Phi_t\bm b_t$. Condition \eqref{eq:admissible} is thus the same thing as the existence of a structured specification of $(\bm\mu_t,\bm Q_t)$ built on the loading matrix. This is a reading of \eqref{eq:admissible} rather than a result: the mean restriction is the risk-premium condition \eqref{eq:arpspanning} verbatim, and by Lemma~\ref{lem:regdec} the existence of a regular decomposition is the support condition \eqref{eq:Phistarass} verbatim, so the equivalence carries no content beyond \eqref{eq:admissible} itself.

Proposition~\ref{prop:COMO} removes the pseudoinverse of $\bm Q_t$ in favor of that of the structured residual part. The two results that follow identify when the moment weighting collapses to a weighting the empirical literature already computes: to GLS, and to OLS.

\begin{corollary}[Coincidence with GLS]\label{prop:MWGLS}
Assume the law of one price \eqref{eq:LoP} holds and the loading matrix $\bm\Phi_t$ is admissible, \eqref{eq:admissible}. Then the moment-weighted and the GLS weight matrices coincide,
\begin{equation}\label{eq:MWGLS}
(\bm\Phi_t^\top\bm Q_t^+\bm\Phi_t)^+\bm\Phi_t^\top\bm Q_t^+=(\bm\Phi_t^\top\bm\Sigma_t^+\bm\Phi_t)^+\bm\Phi_t^\top\bm\Sigma_t^+.
\end{equation}
\end{corollary}
The proof is the invariance once more: under the hypotheses, $\bm\mu_t=\bm\Phi_t\bm b_t$ and the pair $(\bm b_t\bm b_t^\top,\bm\Sigma_t)$ is a regular decomposition of $\bm Q_t=\bm\Phi_t\bm b_t\bm b_t^\top\bm\Phi_t^\top+\bm\Sigma_t$, so stripping the mean part of $\bm Q_t$ leaves the GLS weighting. This explains why the GLS cross-sectional regression slopes of the empirical literature, which use $\bm\Sigma_t$ rather than $\bm Q_t$, deliver the coherent factor representation on the admissible loading matrices.

The second removes the covariance information entirely, at the cost of structure on the second moment matrix.

\begin{proposition}[Coincidence with OLS]\label{lem:propGLSsimple}
Assume
\begin{equation}\label{eq:condOLS}
\bm Q_t=\bm\Phi_t\bm C_t\bm\Phi_t^\top+\bm R_t\bm L_t\bm R_t^\top+\sigma_t^2\bm I_{n_t}
\end{equation}
for $\Fcal_t$-measurable symmetric positive semidefinite matrices $\bm C_t$ and $\bm L_t$, a matrix $\bm R_t$ with $\bm R_t^\top\bm\Phi_t=\bm 0$, and a scalar $\sigma_t\ge 0$. If $\sigma_t=0$, assume in addition $\Ima(\bm\Phi_t\bm C_t)=\Ima\bm\Phi_t$, which holds in particular if $\bm C_t$ is nonsingular. Then the moment-weighted and the OLS weight matrices coincide,
\begin{equation}\label{eq:condOLSid}
(\bm\Phi_t^\top\bm Q_t^+\bm\Phi_t)^+\bm\Phi_t^\top\bm Q_t^+=\bm\Phi_t^+.
\end{equation}
\end{proposition}
For $\sigma_t>0$ this is the invariance a third time: $(\bm C_t,\bm R_t\bm L_t\bm R_t^\top+\sigma_t^2\bm I_{n_t})$ is a regular decomposition of $\bm Q_t$, and the remaining weighting matrix treats $\Ima\bm\Phi_t$ isotropically, which is what OLS assumes. The case $\sigma_t=0$ is the shape of the decomposition in \citet[Proposition~2]{kozaknagel23}, see Remark~\ref{rem:KNhedged}. Under \eqref{eq:condOLS} we have $\bm Q_t\bm\Phi_t=\bm\Phi_t(\bm C_t\bm\Phi_t^\top\bm\Phi_t+\sigma_t^2\bm I_{m_t})$, so $\Ima\bm\Phi_t$ is $\bm Q_t$-invariant: the structure is the concrete form of the invariance criterion of Corollary~\ref{cor:OLScoh}. The GLS factor portfolios join the other two whenever, in addition, the loading matrix is admissible and the law of one price holds, by Corollary~\ref{prop:MWGLS}.

\section{Empirical Evidence}\label{sec:empirical}
Our population theory indicates two quantities a researcher can estimate in any sample, at no cost beyond a singular value decomposition: how far the reported loadings sit from the regression loadings of the reported factors, and how much of the risk premia the loadings fail to span. This section computes both and asks whether the weighting the theory singles out has any bearing out of sample. The panels here are portfolio panels, and the exercise is compact by design.\footnote{A study on individual stocks, with the large characteristic sets of the factor representations of Section~\ref{sec:representations} and the singular estimation theory that large unbalanced panels require, is beyond the scope of this paper. \citet{filipovic2025jointestimationconditionalmean} develop the joint conditional mean and covariance estimator such a study needs.}

\subsection{Data and design}\label{sec:datadesign}
We use two balanced panels of monthly value-weighted US portfolio excess returns from the Kenneth French data library: the 25 portfolios sorted on size and book-to-market (FF25), and the 100 portfolios sorted on size and book-to-market, where we remove the 6 portfolios whose return history is incomplete from January 1963, leaving $n=94$ (FF100). The sample runs from July 1963 to June 2026, in total $T=756$ months. The loading matrix $\bm\Phi_t$ consists of a constant column and up to three portfolio-level characteristics: log market capitalization, log book-to-market, and the prior 2--12 month return,\footnote{Market capitalization is the library's average market capitalization of the portfolio's firms. Book-to-market is the library's portfolio-level value-weighted BE/ME series. The prior 2--12 month return compounds the portfolio's simple returns over the eleven months ending in month $t-1$. The replication package documents every convention.} so that $\bm\Phi_t$ is an $n\times m$ matrix with $m\le 4$ that varies with the conditioning information. Each characteristic is standardized with trailing information: the month-$t$ cross-section is centered and scaled by the mean and standard deviation of the pooled characteristic values over the $\Twin$ months ending at $t-1$, so that only information through $t-1$ enters the standardization.

The design is a conditional implementation of the theory with a rolling information set. At each month $t$ conditional moments are estimated from a trailing window of $\Twin$ months, $\hat{\bm\mu}_t=\frac1\Twin\sum_{s=t-\Twin+1}^{t}\bm x_s$, $\hat{\bm Q}_t=\frac1\Twin\sum_{s=t-\Twin+1}^{t}\bm x_s\bm x_s^\top$, the loading matrix $\bm\Phi_t$ uses the month-$t$ standardized characteristics, and every portfolio weight is a function of $(\hat{\bm\mu}_t,\hat{\bm Q}_t,\hat{\bm\Sigma}_t,\bm\Phi_t)$ applied to next month's return $\bm x_{t+1}$. We run the design at window lengths $\Twin=24$, $48$, $96$, $120$, $180$ and $240$ chosen so that the window matrices sit on both sides of the rank boundary: with $\Twin=24$ the matrices $\hat{\bm Q}_t$ and $\hat{\bm\Sigma}_t$ are singular on both panels in every month, with $\Twin=48$ they are singular on FF100 only, where $\hat{\bm Q}_t$ loses 46 directions and $\hat{\bm\Sigma}_t$ loses 47, and with $\Twin=96$ and above they are nonsingular on both. Each design is evaluated on its own sample, from the first month with a full estimation and standardization window through June 2026, which gives 731, 707, 659, 635, 575 and 515 months at $\Twin=24$, $48$, $96$, $120$, $180$ and $240$. The Sharpe ratios of Section~\ref{sec:admissibilityspanning} apply weights fixed at $t$ to $\bm x_{t+1}$ and are out of sample, while the two diagnostics of Section~\ref{sec:diagnostics} are functions of $(\bm\Phi_t,\hat{\bm\mu}_t,\hat{\bm Q}_t)$ inside the window, exact statements about estimated moments and read as descriptive.

Pseudoinverses are computed by singular value decomposition at relative tolerance $10^{-10}$. The retained directions decay steeply, with the largest eigenvalue carrying on average 85\% and 77\% of the trace of $\hat{\bm Q}_t$ on FF25 and FF100.
Every number is deterministically reproducible, with no estimation step to tune and no randomness.

\subsection{The loading discrepancy}\label{sec:diagnostics}
\begin{figure}[tbp]
\centering
\includegraphics[width=\textwidth]{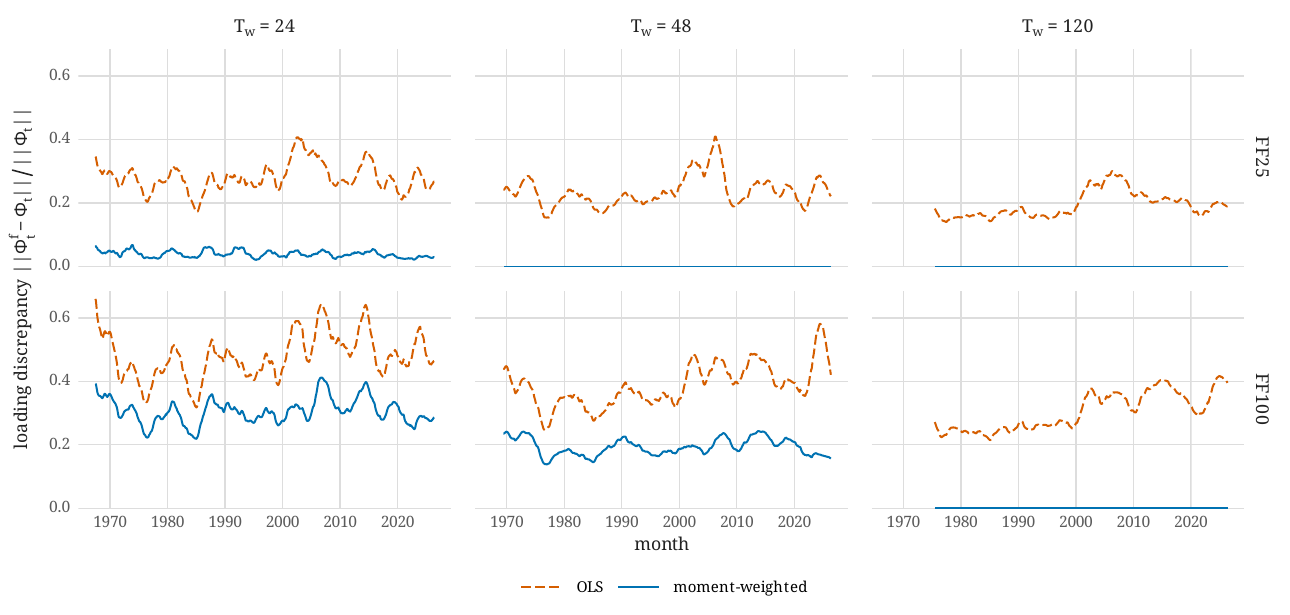}
\caption{The loading discrepancy $\|\bm\Phi^{\bm f_{t+1}}_t-\bm\Phi_t\|/\|\bm\Phi_t\|$ of the OLS and the moment-weighted factor portfolios, with the constant column and the three standardized characteristics as loadings, $m=4$, as 24-month trailing moving averages. Columns are the estimation windows $\Twin=24$, $48$ and $120$, rows the two panels. The moment-weighted series coincides with the loading gap of Corollary~\ref{cor:cohgap} and is zero wherever $\Twin\ge n$.}
\label{fig:loadinggap}
\end{figure}

Corollary~\ref{cor:cohgap} measures a factor representation by the distance between the loadings and the regression coefficients of the factors it reports, $\|\bm\Phi^{\bm f_{t+1}}_t-\bm\Phi_t\|/\|\bm\Phi_t\|$ in the Frobenius norm, and states that the minimum of that distance over all traded factors is the relative loading gap $\|(\bm I_n-\hat{\bm Q}_t\hat{\bm Q}_t^+)\bm\Phi_t\|/\|\bm\Phi_t\|$, attained by the moment-weighted factor portfolios. Figure~\ref{fig:loadinggap} plots the discrepancy month by month for the OLS and the moment-weighted factors, with the constant column included, $m=4$, on both panels and at the windows $\Twin=24$, $48$ and $120$.

The moment-weighted series is positive exactly where the support condition \eqref{eq:Phistarass} fails, averaging 0.04 on FF25 at $\Twin=24$ and 0.31 and 0.19 on FF100 at $\Twin=24$ and $48$, where the moment-weighted factors carry the effective loading matrix $\hat{\bm Q}_t\hat{\bm Q}_t^+\bm\Phi_t$, the part of $\bm\Phi_t$ the window can identify, and it vanishes identically wherever $\Twin\ge n$, in particular at $\Twin=120$, where the window covers both cross-sections. As anticipated by the theory, the OLS discrepancy behaves differently. It sits well above the floor in every month, averaging 0.28, 0.24 and 0.19 on FF25 and 0.48, 0.39 and 0.31 on FF100 across these three windows, and it does not follow the floor to zero as the window lengthens. The distance is the failure of invariance and not the rank of the window: a matrix of firm characteristics is not an invariant subspace of $\hat{\bm Q}_t$, which is Corollary~\ref{cor:OLScoh} in the data. The moment-weighted series attains the floor to numerical precision, with a maximal deviation of $5.5\times10^{-12}$ across both panels and all windows. The magnitude of the discrepancy depends on the standardization of the characteristics, while whether it vanishes does not, since any rescaling that preserves the column span of $\bm\Phi_t$ leaves \eqref{eq:coh1} unchanged, by Remark~\ref{rem:onlyimage}. Section~\ref{sec:guidelines} takes the floor up as a guideline for design.

\begin{figure}[tbp]
\centering
\includegraphics[width=\textwidth]{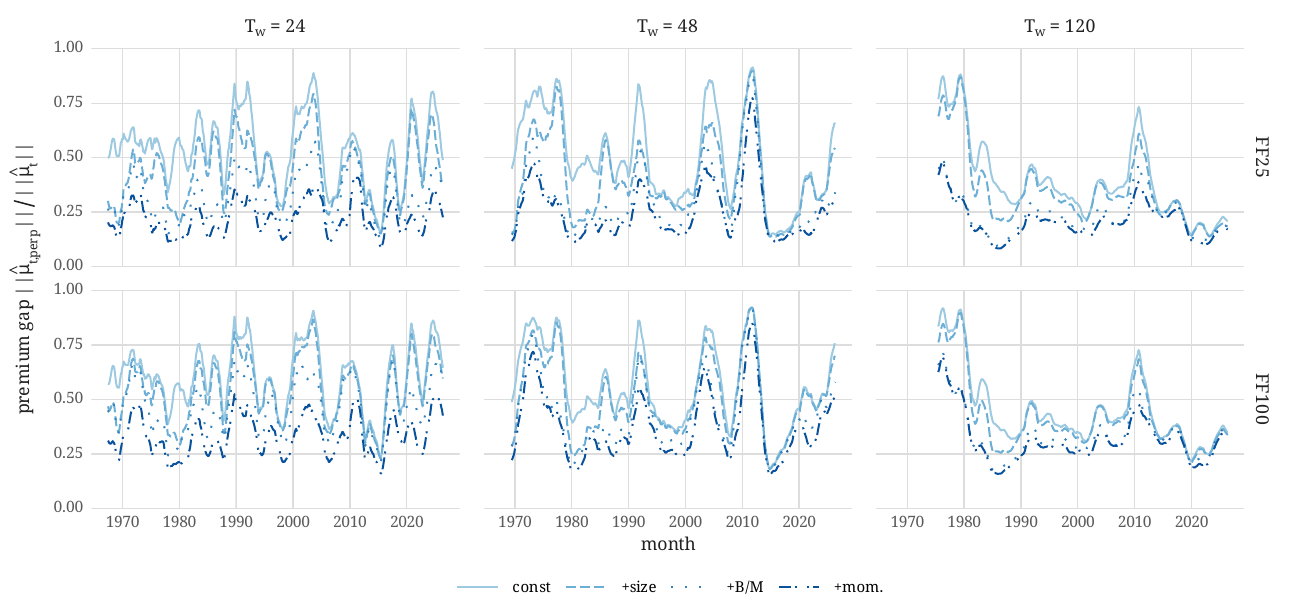}
\caption{The estimated risk-premium gap $\|\hat{\bm\mu}_{t,\perp}\|/\|\hat{\bm\mu}_t\|$ month by month, as a 24-month trailing moving average, for the four nested characteristic sets. Columns are the estimation windows $\Twin=24$, $48$ and $120$, rows the two panels.}
\label{fig:premiumgap}
\end{figure}

We also assess the risk-premium gap, which measures the share of the estimated conditional mean that the loadings fail to span, the failure of \eqref{eq:arpspanning}. Each month we split $\hat{\bm\mu}_t$ orthogonally along $\Ima\bm\Phi_t$ and record the share $\|\hat{\bm\mu}_{t,\perp}\|/\|\hat{\bm\mu}_t\|$ that the loading matrix cannot reach, the distance to the lower endpoint of \eqref{eq:admissible}. Figure~\ref{fig:premiumgap} plots it for the nested characteristic sets, the constant alone, then adding size, book-to-market and momentum. The gap falls as characteristics are added, on average from 0.50 to 0.25 on FF25 and from 0.56 to 0.38 on FF100 at $\Twin=48$, with the same ordering at the other windows, and adding a column can only shrink it, since it enlarges $\Ima\bm\Phi_t$. It also falls as the window lengthens from $\Twin=48$ on, to 0.21 and 0.32 at the full set at $\Twin=120$. The gap is episodic rather than trending. Long stretches at moderate levels alternate with pronounced excursions, most visibly in the early 1970s, in 2003--04, and, largest of all, in 2010--11, consistent with the time variation in characteristic premia documented by \citet{mcleanpontiff16}. The framework permits this: our results restrict the joint specification within a period and not how the mapping from characteristics to premia moves across periods. The gap is informative as a monitor of when a characteristic set loses the premia. It also has a second footprint: the moment-weighted factor representation has residuals orthogonal to its factors, so $\cov_t[\bm\epsilon_{t+1},\bm f_{t+1}]=-\E_t[\bm\epsilon_{t+1}]\bm\mu_{\bm f,t}^\top$, and the residuals are correlated with the factors whenever the gap is positive and the factors carry premia.


\subsection{Out-of-sample performance}\label{sec:admissibilityspanning}
In this section we investigate how, if at all, the estimated loading and risk-premium gaps affect Sharpe ratios. The theory awards the maximal Sharpe ratio \eqref{eq:maxSR} to coherent factor representations only, and coherence needs the risk-premium condition \eqref{eq:arpspanning}, a property of the loading matrix that no weighting supplies as Figure~\ref{fig:premiumgap} indicates. When considering Sharpe ratios, there is a second effect of the sample size beyond estimation error from the design of the study. For $\Twin\le n$ the rank of $\hat{\bm\Sigma}_t$ is $\Twin-1<n$, so $\hat{\bm\mu}_t\notin\Ima\hat{\bm\Sigma}_t$ and the law of one price \eqref{eq:LoP} fails for the window moments: $(\bm I_n-\hat{\bm\Sigma}_t\hat{\bm\Sigma}_t^+)\hat{\bm\mu}_t$ has zero in-window variance and a positive in-window mean, an exact in-sample arbitrage absorbing 27\% of $\|\hat{\bm\mu}_t\|$ on FF100 at $\Twin=48$, so the in-window maximal Sharpe ratio is infinite and Proposition~\ref{prop:span} and Remark~\ref{rem:SharpeCU} are unavailable. To discern the loading gap from law of one price violations, we apply the linear shrinkage of \citet{ledoitwolf04} (``LW'') to the estimated conditional covariance matrix, to make the covariance matrix invertible for the law of one price to hold.

\begin{figure}[tbp]
\centering
\includegraphics[width=\textwidth]{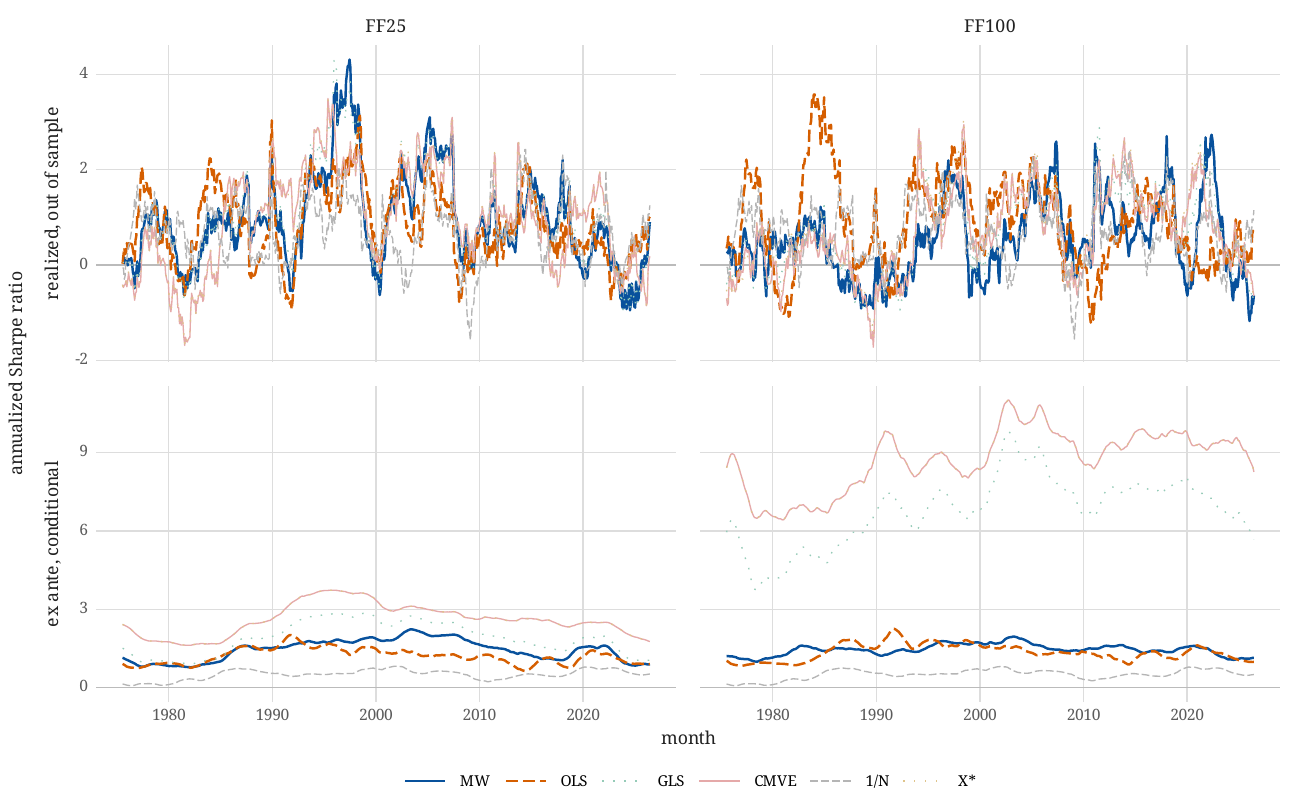}
\caption{Annualized Sharpe ratios at $\Twin=120$ for the moment-weighted, OLS and GLS factor representations at $m=4$ and the CMVE, $1/N$ and $X^\star$ benchmarks. The upper row is the realized out-of-sample ratio over rolling trailing 24-month windows. The lower row is the ex ante ratio $(\bm w_t^\top\hat{\bm\mu}_t)/(\bm w_t^\top\hat{\bm\Sigma}_t\bm w_t)^{1/2}$ the same weights promise from the window moments at $t$, as a 24-month moving average. The rows carry separate vertical scales. There $X^\star$ coincides with CMVE, the two differing by the positive conditional scalar of Corollary~\ref{cor:SMQ}, which the realized ratio does not preserve.}
\label{fig:srrolling}
\end{figure}

Figure~\ref{fig:srrolling} shows ex ante Sharpe ratios in the bottom row and rolling estimates of ex post Sharpe ratios in the top, both at $\Twin=120$. Table~\ref{tab:spanning} pools the ratios for the factor representations and the $X^\star$ benchmark across all six windows.
Performance is not monotone in the window. A striking data point is at $n/\Twin\approx1$, where the Sharpe ratio collapses to $0.01$ for the moment weighting and $-0.03$ for GLS on FF100 at $\Twin=96$, where the in-sample squared Sharpe ratio averages $289$ monthly, while OLS, which inverts no return moment, holds at $0.61$. Ledoit--Wolf shrinkage removes the collapse, and the moment weighting then exceeds OLS in every window with $\Twin>n$, where \eqref{eq:LoP} holds in sample under the plain window moments, on both panels, by $0.07$ to $0.21$. GLS leads it throughout those rows by $0.05$ to $0.18$, which is the cost of coherence: $\hat{\bm Q}_t=\hat{\bm\Sigma}_t+\hat{\bm\mu}_t\hat{\bm\mu}_t^\top$ carries the estimated premia into the moment-weighted span and $\hat{\bm\Sigma}_t$ alone does not, the two coinciding when the gap closes.

\begin{table}[tbp]
\centering
\footnotesize
\setlength{\tabcolsep}{4.5pt}
\renewcommand{\arraystretch}{0.92}
\begin{tabular}{llccccccccc}
\toprule
 & & & & & \multicolumn{2}{c}{$\text{SR}$ MW} & \multicolumn{2}{c}{$\text{SR}$ GLS} & \multicolumn{2}{c}{$\text{SR}$ $X^\star$} \\
\cmidrule(lr){6-7}\cmidrule(lr){8-9}\cmidrule(lr){10-11}
Panel & $\Twin$ & $n/\Twin$ & $\overline{\text{RP gap}}$ & $\text{SR}$ OLS & none & LW & none & LW & none & LW \\
\midrule
FF25 & 24$^\dagger$ & 1.04 & 0.22 & 0.54 & 0.12 & 0.66 & 0.26 & 0.66 & 0.08 & 0.77 \\
 & 48 & 0.52 & 0.25 & 0.62 & 0.72 & 0.78 & 0.77 & 0.86 & 0.66 & 0.90 \\
 & 96 & 0.26 & 0.23 & 0.66 & 0.69 & 0.75 & 0.79 & 0.83 & 0.83 & 0.94 \\
 & 120 & 0.21 & 0.21 & 0.73 & 0.86 & 0.85 & 0.99 & 0.91 & 0.93 & 0.98 \\
 & 180 & 0.14 & 0.17 & 0.83 & 0.92 & 0.94 & 1.00 & 1.00 & 0.99 & 1.07 \\
 & 240 & 0.10 & 0.16 & 0.84 & 0.99 & 0.97 & 1.06 & 1.02 & 1.12 & 1.17 \\
\addlinespace
FF100 & 24$^\dagger$ & 3.92 & 0.33 & 0.56 & 0.81 & 0.52 & 0.63 & 0.48 & 0.80 & 0.83 \\
 & 48$^\dagger$ & 1.96 & 0.38 & 0.62 & 0.86 & 0.74 & 0.76 & 0.91 & 0.67 & 0.82 \\
 & 96 & 0.98 & 0.34 & 0.61 & 0.01 & 0.69 & $-0.03$ & 0.87 & 0.12 & 0.98 \\
 & 120 & 0.78 & 0.32 & 0.72 & 0.42 & 0.79 & 0.62 & 0.95 & 0.58 & 1.08 \\
 & 180 & 0.52 & 0.26 & 0.76 & 0.76 & 0.95 & 0.99 & 1.12 & 0.85 & 1.08 \\
 & 240 & 0.39 & 0.24 & 0.80 & 0.99 & 1.01 & 1.16 & 1.16 & 1.07 & 1.21 \\
\bottomrule
\end{tabular}
\caption{Annualized out-of-sample Sharpe ratios by estimation window at the full characteristic set, $m=4$, each window evaluated one step ahead on its own sample with weights formed from information through $t$. ``MW'' denotes the moment-weighted factor portfolios of Theorem~\ref{thm:rigid}, ``$\overline{\text{RP gap}}$'' is the time-averaged share of the window risk premia outside $\Ima\bm\Phi_t$, and ``SR $X^\star$'' is the efficient portfolio \eqref{eq:defXstar}, with weights $\hat{\bm Q}_t^+\hat{\bm\mu}_t$. Each ratio is reported under the plain window moments (``none'') and under the linear shrinkage of \citet{ledoitwolf04} (``LW''), applied to $\hat{\bm\Sigma}_t$, from which we rebuild $\tilde{\bm Q}_t=\tilde{\bm\Sigma}_t+\hat{\bm\mu}_t\hat{\bm\mu}_t^\top$. The intensity $\delta_t$ is analytic, uses window data alone, and involves no tuning. As $\bm\Phi_t^+$ uses no return moment, the OLS column is common to both. $^\dagger$ marks windows with $\Twin\le n$, where $\rank\hat{\bm\Sigma}_t=\Twin-1<n$, the law of one price \eqref{eq:LoP} fails for the window moments and the in-window maximal Sharpe ratio is infinite.}
\label{tab:spanning}
\end{table}

Two further readings of the $\Twin=120$ design complete the picture. First, pricing: the norm of the time-averaged out-of-sample residual of the moment-weighted factor representation, relative to the norm of time-averaged returns, falls from $0.21$ at the constant alone to $0.12$ at the full characteristic set on FF25, and from $0.30$ to $0.18$ on FF100, so the characteristic sets that shrink the in-window risk-premium gap of Figure~\ref{fig:premiumgap} also shrink the out-of-sample pricing errors, in the same order. Second, an invariant benchmark: principal components of $\hat{\bm Q}_t$ with the same number of factors satisfy \eqref{eq:coh1} by construction, by Corollary~\ref{cor:OLScoh}, and reach $0.69$ on FF25 and $0.64$ on FF100 at the full set, below the moment weighting on FF25 and above it on FF100, where the window sits near the rank boundary: an invariant span buys consistency with the reported loadings but gives up the choice of column space, as Section~\ref{sec:leadingreps} discusses.

One reading must be resisted. The singularity this paper is about lives in the return distribution: Proposition~\ref{prop:rankdef} and Theorem~\ref{thm:rigid} are population statements, hold for every $n_t$, and use pseudoinverses so that no rank condition is needed. The deficiency in Table~\ref{tab:spanning} has the opposite source, too few observations in a rolling window, and the population second moment matrix of $25$ or $94$ portfolios might well be non-singular. Estimation and inference under a potentially rank-deficient design with a potentially singular second moment matrix is a separate problem, treated by \citet{quainitrojani26}. We attempt no within-window inference, and the table is to be interpreted as a design diagnostic, rather than a formal statistical test of the theory.

\subsection{Guidelines}\label{sec:guidelines}
The analysis suggests five guidelines for specifying, implementing, and evaluating characteristic-based factor models in a conditional design. The first is to specify the triple $(\bm\Phi_t,\bm\mu_t,\bm Q_t)$ jointly: \eqref{eq:admissible} restricts all three at once, and the risk-premium gap of Section~\ref{sec:diagnostics} makes the mean restriction measurable in every window at no cost. The second is to implement factor portfolios with pseudoinverses, at least as a validation and parameter-free benchmark for the many types of shrinkage in the literature: the moment-weighted portfolios of Theorem~\ref{thm:rigid} are well defined at any rank, the computation is a singular value decomposition whose tolerance should be reported, and any window shorter than the cross-section places $\hat{\bm Q}_t$ beyond the rank boundary by construction. The third is to check the support condition and report the effective loadings: with $\rank\hat{\bm Q}_t<n$ the factors load only on $\hat{\bm Q}_t\hat{\bm Q}_t^+\bm\Phi_t$, and the discarded share, the floor of Section~\ref{sec:diagnostics}, measures the window-versus-dimension trade-off of the design rather than the quality of the specification, provided the deficient rank is the window's and not the population's. A related zero-cost check is invariance: with a constant column in $\bm\Phi_t$, cross-sectional standardization must leave factor spans, pricing errors and Sharpe ratios exactly unchanged, by Remark~\ref{rem:onlyimage}, so any difference beyond the numerical tolerance of the implementation, $2\times10^{-12}$ in ours, reveals an implementation error. The fourth is to know when the weightings coincide: when the second moment matrix has the structure \eqref{eq:condOLS}, moment weighting reduces to OLS, Proposition~\ref{lem:propGLSsimple}, GLS joins them for admissible loadings under the law of one price, Corollary~\ref{prop:MWGLS}, and the specification \eqref{eq:running} meets all of these, so there the three weightings produce the same factor portfolios, see Section~\ref{sec:computation}. Observed differences between them are themselves diagnostic of departures from that structure. The fifth is to compute and report the loading discrepancy of Section~\ref{sec:diagnostics}: the loadings a factor representation reports are the regression loadings of its factors only under the invariance of Corollary~\ref{cor:OLScoh}, the variance decompositions and risk attributions built on the reported loadings inherit the discrepancy, and the discrepancy measures the distance from that invariance. The computation is a singular value decomposition the representation already performs, and Corollary~\ref{cor:cohgap} supplies the floor below which no traded factor representation carrying the given loadings can go.

\section{Conclusion}\label{sec:conclusion}
This paper investigates characteristics-based factor representations \eqref{eq:linfacmodelPhi} of finite return panels in population, through the four properties of the introduction: \ref{it:prop1} the loadings are the regression loadings of the factors, \ref{it:prop2} residual risk is unpriced, \ref{it:prop3} the factors span the maximal Sharpe ratio attainable from the assets, and \ref{it:prop4} the residuals are idiosyncratic. Property~\ref{it:prop1} is attainable by exactly one traded factor representation, the moment-weighted factor portfolios. Property~\ref{it:prop2} restricts the loading matrix. Property~\ref{it:prop3} follows from properties \ref{it:prop1} and \ref{it:prop2} under the law of one price, and property~\ref{it:prop4} is unattainable for traded factors.

Our analysis has direct implications for practice. A factor representation specifies not only loadings but, at least implicitly, also conditional first and second moments of returns: only certain combinations of the three allow characteristics to be covariances, in the sense that the loading matrix is the beta matrix in a return regression. The distance between the reported loadings and the regression loadings of the reported factors is a natural measure of misspecification, and we discuss instrumented and projected principal component analysis, among other representations, in this light. The reported loadings are not the regression loadings of the reported factors unless the invariance of Corollary~\ref{cor:OLScoh} holds, so the discrepancy should be computed and reported, as Section~\ref{sec:guidelines} sets out. The factor--residual correlations that characteristic-based portfolios exhibit are implied rather than anomalous, Proposition~\ref{prop:rankdef}, and cannot be removed by any choice of traded factors that retains a full-rank residual covariance matrix.

For implementation, we provide a factor modeling framework based on pseudoinverses. It accommodates recent developments in which the number of factors or features exceeds the number of returns to be explained, and it is a reproducible benchmark for the many types of shrinkage used in the empirical literature: the pseudoinverse carries no tuning parameter, so the benchmark needs no validation.

We leave two directions to future research. Estimation theory for the conditional moments entering the characterization, in particular for large unbalanced panels, is beyond the scope of this paper. \citet{filipovic2025jointestimationconditionalmean} develop a joint estimator of the conditional mean and covariance in that setting. And the large-economy discussion of Section~\ref{sec:largeeconomies} is deliberately elementary. A full asymptotic theory of \eqref{eq:admissible} and factor spanning as the cross-section grows remains to be developed.


\bibliographystyle{apalike}
\bibliography{master}

\begin{thebibliography}{}

\bibitem[Admati and Pfleiderer, 1985]{admatipfleiderer85}
Admati, A.~R. and Pfleiderer, P. (1985).
\newblock Interpreting the factor risk premia in the arbitrage pricing theory.
\newblock {\em Journal of Economic Theory}, 35(1):191--195.

\bibitem[Albert, 1969]{alb_69}
Albert, A. (1969).
\newblock Conditions for positive and nonnegative definiteness in terms of
  pseudoinverses.
\newblock {\em SIAM Journal on Applied Mathematics}, 17(2):434--440.

\bibitem[Anderson and Duffin, 1969]{and_duf_69}
Anderson, W. and Duffin, R. (1969).
\newblock Series and parallel addition of matrices.
\newblock {\em Journal of Mathematical Analysis and Applications},
  26(3):576--594.

\bibitem[Bai, 2003]{bai03}
Bai, J. (2003).
\newblock Inferential theory for factor models of large dimensions.
\newblock {\em Econometrica}, 71(1):135--171.

\bibitem[Balduzzi and Robotti, 2008]{bal_rob_08}
Balduzzi, P. and Robotti, C. (2008).
\newblock Mimicking portfolios, economic risk premia, and tests of multi-beta
  models.
\newblock {\em Journal of Business \& Economic Statistics}, 26(3):354--368.

\bibitem[Bernstein, 2018]{ber_18}
Bernstein, D.~S. (2018).
\newblock {\em Scalar, Vector, and Matrix Mathematics: Theory, Facts, and
  Formulas - Revised and Expanded Edition}.
\newblock Princeton University Press, revised edition.

\bibitem[Breeden et~al., 1989]{bre_gib_lit_89}
Breeden, D.~T., Gibbons, M.~R., and Litzenberger, R.~H. (1989).
\newblock Empirical tests of the consumption-oriented {CAPM}.
\newblock {\em The Journal of Finance}, 44(2):231--262.

\bibitem[Chamberlain and Rothschild, 1983]{chamberlainrothschild83}
Chamberlain, G. and Rothschild, M. (1983).
\newblock Arbitrage, factor structure, and mean-variance analysis on large
  asset markets.
\newblock {\em Econometrica}, 51(5):1281--1304.

\bibitem[Chen et~al., 2023]{chenroussanovwang23}
Chen, Q., Roussanov, N., and Wang, X. (2023).
\newblock Semiparametric conditional factor models: Estimation and inference.
\newblock Working Paper 31817, National Bureau of Economic Research.

\bibitem[Cochrane, 2005]{Cochranebook2005}
Cochrane, J. (2005).
\newblock {\em Asset Pricing}.
\newblock Princeton University Press, revised edition.

\bibitem[Connor, 1984]{connor84}
Connor, G. (1984).
\newblock A unified beta pricing theory.
\newblock {\em Journal of Economic Theory}, 34(1):13--31.

\bibitem[Connor et~al., 2012]{connorhagmannlinton12}
Connor, G., Hagmann, M., and Linton, O. (2012).
\newblock Efficient semiparametric estimation of the fama-french model and
  extensions.
\newblock {\em Econometrica}, 80(2):713--754.

\bibitem[Connor and Korajczyk, 1986]{ConnorKorajckzyk1986}
Connor, G. and Korajczyk, R.~A. (1986).
\newblock Performance measurement with the arbitrage pricing theory: A new
  framework for analysis.
\newblock {\em Journal of Financial Economics}, 15(3):373--394.

\bibitem[Daniel and Titman, 1997]{danieltitman97}
Daniel, K. and Titman, S. (1997).
\newblock Evidence on the characteristics of cross sectional variation in stock
  returns.
\newblock {\em Journal of Finance}, 52(1):1--33.

\bibitem[Dudley, 2002]{dud_02}
Dudley, R.~M. (2002).
\newblock {\em Real analysis and probability}, volume~74 of {\em Cambridge
  Studies in Advanced Mathematics}.
\newblock Cambridge University Press, Cambridge.
\newblock Revised reprint of the 1989 original.

\bibitem[Fama and MacBeth, 1973]{famamacbeth73}
Fama, E.~F. and MacBeth, J.~D. (1973).
\newblock Risk, return, and equilibrium: Empirical tests.
\newblock {\em Journal of Political Economy}, 81(3):607--636.

\bibitem[Fan et~al., 2022]{fankeliaoneuhierl22}
Fan, J., Ke, Z.~T., Liao, Y., and Neuhierl, A. (2022).
\newblock Structural deep learning in conditional asset pricing.
\newblock Available at SSRN 4117882.

\bibitem[Fan et~al., 2016]{fanliaowang16}
Fan, J., Liao, Y., and Wang, W. (2016).
\newblock Projected principal component analysis in factor models.
\newblock {\em The Annals of Statistics}, 44(1):219--254.

\bibitem[Fan, 1949]{fan1949}
Fan, K. (1949).
\newblock On a theorem of {W}eyl concerning eigenvalues of linear
  transformations {I}.
\newblock {\em Proceedings of the National Academy of Sciences},
  35(11):652--655.

\bibitem[Filipovic et~al., 2009]{fil_kup_vog_09}
Filipovic, D., Kupper, M., and Vogelpoth, N. (2009).
\newblock Separation and duality in locally {$L^0$}-convex modules.
\newblock {\em J. Funct. Anal.}, 256(12):3996--4029.

\bibitem[Filipovic et~al., 2012]{fil_kup_12}
Filipovic, D., Kupper, M., and Vogelpoth, N. (2012).
\newblock Approaches to conditional risk.
\newblock {\em SIAM J. Financial Math.}, 3(1):402--432.

\bibitem[Filipovic and Schneider,
  2025]{filipovic2025jointestimationconditionalmean}
Filipovic, D. and Schneider, P. (2025).
\newblock Joint estimation of conditional mean and covariance for unbalanced
  panels.
\newblock working paper, EPFL, Universit\`a della Svizzera italiana, and SFI.

\bibitem[F\"ollmer and Schied, 2016]{foe_sch_16}
F\"ollmer, H. and Schied, A. (2016).
\newblock {\em Stochastic finance}.
\newblock De Gruyter Graduate. De Gruyter, Berlin, extended edition.
\newblock An introduction in discrete time.

\bibitem[Fortin et~al., 2025]{fortin2025eigenvalue}
Fortin, A.-P., Gagliardini, P., and Scaillet, O. (2025).
\newblock Eigenvalue tests for the number of latent factors in short panels.
\newblock {\em Journal of Financial Econometrics}, 23(1):nbad024.

\bibitem[Fortin et~al., 2026]{fortingagliardiniscaillet24}
Fortin, A.-P., Gagliardini, P., and Scaillet, O. (2026).
\newblock Latent factor analysis in short panels.
\newblock {\em Journal of Econometrics}, 255:106249.

\bibitem[Freyberger et~al., 2020]{freybergerneuhierlweber20}
Freyberger, J., Neuhierl, A., and Weber, M. (2020).
\newblock {Dissecting Characteristics Nonparametrically}.
\newblock {\em Review of Financial Studies}, 33(5):2326--2377.

\bibitem[Gagliardini et~al., 2016]{gagliardiniossolascaillet16}
Gagliardini, P., Ossola, E., and Scaillet, O. (2016).
\newblock Time-varying risk premium in large cross-sectional equity data sets.
\newblock {\em Econometrica}, 84(3):985--1046.

\bibitem[Gallant et~al., 1990]{gal_han_tau_90}
Gallant, A.~R., Hansen, L.~P., and Tauchen, G. (1990).
\newblock Using conditional moments of asset payoffs to infer the volatility of
  intertemporal marginal rates of substitution.
\newblock {\em Journal of Econometrics}, 45(1-2):141--179.

\bibitem[Gibbons et~al., 1989]{gib_ros_sha_89}
Gibbons, M.~R., Ross, S.~A., and Shanken, J. (1989).
\newblock A test of the efficiency of a given portfolio.
\newblock {\em Econometrica}, 57(5):1121--1152.

\bibitem[Gu et~al., 2020]{gukellyxiu20}
Gu, S., Kelly, B., and Xiu, D. (2020).
\newblock Autoencoder asset pricing models.
\newblock {\em Journal of Econometrics}.

\bibitem[Hansen and Jagannathan, 1991]{han_jag_91}
Hansen, L.~P. and Jagannathan, R. (1991).
\newblock Implications of security market data for models of dynamic economies.
\newblock {\em Journal of Political Economy}, 99(2):225--262.

\bibitem[Hansen and Richard, 1987]{han_ric_87}
Hansen, L.~P. and Richard, S.~F. (1987).
\newblock The role of conditioning information in deducing testable
  restrictions implied by dynamic asset pricing models.
\newblock {\em Econometrica}, 55(3):587--613.

\bibitem[Horn and Johnson, 1990]{hor_joh_85}
Horn, R.~A. and Johnson, C.~R. (1990).
\newblock {\em Matrix analysis}.
\newblock Cambridge University Press, Cambridge.
\newblock Corrected reprint of the 1985 original.

\bibitem[Huberman et~al., 1987]{hub_kan_sta_87}
Huberman, G., Kandel, S., and Stambaugh, R.~F. (1987).
\newblock Mimicking portfolios and exact arbitrage pricing.
\newblock {\em The Journal of Finance}, 42(1):1--9.

\bibitem[Kelly et~al., 2026]{kellykuznetsovmalamudzhang26}
Kelly, B., Kuznetsov, B., Malamud, S., and Zhang, Y. (2026).
\newblock Large and deep factor models.
\newblock arXiv:2402.06635.

\bibitem[Kelly et~al., 2019]{kellypruittsu19}
Kelly, B.~T., Pruitt, S., and Su, Y. (2019).
\newblock Characteristics are covariances: A unified model of risk and return.
\newblock {\em Journal of Financial Economics}, 134(3):501--524.

\bibitem[Kozak and Nagel, 2023]{kozaknagel23}
Kozak, S. and Nagel, S. (2023).
\newblock When do cross-sectional asset pricing factors span the stochastic
  discount factor?
\newblock Working Paper 31275, National Bureau of Economic Research.
\newblock Revised November 2024; equation and result numbers cited in this
  paper are those of the November 2024 version.

\bibitem[Ledoit and Wolf, 2004]{ledoitwolf04}
Ledoit, O. and Wolf, M. (2004).
\newblock A well-conditioned estimator for large-dimensional covariance
  matrices.
\newblock {\em Journal of Multivariate Analysis}, 88(2):365--411.

\bibitem[Ledoit and Wolf, 2022]{ledoitwolf20}
Ledoit, O. and Wolf, M. (2022).
\newblock {The Power of (Non-)Linear Shrinking: A Review and Guide to
  Covariance Matrix Estimation}.
\newblock {\em Journal of Financial Econometrics}, 20(1):187--218.

\bibitem[Lehmann and Modest, 1988]{leh_mod_88}
Lehmann, B.~N. and Modest, D.~M. (1988).
\newblock The empirical foundations of the arbitrage pricing theory.
\newblock {\em Journal of Financial Economics}, 21(2):213--254.

\bibitem[Lettau and Pelger, 2020]{lettaupelger20}
Lettau, M. and Pelger, M. (2020).
\newblock Estimating latent asset-pricing factors.
\newblock {\em Journal of Econometrics}, 218(1):1--31.

\bibitem[Lu and Schmidt, 2012]{lu_sch_12}
Lu, C. and Schmidt, P. (2012).
\newblock Conditions for the numerical equality of the {OLS}, {GLS} and
  {A}memiya–{C}ragg estimators.
\newblock {\em Economics Letters}, 116(3):538--540.

\bibitem[Mardia et~al., 1979]{mar_ken_bib_79}
Mardia, K.~V., Kent, J.~T., and Bibby, J.~M. (1979).
\newblock {\em Multivariate analysis}.
\newblock Probability and Mathematical Statistics: A Series of Monographs and
  Textbooks. Academic Press [Harcourt Brace Jovanovich, Publishers], London-New
  York-Toronto.

\bibitem[McLean and Pontiff, 2016]{mcleanpontiff16}
McLean, R.~D. and Pontiff, J. (2016).
\newblock Does academic research destroy stock return predictability?
\newblock {\em Journal of Finance}, 71(1):5--32.

\bibitem[Minamide, 1985]{min_85}
Minamide, N. (1985).
\newblock An extension of the matrix inversion lemma.
\newblock {\em SIAM J. Algebraic Discrete Methods}, 6(3):371--377.

\bibitem[Ogawa, 1988]{oga_88}
Ogawa, H. (1988).
\newblock An operator pseudo-inversion lemma.
\newblock {\em SIAM J. Appl. Math.}, 48(6):1527--1531.

\bibitem[Quaini and Trojani, 2026]{quainitrojani26}
Quaini, A. and Trojani, F. (2026).
\newblock Proximal estimation and inference.
\newblock {\em Econometric Theory}.
\newblock Forthcoming.

\bibitem[Raponi et~al., 2020]{raponirobottizaffaroni19}
Raponi, V., Robotti, C., and Zaffaroni, P. (2020).
\newblock {Testing Beta-Pricing Models Using Large Cross-Sections}.
\newblock {\em The Review of Financial Studies}, 33(6):2796--2842.
\newblock Advance Access publication 10 September 2019.

\bibitem[Ross, 1976]{ross76}
Ross, S.~A. (1976).
\newblock The arbitrage theory of capital asset pricing.
\newblock {\em Journal of Economic Theory}, 13(3):341--360.

\bibitem[Schott, 2017]{sch_17}
Schott, J.~R. (2017).
\newblock {\em Matrix analysis for statistics}.
\newblock Wiley Series in Probability and Statistics. John Wiley \& Sons, Inc.,
  Hoboken, NJ, third edition.

\bibitem[Zaffaroni, 2025]{zaffaroni25}
Zaffaroni, P. (2025).
\newblock Factor models for conditional asset pricing.
\newblock {\em Journal of Political Economy}, 133(8):2615--2642.

\end{thebibliography}

\begin{appendix}

\section{Matrix Preliminaries}\label{app:matrix}
We collect the linear algebra used throughout the paper. All new statements are elementary.

For an $n\times m$ real matrix $\bm A$, we write $\Ima\bm A\coloneqq\{\bm A\bm v:\bm v\in\R^m\}\subseteq\R^n$ for its image (column space) and $\ker\bm A\coloneqq\{\bm v\in\R^m:\bm A\bm v=\bm 0\}\subseteq\R^m$ for its kernel (null space). The fundamental orthogonal decompositions are $\R^n=\Ima\bm A\oplus\ker\bm A^\top$ and $\R^m=\Ima\bm A^\top\oplus\ker\bm A$, where $\oplus$ denotes the direct sum of two subspaces $U,V$ with $U\cap V=\{\bm 0\}$, which is orthogonal in these two cases. For subspaces $U,V\subseteq\R^n$ we write $U+V\coloneqq\{\bm u+\bm v:\bm u\in U,\,\bm v\in V\}$ for their sum and $U^\perp$ for the orthogonal complement, with $(U\cap V)^\perp=U^\perp+V^\perp$. For symmetric positive semidefinite matrices $\bm A,\bm B$ we have $\Ima(\bm A+\bm B)=\Ima\bm A+\Ima\bm B$. For symmetric matrices $\bm A,\bm B$ we write $\bm A\preceq\bm B$, equivalently $\bm B\succeq\bm A$, the Loewner order, if $\bm B-\bm A$ is positive semidefinite.

The pseudoinverse $\bm A^+$ of a matrix $\bm A$, also called the Moore--Penrose generalized inverse, is defined as follows. Let $\bm A=\bm V \bm D \bm W^\top$ be the singular value decomposition of $\bm A$, with orthogonal matrices $\bm V$ and $\bm W$ and a (possibly rectangular) diagonal matrix $\bm D$. Then
\[
\bm A^+ = \bm W \bm D^+ \bm V^\top,
\]
where $\bm D^+$ is obtained from $\bm D$ by transposing and replacing the positive singular values by their reciprocals. See \cite[Problem~7.3.7]{hor_joh_85} or \cite[Chapter~5]{sch_17}. Consequently, $\bm A^+\bm A$ and $\bm A\bm A^+$ are the orthogonal projections onto $\Ima\bm A^\top$ and $\Ima\bm A$, respectively. In particular, $\bm A^+=\bm A^{-1}$ whenever $\bm A$ is invertible. The following identities are immediate from the singular value decomposition:
\begin{equation}\label{eq:pinvid}
\begin{gathered}
(\bm A^+)^\top=(\bm A^\top)^+,\qquad
(\bm A^+)^+=\bm A,\qquad
\Ima\bm A^+=\Ima\bm A^\top ,\\
\bm A^+=(\bm A^\top\bm A)^+\bm A^\top=\bm A^\top(\bm A\bm A^\top)^+.
\end{gathered}
\end{equation}
Transposing $\bm A^+=(\bm A^\top\bm A)^+\bm A^\top$ and using the symmetry of $(\bm A^\top\bm A)^+$ gives the key relation
\begin{equation}\label{eq:pinvkey}
\bm A(\bm A^\top\bm A)^+=(\bm A^+)^\top,
\end{equation}
with no rank condition on $\bm A$, and \eqref{eq:pinvid} also gives
\begin{equation}\label{eq:pinvim}
\Ima\big((\bm A^+)^\top\big)=\Ima\bm A.
\end{equation}
Moreover, $\bm A^+=\lim_{\lambda\downarrow 0}(\bm A^\top\bm A+\lambda\bm I)^{-1}\bm A^\top$, so $\bm A^+$ is a measurable function of $\bm A$. \cite[Chapter~8]{ber_18} contains a valuable collection of pseudoinverse formulas, though the stated results should be cross-checked with the original sources.

For a symmetric positive semidefinite matrix $\bm A$, we write $\bm A^{1/2}$ for its symmetric positive semidefinite square root and $\bm A^{+1/2}\coloneqq(\bm A^{1/2})^+$ for the induced square root of the pseudoinverse. For the return second moment matrix $\bm Q_t$ we write $\bm P_{\bm Q,t}\coloneqq\bm Q_t\bm Q_t^+$ for the orthogonal projection onto $\Ima\bm Q_t$ and record
\begin{equation}\label{eq:sqrtproj}
\bm Q_t^{+1/2}\bm Q_t^{1/2}=\bm Q_t^{1/2}\bm Q_t^{+1/2}=\bm P_{\bm Q,t},
\qquad \Ima\bm Q_t^{1/2}=\Ima\bm Q_t^{+1/2}=\Ima\bm Q_t.
\end{equation}

\section{Conditional \texorpdfstring{$L^2$}{L2} Spaces and Projections}\label{app:probsetup}
Random variables are modeled on a probability space $(\Omega,\Fcal,\Pa)$ together with a sequence of information sets (sub-$\sigma$-algebras) $\Fcal_t\subseteq \Fcal_{t+1}\subseteq\cdots\subseteq\Fcal$. We write $\E_t[\cdot]$, $\var_t[\cdot]$, and $\cov_t[\cdot]$ for conditional mean, variance, and covariance given $\Fcal_t$. The unconditional case is included by taking $\Fcal_t=\{\emptyset,\Omega\}$.

We denote by $L^0$ the space of (almost surely finite) random variables defined up to almost sure equality. Formally, $L^0$ consists of equivalence classes $[X]$ of measurable mappings $X:\Omega\to[-\infty,+\infty]$ with $X\in\R$ a.s., where two random variables are identified if they coincide almost surely. Any measurable representative $X'$ of $[X]$ is called a version of $X$. By convention, random variables and events that agree almost surely are identified, and we use the same symbol for $X$ and $[X]$. See \cite[Appendix~A.7]{foe_sch_16}.

We denote by $L^2$ the subspace of $L^0$ consisting of square-integrable random variables. Endowed with the norm $\|X\|_{L^2}\coloneqq \E[X^2]^{1/2}$, it is a Hilbert space. We further define the conditional $L^2$-space
\[
L^2_t \coloneqq \{X\in L^0:\E_t[X^2]<\infty \text{ a.s.}\}.
\]
Equivalently, $X\in L^2_t$ if and only if $\E_t[X^2]\in L^0(\Fcal_t)$, where $L^0(\Fcal_t)$ denotes the space of equivalence classes of $\Fcal_t$-measurable random variables. In particular, for every $X\in L^2_t$ there exists a version $X'$ with $\E_t[X'^2]<\infty$ everywhere, for instance $X'=X\,1_{\{\E_t[X^2]<\infty\}}$.

In general, $L^2_t$ does not admit a natural inner product and hence is not a Hilbert space. Nevertheless, the seminal result of \cite{han_ric_87} shows that $L^2_t$ is conditionally closed and complete with respect to convergence in probability: a sequence $(X_n)_{n\ge 1}\subset L^2_t$ converges to $X\in L^2_t$ if $\E_t[(X_n-X)^2]\to 0$ in probability.\footnote{With a stronger topology, $L^2_t$ can be viewed as an $L^0$-normed module. See \cite{fil_kup_vog_09,fil_kup_12}.}

\begin{remark}[Support]\label{rem:support}
For any $\bm y_{t+1}\in L^2_t$, the support of the conditional distribution of $\bm y_{t+1}$ is contained in $\Ima\bm Q_{\bm y,t}$, and $\Ima\bm Q_{\bm y,t}$ is the smallest subspace with this property. Indeed, for $\bm v_t\perp\Ima\bm Q_{\bm y,t}$ we have $\E_t[(\bm v_t^\top\bm y_{t+1})^2]=\bm v_t^\top\bm Q_{\bm y,t}\bm v_t=0$ and thus $\bm v_t^\top\bm y_{t+1}=0$ almost surely. Conversely, if $\bm y_{t+1}\in V_t$ almost surely for an $\Fcal_t$-measurable subspace $V_t$, then $\bm v_t^\top\bm Q_{\bm y,t}\bm v_t=\E_t[(\bm v_t^\top\bm y_{t+1})^2]=0$ for $\bm v_t\perp V_t$, so that $\Ima\bm Q_{\bm y,t}\subseteq V_t$. Two consequences are used repeatedly. First, $\bm Q_{\bm y,t}\bm Q_{\bm y,t}^+\bm y_{t+1}=\bm y_{t+1}$ almost surely, as $\bm Q_{\bm y,t}\bm Q_{\bm y,t}^+$ is the orthogonal projection onto $\Ima\bm Q_{\bm y,t}$. Second, $\Ima\E_t[\bm y_{t+1}\bm z_{t+1}^\top]\subseteq\Ima\bm Q_{\bm y,t}$ for any $\bm z_{t+1}\in L^2_t$.
\end{remark}

\medskip

The following proposition defines and characterizes the space of excess returns of managed portfolios in a given universe of assets. Let
\[
\R^\infty \coloneqq \bigcup_{n\in\N}\R^n\times\{(0,0,\dots)\}
\]
denote the space of finite real sequences. It is a linear vector space consisting of elements of the form $\bm y=(y_1,\dots,y_n,0,0,\dots)$ for some $n\in\N$ and $y_i\in\R$. We identify each such $\bm y$ with the corresponding vector $\bm y\equiv [y_1,\dots,y_n]^\top\in\R^n$, and write
\[
\length(\bm y)\coloneqq \max\{i:y_i\neq 0\}
\]
for its length. This convention allows us to treat $\bm v,\bm y\in\R^\infty$ as Euclidean vectors (via $\bm v^\top\bm y$ and $\|\bm y\|_2^2$) while keeping the cross-sectional dimension flexible. For readability, one may replace $\R^\infty$ by $\R^n$ with fixed $n$ throughout.

\begin{remark}
The situation assumed in Proposition~\ref{prop:Sty} can be represented by the $\Fcal_t$-measurable partition $\Omega=\cup_{n\in\N}\{ n_t=n\}$ such that $\length(\bm y_{t+1})\le n$ on $\{ n_t=n\}$. Note that $[y_{t+1,1},\dots,y_{t+1,n}]$ can take arbitrary values in $\R^n$ including $\length(\bm y_{t+1})<n$ on $\{ n_t=n\}$.
\end{remark}

\begin{proposition}\label{prop:Sty}
Let $\bm y_{t+1}$ be an $\R^\infty$-valued random vector with $\E_t[\|\bm y_{t+1}\|_2^2]<\infty$ everywhere, and suppose there exists an $\Fcal_t$-measurable $\N$-valued random variable $n_t$ such that $\length(\bm y_{t+1})\le n_t$ everywhere. That is, conditional on $\Fcal_t$, only $n_t$ assets are traded over $[t,t+1]$. Define
\[
\Scal_t(\bm y_{t+1}) \coloneqq \Big\{ \bm v_t^\top \bm y_{t+1}:\ \text{$\bm v_t$ is an $\Fcal_t$-measurable $\R^\infty$-valued random vector}\Big\},
\]
and
\[
\Scal(\bm y_{t+1}) \coloneqq \Scal_t(\bm y_{t+1})\cap L^2.
\]
Then:
\begin{enumerate}
\item\label{it:Sty1} The conditional second moment matrix $\bm Q_{\bm y,t}$ is well defined, with $\size\bm Q_{\bm y,t}=n_t\times n_t$ everywhere.

\item\label{it:Sty2} For any $Y=\bm v_t^\top \bm y_{t+1}\in \Scal_t(\bm y_{t+1})$,
\[
\E_t[Y^2]=\bm v_t^\top \bm Q_{\bm y,t}\bm v_t=\|\bm Q_{\bm y,t}^{1/2}\bm v_t\|_2^2<\infty
\quad\text{everywhere}.
\]
In particular, $\Scal_t(\bm y_{t+1})\subseteq L^2_t$, and, with both sides read in $[0,\infty]$,
\begin{equation}\label{eq:Styeq1}
\|Y\|_{L^2} = \|\bm Q_{\bm y,t}^{1/2}\bm v_t\|_{L^2},
\end{equation}
so that $Y\in L^2$ exactly when the right-hand side is finite.

\item\label{it:Sty2CC} $\Scal_t(\bm y_{t+1})$ is conditionally complete: every conditional Cauchy sequence $(Y_j)$ in $\Scal_t(\bm y_{t+1})$ (that is, $\lim_{j,k\to\infty}\Pa[\E_t[(Y_j-Y_k)^2]>\epsilon]=0$ for all $\epsilon>0$) converges conditionally to some $Y\in\Scal_t(\bm y_{t+1})$ (that is, $\lim_{j\to\infty}\Pa[\E_t[(Y_j-Y)^2]>\epsilon]=0$ for all $\epsilon>0$).

\item\label{it:Sty3} For any $X\in L^2_t$, the conditional least-squares projection onto $\Scal_t(\bm y_{t+1})$,
\begin{equation}\label{eq:Styeq2}
P_t[X\mid\bm y_{t+1}]\coloneqq \E_t[X\bm y_{t+1}^\top]\bm Q_{\bm y,t}^+\bm y_{t+1},
\end{equation}
is the unique element of $\Scal_t(\bm y_{t+1})$ satisfying the conditional orthogonality condition
\begin{equation}\label{eq:L2tortho}
\E_t[(X-P_t[X\mid \bm y_{t+1}])Y]=0
\quad\text{for all }Y\in\Scal_t(\bm y_{t+1}),
\end{equation}
or, equivalently, the conditional least-squares condition
\begin{equation}\label{eq:L2tmini}
\E_t[(X-P_t[X\mid \bm y_{t+1}])^2]
\le\E_t[(X-Y)^2]
\quad\text{for all }Y\in\Scal_t(\bm y_{t+1}).
\end{equation}
In addition, it satisfies the idempotence property
\begin{equation}\label{eq:L2tprojprop}
P_t[X\mid \bm y_{t+1}]=X
\quad\text{if and only if}\quad
X\in\Scal_t(\bm y_{t+1}),
\end{equation}
the Pythagorean identity
\begin{equation}\label{eq:L2tprojpyth}
\E_t[X^2]
=\E_t[(X-P_t[X\mid \bm y_{t+1}])^2]
+\E_t[P_t[X\mid \bm y_{t+1}]^2],
\end{equation}
and the norm-contraction bound
\begin{equation}\label{eq:L2tprojbounds}
\E_t[P_t[X\mid \bm y_{t+1}]^2]\le\E_t[X^2],
\qquad
\|P_t[X\mid \bm y_{t+1}]\|_{L^2}\le\|X\|_{L^2},
\end{equation}
the second bound read in $[0,\infty]$. In particular $P_t[X\mid\bm y_{t+1}]\in L^2$ whenever $X\in L^2$.

\item\label{it:Sty4} $\Scal(\bm y_{t+1})$ is a closed subspace of $L^2$, so the $L^2$-orthogonal projection of any $X\in L^2$ onto $\Scal(\bm y_{t+1})$, denoted by $P[X\mid\bm y_{t+1}]$, is well defined. Furthermore, it coincides with the conditional least-squares projection onto $\Scal_t(\bm y_{t+1})$,
\begin{equation}\label{eq:Styeq4}
P[X\mid\bm y_{t+1}] = P_t[X\mid\bm y_{t+1}],
\end{equation}
and $\Scal(\bm y_{t+1})=P_t[L^2\mid\bm y_{t+1}]$.
\end{enumerate}
\end{proposition}

In Proposition~\ref{prop:Sty} we do not assume that $\bm y_{t+1}$ has finite unconditional second moment. Nevertheless, $\Scal(\bm y_{t+1})$ is nontrivial unless $\bm y_{t+1}$ is the zero vector. For example, let $\bm w\in\R^\infty$ be such that $\bm w^\top\bm Q_{\bm y,t}\bm w>0$ with positive probability. Then
$Y\coloneqq (1+\bm w^\top\bm Q_{\bm y,t}\bm w)^{-1/2}\bm w^\top\bm y_{t+1}$
is a nonzero element of $\Scal_t(\bm y_{t+1})$ with $\|Y\|_{L^2}\le 1$.

\begin{remark}
Proposition~\ref{prop:Sty}~\ref{it:Sty2CC} generalizes the conditional completeness statement mentioned below Equation~(4) of \citet{gal_han_tau_90}, where $\length(\bm y_{t+1})$ is assumed to be fixed and deterministic. \citet[Theorem~A.2]{han_ric_87} prove existence and uniqueness of the conditional projection with the orthogonality property \eqref{eq:L2tortho} onto conditionally complete subspaces of $L^2_t$ (they denote $L^2_t$ by $P^+$). In view of Proposition~\ref{prop:Sty}~\ref{it:Sty2CC}, their result applies here. However, they do not provide the explicit expression \eqref{eq:Styeq2}. \citet[Theorem~A.1]{han_ric_87} also establish conditional completeness of $L^2_t$, which we do not use.
\end{remark}

\begin{proof}[Proof of Proposition~\ref{prop:Sty}]
\ref{it:Sty1}: That $\bm Q_{\bm y,t}$ is well defined and finite, with $\size\bm Q_{\bm y,t}=n_t\times n_t$, everywhere follows by the Cauchy--Schwarz inequality and the assumed finiteness of $\E_t[\|\bm y_{t+1}\|_2^2]$.

\ref{it:Sty2}: Given \ref{it:Sty1}, we have $\E_t[(\bm v_t^\top \bm y_{t+1})^2]=\E_t[\bm v_t^\top \bm y_{t+1} \bm y_{t+1}^\top\bm v_t]=\bm v_t^\top\bm Q_{\bm y,t}\bm v_t=\| \bm Q_{\bm y,t}^{1/2}\bm v_t\|_2^2<\infty$ everywhere, which proves the claims.

\ref{it:Sty2CC}: Let $Y_j = (\bm v_j^\top\bm y_{t+1})$ form a conditional Cauchy sequence in $\Scal_t(\bm y_{t+1})$. In view of \ref{it:Sty2} it induces a sequence of $\Fcal_t$-measurable $\ell^2$-valued random vectors $\bm w_j\coloneqq \bm Q_{\bm y, t}^{1/2}\bm v_j$, which is Cauchy in probability, that is $\lim_{j,k\to\infty} \Pa[\| \bm w_j-\bm w_k\|_2>\epsilon]=0$ for every $\epsilon>0$. As $\ell^2$ is complete, by \cite[Theorem 9.2.3]{dud_02}, there exists an $\Fcal_t$-measurable $\ell^2$-valued random vector $\bm w$ such that $\bm w_j\to \bm w$ in probability. As every $\bm w_j(\omega)$ lies in the image of the matrix $\bm Q_{\bm y,t}^{1/2}(\omega)$, which is finite-dimensional and thus closed, we conclude that also $\bm w(\omega)$ lies in $\Ima \bm Q_{\bm y,t}^{1/2}(\omega)$ for almost every $\omega$. Thus, there exists an $\Fcal_t$-measurable random vector $\bm v$ such that $\bm w= \bm Q_{\bm y,t}^{1/2}\bm v$ a.s. Then $Y\coloneqq \bm v^\top\bm y_{t+1}$ is an element of $\Scal_t(\bm y_{t+1})$ and satisfies $\E_t[( Y_j-Y)^2]= \|\bm w_j-\bm w\|_2^2 \to 0$ in probability, which proves the claim.

\ref{it:Sty3}: By the Cauchy--Schwarz inequality, $P_t[X\mid\bm y_{t+1}]$ is well defined and an element of $\Scal_t(\bm y_{t+1})$. Now let $Y=\bm v_t^\top \bm y_{t+1}\in \Scal_t(\bm y_{t+1})$. As the support of the $\Fcal_t$-conditional distribution of $\bm y_{t+1}$ lies in $\Ima\bm Q_{\bm y,t}$, see Remark~\ref{rem:support}, we deduce that $ \E_t[ \bm y_{t+1} X]\in\Ima \bm Q_{\bm y,t}$ and thus
\begin{align*}
 \E_t[Y P_t[X\mid\bm y_{t+1}]  ] &=    \bm v_t^\top \E_t[\bm y_{t+1} \bm y_{t+1}^\top ]  \bm Q_{\bm y,t}^+ \E_t[    \bm y_{t+1} X] \\
 &=  \bm v_t^\top     \bm Q_{\bm y,t}\bm Q_{\bm y,t}^+ \E_t[    \bm y_{t+1} X]  = \bm v_t^\top    \E_t[    \bm y_{t+1} X] =\E_t[Y X],
\end{align*}
which proves that $P_t[X\mid\bm y_{t+1}] $ satisfies \eqref{eq:L2tortho}.

Now let $U,Y\in \Scal_t(\bm y_{t+1})$ be any elements. Expanding the square gives
\begin{align*}
\E_t[ (X-Y)^2] &= \E_t[ (X-U + U - Y)^2] \\
&= \E_t[ (X-U)^2]+ 2\E_t[ (X-U)(U-Y)] + \E_t[ (U - Y)^2]\\
&\ge \E_t[ (X-U)^2]+ 2\E_t[ (X-U)(U-Y)].
\end{align*}
If $U$ satisfies \eqref{eq:L2tortho} in lieu of $P_t[X\mid\bm y_{t+1}]$ then the right hand side is equal to $\E_t[ (X - U)^2]$, which is \eqref{eq:L2tmini}, and equality holds if and only if $Y=U$. Conversely, set $Y=U-Z$ for any $Z\in \Scal_t(\bm y_{t+1})$. The above relation shows that $U$ satisfies \eqref{eq:L2tmini} in lieu of $P_t[X\mid\bm y_{t+1}]$ only if $\E_t[ (X-U) Z]=0$, which is \eqref{eq:L2tortho}. This proves equivalence of \eqref{eq:L2tortho} and \eqref{eq:L2tmini} and uniqueness of $U=P_t[X\mid\bm y_{t+1}]$ in $\Scal_t(\bm y_{t+1})$ with these properties.
It also proves \eqref{eq:L2tprojprop}. Setting $Y=0$ in the above equation proves \eqref{eq:L2tprojpyth} and therefore \eqref{eq:L2tprojbounds}.

\ref{it:Sty4}: It follows by definition that $\Scal(\bm y_{t+1})$ is a subspace of $L^2$. Now consider any sequence $(Y_j)$ in $\Scal(\bm y_{t+1})$ that converges to some $Y$ in $L^2$. This implies that $\E_t[(Y_j-Y)^2]\to 0$ in probability, and thus $(Y_j)$ is a conditional Cauchy sequence, as $(Y_j-Y_k)^2 \le 2 (Y_j-Y)^2 + 2 (Y_k-Y)^2$. By \ref{it:Sty2CC} thus $Y\in\Scal_t(\bm y_{t+1})$, which proves that $\Scal(\bm y_{t+1})$ is closed in $L^2$.

By the projection property, $P[ X\mid\bm y_{t+1}]$ is the unique (modulo a.s.~equality) element of $\Scal(\bm y_{t+1})$ satisfying $\| X - P[ X\mid\bm y_{t+1}]\|_{L^2} = \min_{Y\in \Scal(\bm y_{t+1})} \| X - Y \|_{L^2}$. In view of \ref{it:Sty3}, for any $X\in L^2$, we have that $P_t[X \mid\bm y_{t+1}]\in \Scal_t(\bm y_{t+1})\cap L^2=\Scal(\bm y_{t+1})$ and thus
\begin{align*}
\| X - P[ X\mid\bm y_{t+1}]\|_{L^2}^2 &\le \| X - P_t[ X\mid\bm y_{t+1}]\|_{L^2}^2 = \E[ \E_t[ (X - P_t[ X\mid\bm y_{t+1}])^2]] \\
&\le \E[ \E_t[ (X - P[ X\mid\bm y_{t+1}])^2]] = \| X - P[ X\mid\bm y_{t+1}]\|_{L^2}^2,
  \end{align*}
  which proves \eqref{eq:Styeq4}. The last statement follows because $\Scal(\bm y_{t+1})=P[L^2\mid \bm y_{t+1}]$.
\end{proof}

\section{Conditional Asset Pricing}\label{app:CAP}
This appendix develops the conditional asset pricing basis behind Section~\ref{sec:pricingbenchmarks}, building on Appendix~\ref{app:probsetup}. The notation is that of Section~\ref{sec:pricingbenchmarks}.

To place the analysis in the standard pricing framework, suppose that a risk-free asset is traded over $[t,t+1]$ with $\Fcal_t$-measurable gross return $R_t^f>0$. Excess returns are then given by
\[
x_{t+1,i}=\frac{z_{t+1,i}}{z_{t,i}}-R_t^f,
\]
where $z_{t,i}$ denotes the price of asset $i$ at time $t$. Item~\ref{it:LOP1} of the following proposition states the law of one price: there exists a stochastic discount factor $M_{t+1}$ such that
$z_{t,i}=\E_t[M_{t+1}z_{t+1,i}]$ for all $i=1,\dots,n_t$, with $\E_t[M_{t+1}]=1/R_t^f$, see \cite{Cochranebook2005}.

\begin{proposition}\label{prop:LOP}
The following conditions are equivalent:
\begin{enumerate}
\item\label{it:LOP1} Law of one price: there exists a stochastic discount factor $M_{t+1}\in L_t^2$ satisfying $\E_t[M_{t+1}]>0$ and $\E_t[M_{t+1}\bm x_{t+1}]=\bm 0$.

\item\label{it:LOP21} Conditional Sharpe ratios are uniformly bounded, that is, $\text{SR}_t^\star<\infty$.

\item\label{it:LOP22} Returns with nonzero conditional mean are risky:
\begin{equation}\label{eq:riskyreturns}
\text{$\var_t[X]>0$ whenever $\E_t[X]\neq 0$, for all $X\in\Scal_t(\bm x_{t+1})$.}
\end{equation}

\item\label{it:LOP31} The projection of the constant strictly reduces its mean:
$\E_t[X^\star_{t+1}]<1$, with $X^\star_{t+1}=P_t[1\mid\bm x_{t+1}]$ given in \eqref{eq:defXstar}.

\item\label{it:LOP32} $M^\star_{t+1}= 1-X^\star_{t+1}$ is a stochastic discount factor.

\item\label{it:LOP4} $X_{t+1}^\star$ has finite conditional Sharpe ratio.

\item\label{it:LOP5} Asset risk premia are spanned by covariances, \eqref{eq:LoP}: $\bm\mu_t\in\Ima\bm\Sigma_t$.

\item\label{it:LOP6} The extended Sherman--Morrison formula holds:
\begin{equation}\label{eq:SMformula}
 \bm Q_t^+=(\bm\Sigma_t+\bm\mu_t\bm\mu_t^\top)^+
 = \bm\Sigma_t^+
   - (1+\bm\mu_t^\top\bm\Sigma_t^+\bm\mu_t)^{-1}
     \bm\Sigma_t^+\bm\mu_t\bm\mu_t^\top\bm\Sigma_t^+.
\end{equation}
\end{enumerate}
\end{proposition}

\begin{proof}
   \ref{it:LOP1}$\Rightarrow$\ref{it:LOP21}: Let $M_{t+1}$ be an SDF. For any $X\in\Scal_t(\bm x_{t+1})$ we have $0=\E_t[X M_{t+1}] = \cov_t[X, M_{t+1} ] + \E_t[X] \E_t[M_{t+1}]$. As $\E_t[M_{t+1}]>0$, we can divide and obtain, with the convention $\corr_t[X,M_{t+1}]\coloneqq 0$ on the $\Fcal_t$-events where $\var_t[X]$ or $\var_t[M_{t+1}]$ vanishes, on which both sides vanish by the conditional Cauchy--Schwarz inequality,
\begin{equation}\label{eq:preHJbound}
\E_t[X] = -\sqrt{\var_t[X]} \corr_t[X,M_{t+1}] \frac{\sqrt{\var_t[M_{t+1}]}}{\E_t[M_{t+1}]}.
\end{equation}
As the correlation is uniformly bounded by one, we obtain the Hansen--Jagannathan bound \citep{han_jag_91}
\begin{equation}\label{eq:HJbound}
\text{SR}_t^\star\le \frac{\sqrt{\var_t[M_{t+1}]}}{\E_t[M_{t+1}]}<\infty,
\end{equation}
which proves the claim.

\ref{it:LOP21}$\Rightarrow$\ref{it:LOP22}: Let $X\in\Scal_t(\bm x_{t+1})$ and define the events $A\coloneqq\{ \E_t[X]\neq 0\}\in\Fcal_t$ and $B\coloneqq \{ \var_t[X]=0\}\in\Fcal_t$. We have $|\text{SR}_t(X)|=\infty$ on $A\cap B$. By assumption hence $\Pa[A\cap B]=0$, which proves \eqref{eq:riskyreturns}.

\ref{it:LOP22}$\Rightarrow$\ref{it:LOP31}: From \eqref{eq:L2tortho}, with $X=1$ and $Y= X^\star_{t+1}$, we deduce that $\E_t [(X^\star_{t+1})^2]=\E_t [X^\star_{t+1}]$. Therefore $0\le \E_t [X^\star_{t+1}]\le 1$, and we can expand
\begin{equation}\label{eq:varPEP}
\var_t[ X^\star_{t+1}] = \E_t [X^\star_{t+1}] ( 1 - \E_t [X^\star_{t+1}]).
\end{equation}
As, by assumption, $\E_t[X^\star_{t+1}]=0$ if $\var_t[ X^\star_{t+1}] =0$, we infer that $1 - \E_t [X^\star_{t+1}] >0$, as desired.

\ref{it:LOP31}$\Leftrightarrow$\ref{it:LOP32}: Clearly, $M_{t+1}^\star\in L^2_t$, and \eqref{eq:L2tortho} implies that $\E_t[M_{t+1}^\star\bm x_{t+1}]=\bm 0$. Hence $M_{t+1}^\star$ is an SDF if and only if $\E_t[M_{t+1}^\star]>0$.

\ref{it:LOP32}$\Rightarrow$\ref{it:LOP1}: is trivial.

\ref{it:LOP31}$\Leftrightarrow$\ref{it:LOP4}: From \eqref{eq:varPEP} we see that $\text{SR}_t(X_{t+1}^\star)<\infty$ if and only if $1 - \E_t [X^\star_{t+1}]>0$, which proves the claim.

\ref{it:LOP22}$\Rightarrow$\ref{it:LOP5}: We decompose $\bm\mu_t = \bm \mu^1_t + \bm \mu^2_t$ according to the direct sum $\R^{n_t}=\Ima\bm\Sigma_t\oplus\ker\bm\Sigma_t$. We define the portfolio return $X\coloneqq \bm\mu^{2\top}_t\bm x_{t+1}\in\Scal_t(\bm x_{t+1})$. Its mean and variance are given by $\E_t[X]=\|\bm\mu^2_t\|_2^2$ and $\var_t[X]=0$, hence $X$ is risk-free. By assumption this implies $\bm\mu_t^2=\bm 0$, which proves \eqref{eq:LoP}.

\ref{it:LOP5}$\Rightarrow$\ref{it:LOP6}: That \eqref{eq:LoP} implies \eqref{eq:SMformula} is the rank-one case of the matrix pseudoinversion lemma of \citet[Theorem~2.3]{min_85}, worked out in the example of his Section~2.4. In fact \eqref{eq:LoP} is also necessary: by \citet[Theorem~1]{oga_88}, for a positive semidefinite $\bm A$ and any $\bm B$ the identity $(\bm A+\bm B\bm B^\top)^+=\bm A^+-\bm A^+\bm B(\bm I+\bm B^\top\bm A^+\bm B)^{-1}\bm B^\top\bm A^+$ holds if and only if $\Ima\bm A\supseteq\Ima\bm B$, which for $\bm B=\bm\mu_t$ is \eqref{eq:LoP}.

\ref{it:LOP6}$\Rightarrow$\ref{it:LOP31}: Using \eqref{eq:defXstar} and \eqref{eq:SMformula}, we obtain $ \E_t[X^\star_{t+1}] =\bm\mu_t^\top\bm Q_t^+\bm\mu_t= \frac{\bm\mu_t^\top\bm\Sigma_t^+\bm\mu_t}{1+\bm\mu_t^\top\bm\Sigma_t^+\bm\mu_t} < 1$.
 \end{proof}

The law of one price is strictly weaker than the absence of arbitrage, as illustrated by the following example.

\begin{example}\label{ex:LOOPnoarb}
Assume $R_t^f=1$ and $n_t=1$. Suppose the risky asset price satisfies $z_{t+1}\ge z_t$ almost surely, with $\E_t[z_{t+1}]>\E_t[z_t]$ and $\var_t[z_{t+1}]>0$. Buying one share of the risky asset and shorting $z_t$ units of the risk-free asset at time $t$ has zero cost and yields the nonnegative excess return
$x_{t+1}=z_{t+1}/z_t-1\ge 0$ with strictly positive mean. This constitutes an arbitrage. At the same time, the law of one price holds, since $\Sigma_t=\var_t[x_{t+1}]>0$ and hence \eqref{eq:LoP} is satisfied.
\end{example}

The following corollary makes explicit that all pricing objects of Section~\ref{sec:pricingbenchmarks} are $\bm Q_t$-objects. It backs the Sherman--Morrison consequence quoted after Proposition~\ref{prop:LOPmain}.

\begin{corollary}\label{cor:SMQ}
Under the law of one price \eqref{eq:LoP},
\[
\bm\mu_t^\top\bm Q_t^+=(1+\bm\mu_t^\top\bm\Sigma_t^+\bm\mu_t)^{-1}\bm\mu_t^\top\bm\Sigma_t^+,
\]
so that $X^\star_{t+1}=\bm\mu_t^\top\bm Q_t^+\bm x_{t+1}=(1+\bm\mu_t^\top\bm\Sigma_t^+\bm\mu_t)^{-1}\bm\mu_t^\top\bm\Sigma_t^+\bm x_{t+1}$.
\end{corollary}

\begin{proof}
Right multiply \eqref{eq:SMformula} by $\bm\mu_t$, use $\bm\Sigma_t^+\bm\Sigma_t\bm\Sigma_t^+=\bm\Sigma_t^+$, and simplify. The expression for $X^\star_{t+1}$ follows from \eqref{eq:defXstar}.
\end{proof}

A portfolio $X\in\Scal_t(\bm x_{t+1})$ is CMVE if it attains the maximal conditional Sharpe ratio, $\text{SR}_t(X)=\text{SR}_t^\star$, and $X\in\Scal(\bm x_{t+1})$ is UMVE if $\text{SR}(X)=\text{SR}^\star$. By definition, the CMVE (UMVE) portfolios trace out the conditional (unconditional) mean--variance frontier: for any $X\in\Scal_t(\bm x_{t+1})$ (resp.\ $X\in\Scal(\bm x_{t+1})$), the point $(\sqrt{\var_t[X]},\E_t[X])$ (resp.\ $(\sqrt{\var[X]},\E[X])$) lies in the cone bounded by the two rays through the origin with slopes $\pm\text{SR}_t^\star$ (resp.\ $\pm\text{SR}^\star$), and the frontier is the upper ray, attained exactly by the CMVE (resp.\ UMVE) portfolios.

\begin{proposition}\label{prop:Xstar}
Under any of the equivalent conditions in Proposition~\ref{prop:LOP}, the following properties hold for the portfolio $X^\star_{t+1}$ defined in \eqref{eq:defXstar}:
\begin{enumerate}
\item\label{it:Xstar1} The weights of $X^\star_{t+1}=\bm w_t^{\star\top}\bm x_{t+1}$, and its conditional mean, second moment, and variance are given by
\begin{align*}
 \bm w_t^\star
 &= \bm Q_t^+\bm\mu_t
 = (1+\bm\mu_t^\top\bm\Sigma_t^+\bm\mu_t)^{-1}\bm\Sigma_t^+\bm\mu_t,\\
 \E_t[X^\star_{t+1}]
 &= \E_t[(X^\star_{t+1})^2]
 = \frac{\bm\mu_t^\top\bm\Sigma_t^+\bm\mu_t}
        {1+\bm\mu_t^\top\bm\Sigma_t^+\bm\mu_t},\\
 \var_t[X^\star_{t+1}]
 &= \frac{\bm\mu_t^\top\bm\Sigma_t^+\bm\mu_t}
        {(1+\bm\mu_t^\top\bm\Sigma_t^+\bm\mu_t)^2}.
\end{align*}

\item\label{it:Xstar2} $X^\star_{t+1}$ is CMVE, and
\begin{equation}\label{eq:SReq}
  \text{SR}_t^\star = \sqrt{\bm\mu_t^\top\bm\Sigma_t^+\bm\mu_t}.
\end{equation}

\item\label{it:Xstar3} If $\bm\mu_t\neq\bm 0$ almost surely, the CMVE portfolios are exactly the conditional scalings of $X^\star_{t+1}$:
\begin{equation}\label{eq:CMVEfrontier}
\text{$X=s_tX^\star_{t+1}$ for some $\Fcal_t$-measurable $s_t>0$.}
\end{equation}

\item\label{it:Xstar4} $X^\star_{t+1}$ is UMVE.

\item\label{it:Xstar5} If $\bm\mu_t\neq\bm 0$ almost surely, the UMVE portfolios are exactly the constant scalings of $X^\star_{t+1}$:
\begin{equation}\label{eq:UMVEfrontier}
\text{$X=sX^\star_{t+1}$ for some constant $s>0$.}
\end{equation}

\item\label{it:Xstar5.1} Every UMVE portfolio is CMVE, but not conversely in general.

\item\label{it:Xstar6} The return vector admits the one-factor representation
\[
\bm x_{t+1}
= \bm\beta_t^\star X^\star_{t+1}
  + \bm\epsilon^\star_{t+1},
\]
with loading matrix
\[\bm\beta_t^\star\coloneqq \cov_t[\bm x_{t+1},X^\star_{t+1}]\var_t[X^\star_{t+1}]^+=\bm\mu_t(\bm\mu_t^\top\bm Q_t^+\bm\mu_t)^+,\]
where the residuals satisfy $\E_t[\bm\epsilon^\star_{t+1}]=\bm 0$ and
$\cov_t[\bm\epsilon^\star_{t+1},X^\star_{t+1}]=\bm 0$. The covariance matrix of the factor component equals
\[ \cov_t\big[\bm\beta^\star_t X^\star_{t+1}\big] = (\bm\mu_t^\top\bm\Sigma_t^+\bm\mu_t)^+ \bm\mu_t\bm\mu_t^\top.\]
\end{enumerate}
\end{proposition}

\begin{proof}
\ref{it:Xstar1}: These expressions follow by elementary calculations using \eqref{eq:defXstar}, \eqref{eq:SMformula}, and Corollary~\ref{cor:SMQ}.

\ref{it:Xstar2}: Equation \eqref{eq:varPEP} implies that
\begin{equation}\label{eq:SRXstar}
\text{SR}_t(X_{t+1}^\star)= \frac{\sqrt{\var_t[X_{t+1}^\star]}}{\E_t[1-X_{t+1}^\star]}=\frac{\sqrt{\var_t[M_{t+1}^\star]}}{\E_t[M_{t+1}^\star]},
\end{equation}
where the second equality follows as $M^\star_{t+1}=1-X^\star_{t+1}$, by definition. Now \eqref{eq:HJbound} proves that $X_{t+1}^\star$ is CMVE and the expression for $\text{SR}_t(X_{t+1}^\star)$ follows from \ref{it:Xstar1}.

\ref{it:Xstar3}: For any $X\in\Scal_t(\bm x_{t+1})$, combining \eqref{eq:preHJbound} and \eqref{eq:SRXstar}, we obtain
\[ \E_t[X] = \sqrt{\var_t[X]}\corr_t[X,X_{t+1}^\star] \text{SR}_t(X_{t+1}^\star) .\]
Hence $\text{SR}_t(X) = \text{SR}_t(X_{t+1}^\star) $ if and only if $\corr_t[X,X_{t+1}^\star] =1$. This again is equivalent to \eqref{eq:CMVEfrontier}, as $\Scal_t(\bm x_{t+1})$ does not contain any nonzero $\Fcal_t$-measurable random variables in view of \eqref{eq:riskyreturns}.

\ref{it:Xstar4}: Using similar arguments, with conditional replaced by unconditional expectations, as in the proof of Proposition~\ref{prop:LOP}~\ref{it:LOP1}$\Rightarrow$\ref{it:LOP21}, we see that the unconditional version of \eqref{eq:preHJbound} and the unconditional Hansen--Jagannathan bound,
\begin{equation}\label{eq:uncHJbound}
\text{SR}^\star\le \frac{\sqrt{\var[M_{t+1}]}}{\E[M_{t+1}]}<\infty,
\end{equation}
hold for any SDF $M_{t+1}$ in $L^2$. On the other hand, using \eqref{eq:Styeq4} we see that \eqref{eq:varPEP} also holds for unconditional variance and mean. Hence, arguing as in \ref{it:Xstar2}, we obtain the unconditional version of \eqref{eq:SRXstar},
\begin{equation}\label{eq:SRXstarU}
\text{SR}(X^\star_{t+1})= \frac{\sqrt{\var[X_{t+1}^\star]}}{\E[1-X_{t+1}^\star]} = \frac{\sqrt{\var[M_{t+1}^\star]}}{\E[M_{t+1}^\star]}.
\end{equation}
Combining \eqref{eq:uncHJbound} and \eqref{eq:SRXstarU} proves the claim.

\ref{it:Xstar5}: For any $X\in\Scal(\bm x_{t+1})$, combining the unconditional version of \eqref{eq:preHJbound} and \eqref{eq:SRXstarU}, we obtain
\[ \E[X] = \sqrt{\var[X]}\corr[X,X_{t+1}^\star] \text{SR}(X_{t+1}^\star) .\]
Hence $\text{SR}(X) = \text{SR}(X_{t+1}^\star) $ if and only if $\corr[X,X_{t+1}^\star] =1$. This again is equivalent to \eqref{eq:UMVEfrontier}, as $\Scal_t(\bm x_{t+1})$ does not contain any nonzero constant in view of \eqref{eq:riskyreturns}.

\ref{it:Xstar5.1}: This follows from \ref{it:Xstar3} and \ref{it:Xstar5}.

\ref{it:Xstar6}: By \eqref{eq:defXstar}, and as $\bm\mu_t\in\Ima\bm Q_t$, we have $\E_t[\bm x_{t+1}X^\star_{t+1}]=\bm Q_t\bm Q_t^+\bm\mu_t=\bm\mu_t$ and $\E_t[(X^\star_{t+1})^2]=\bm\mu_t^\top\bm Q_t^+\bm\mu_t$. The componentwise conditional least-squares projection of $\bm x_{t+1}$ onto $\Scal_t(X^\star_{t+1})\subseteq\Scal_t(\bm x_{t+1})$ from Proposition~\ref{prop:Sty}~\ref{it:Sty3} is thus
\[ P_t[\bm x_{t+1}\mid X^\star_{t+1}] = \E_t[\bm x_{t+1} X^\star_{t+1}]\E_t[(X^\star_{t+1})^2]^+ X^\star_{t+1} = \bm\beta^\star_t X^\star_{t+1},\]
with $\bm\beta^\star_t=\bm\mu_t(\bm\mu_t^\top\bm Q_t^+\bm\mu_t)^+$, and the residuals $\bm\epsilon^\star_{t+1}\coloneqq\bm x_{t+1}-\bm\beta^\star_t X^\star_{t+1}$ satisfy $\E_t[\bm\epsilon^\star_{t+1}X^\star_{t+1}]=\bm 0$, by \eqref{eq:L2tortho}. The conditional orthogonality \eqref{eq:L2tortho} of $1-X^\star_{t+1}$ to the components of $\bm\epsilon^\star_{t+1}\in\Scal_t(\bm x_{t+1})$ further gives $\E_t[\bm\epsilon^\star_{t+1}]=\E_t[\bm\epsilon^\star_{t+1}(1-X^\star_{t+1})]=\bm 0$, and thus $\cov_t[\bm\epsilon^\star_{t+1},X^\star_{t+1}]=\bm 0$. For the first expression for $\bm\beta^\star_t$, note that $\cov_t[\bm x_{t+1},X^\star_{t+1}]=\bm\mu_t(1-\bm\mu_t^\top\bm Q_t^+\bm\mu_t)$ and $\var_t[X^\star_{t+1}]=(\bm\mu_t^\top\bm Q_t^+\bm\mu_t)(1-\bm\mu_t^\top\bm Q_t^+\bm\mu_t)$, by \ref{it:Xstar1}, where $1-\bm\mu_t^\top\bm Q_t^+\bm\mu_t=(1+\bm\mu_t^\top\bm\Sigma_t^+\bm\mu_t)^{-1}>0$ by Corollary~\ref{cor:SMQ}. As $\bm\mu_t^\top\bm Q_t^+\bm\mu_t>0$ whenever $\bm\mu_t\neq\bm 0$, by $\bm\mu_t\in\Ima\bm Q_t$, we obtain $\cov_t[\bm x_{t+1},X^\star_{t+1}]\var_t[X^\star_{t+1}]^+=\bm\beta^\star_t$. The expression for the factor component covariance matrix follows from $\cov_t[\bm\beta^\star_tX^\star_{t+1}]=\bm\beta^\star_t\var_t[X^\star_{t+1}]\bm\beta^{\star\top}_t$ and Corollary~\ref{cor:SMQ}.
\end{proof}

\begin{remark}[Conditional and Unconditional Sharpe Equivalence]\label{rem:SharpeCU}
Under the law of one price, the same portfolio $X^\star_{t+1}$ attains both maximal Sharpe ratios, by Proposition~\ref{prop:Xstar}~\ref{it:Xstar2} and \ref{it:Xstar4}, and every UMVE portfolio is CMVE, by \ref{it:Xstar5.1}. For factor portfolios, Proposition~\ref{prop:span}~\ref{it:span1}$\Leftrightarrow$\ref{it:span4} states the corresponding equivalence at the level of spanning. Conditional factor spanning can therefore be assessed through unconditional Sharpe ratios. This justifies the unconditional evaluation of the conditional models in Section~\ref{sec:admissibilityspanning}.
\end{remark}

\section{Optimal Factor Representations}\label{app:optimalfactors}
This appendix develops the conditional weighted mean squared error view of factor representations \eqref{eq:linfacmodelPhi} with general, possibly non-traded, factors $\bm f_{t+1}\in L^2_t$. Appendix~\ref{app:wlscriterion} defines the weighted panel loss, Appendix~\ref{app:TSproblem} treats the time-series problem of the loadings given the factors, Appendix~\ref{app:CSproblem} the cross-sectional problem of the factors given the loadings, Appendix~\ref{app:jointproblem} the joint minimization over loadings and factors, and Appendix~\ref{app:intercept} the regression with an explicit intercept and the zero-alpha restriction. It builds on Appendices~\ref{app:matrix} and~\ref{app:probsetup}.

\subsection{The weighted panel loss}\label{app:wlscriterion}
For an $\Fcal_t$-measurable weight matrix $\bm S_t$ satisfying \eqref{eq:NoNDS}, define the \emph{weighted panel loss} of a pair $(\bm\Phi_t,\bm f_{t+1})$ as the conditional weighted mean squared error
\begin{equation}\label{eq:Ecaldef}
 \Ecal_t( \bm\Phi_t,\bm f_{t+1}) \coloneqq \E_t[ \| \bm S_t(\bm x_{t+1} - \bm\Phi_t\bm f_{t+1})\|_2^2],
\end{equation}
and its unconditional counterpart
\[
\Ecal( \bm\Phi_t,\bm f_{t+1}) \coloneqq \E[\Ecal_t( \bm\Phi_t,\bm f_{t+1})] = \E[\| \bm S_t(\bm x_{t+1} - \bm\Phi_t\bm f_{t+1})\|_2^2],
\]
which can be infinite. Minimization of the two is essentially equivalent.

\begin{lemma}\label{lem:conduncond}
If $(\bm\Phi_t,\bm f_{t+1})$ minimizes $\Ecal_t$ then it minimizes $\Ecal$. The converse is true under the additional assumption that $\Ecal( \bm\Phi_t,\bm f_{t+1})<\infty$.
\end{lemma}

\begin{proof}
If $\Ecal_t( \bm\Phi_t,\bm f_{t+1}) \le \Ecal_t( \bm\Phi'_t,\bm f'_{t+1})$ a.s.\ for all competitors $(\bm\Phi'_t,\bm f'_{t+1})$, then taking expectations gives $\Ecal( \bm\Phi_t,\bm f_{t+1}) \le \Ecal( \bm\Phi'_t,\bm f'_{t+1})$. Conversely, suppose $(\bm\Phi_t,\bm f_{t+1})$ satisfies $\Ecal( \bm\Phi_t,\bm f_{t+1})<\infty$ and minimizes $\Ecal$ but not $\Ecal_t$. Then there exist $(\bm\Phi'_t,\bm f'_{t+1})$ and $A\in\Fcal_t$ with $\Pa(A)>0$ such that $\Ecal_t( \bm\Phi'_t,\bm f'_{t+1}) < \Ecal_t( \bm\Phi_t,\bm f_{t+1})$ on $A$. Define $\widetilde{\bm\Phi}_t = \bm\Phi'_t\bm 1_A + \bm\Phi_t\bm 1_{A^c}$, and analogously $\widetilde{\bm f}_{t+1}$. These are $\Fcal_t$-measurable (resp.\ in $L^2_t$), and $\Ecal( \widetilde{\bm\Phi}_t, \widetilde{\bm f}_{t+1}) < \Ecal( \bm\Phi_t,\bm f_{t+1})$ from the law of total expectations, contradicting unconditional optimality.
\end{proof}

The loss splits along a second seam, orthogonal to the split over its two arguments. Writing $X^0\coloneqq X-\E_t[X]$ for the conditional demeaned component of a random variable $X$, the loss \eqref{eq:Ecaldef} decomposes as
\[ \Ecal_t( \bm\Phi_t,\bm f_{t+1}) = \| \bm S_t(\bm \mu_{t} - \bm\Phi_t\bm \mu_{\bm f,t})\|_2^2 + \E_t[ \| \bm S_t(\bm x^0_{t+1} - \bm \Phi_t\bm f^0_{t+1})\|_2^2],\]
which exhibits the trade-off between asset pricing and risk management objectives. The first term vanishes if factors span the asset risk premia, $\bm\mu_t=\bm\Phi_t\bm\mu_{\bm f,t}$, equivalently if residual risk is unpriced, \eqref{eq:coh2}, by the identity $\bm\mu_t=\bm\Phi_t\bm\mu_{\bm f,t}+\bm\mu_{\bm\epsilon,t}$. The second measures residual risk in terms of weighted residual variance. Condition \eqref{eq:coh2} thus annihilates the first term, but it is not a first-order condition of either the time-series problem of Appendix~\ref{app:TSproblem} or the cross-sectional problem of Appendix~\ref{app:CSproblem}. It is a restriction on the representation.

\begin{remark}[Centering]\label{rem:centering}
The centered panel is itself a factor representation: in the demeaned notation just introduced, any representation \eqref{eq:linfacmodelPhi} induces the representation
\[
\bm x^0_{t+1}=\bm\Phi_t\bm f^0_{t+1}+\bm\epsilon^0_{t+1}
\]
of the centered returns with the same loading matrix. For factor portfolios $\bm f_{t+1}=\bm W_t^\top\bm x_{t+1}$, moreover, $\bm f^0_{t+1}=\bm W_t^\top\bm x^0_{t+1}$ and the factor component operator $\bm\Pi_t$ is unchanged. The second moments of the centered representation are the central moments of the original one,
\[
\bm Q_{\bm x^0,t}=\bm\Sigma_t,\qquad
\bm Q_{\bm f^0,t}=\bm\Sigma_{\bm f,t},\qquad
\bm Q_{\bm\epsilon^0,t}=\bm\Sigma_{\bm\epsilon,t},\qquad
\E_t[\bm\epsilon^0_{t+1}\bm f^{0\top}_{t+1}]=\cov_t[\bm\epsilon_{t+1},\bm f_{t+1}].
\]
Every second-moment statement therefore has a covariance counterpart, obtained by applying it to the centered representation, with hypotheses translated in the same way. For one and the same representation, orthogonality \eqref{eq:ORTHO} and uncorrelatedness \eqref{eq:UNCORR} remain distinct conditions, differing by the term $\bm\mu_{\bm\epsilon,t}\bm\mu_{\bm f,t}^\top$: what the centering transfers are statements, not conditions.
\end{remark}

\subsection{Loadings given factors: the time-series problem}\label{app:TSproblem}
Minimizing \eqref{eq:Ecaldef} over $\bm\Phi_t$ for given factors is componentwise time-series regression.

\begin{lemma}\label{lem:TSregral0}
Given factors $\bm f_{t+1}$, the regression loading matrix $\bm\Phi^{\bm f_{t+1}}_t$ of \eqref{eq:TSreg},
 \begin{equation}\label{eq:hatalphaeqal0}
   \bm\Phi^{\bm f_{t+1}}_t = \E_t[\bm x_{t+1}\bm f_{t+1}^\top]\bm Q_{\bm f,t}^+,
\end{equation} 
is optimal in the sense that
\begin{equation}\label{eq:ortho1al0}
 \Ecal_t(\bm \Phi^{\bm f_{t+1}}_t, \bm f_{t+1})\le \Ecal_t(\bm \Phi_t, \bm f_{t+1}),\quad\text{for all $\Fcal_t$-measurable $ \bm\Phi_t$,}
\end{equation}
simultaneously for every weight matrix $\bm S_t$. The implied residuals $\bm\epsilon^{\bm f_{t+1}}_{t+1}\coloneqq \bm x_{t+1}- \bm\Phi^{\bm f_{t+1}}_t\bm f_{t+1}$ satisfy the orthogonality condition
\begin{equation}\label{eq:ortho3al0}
\E_t[\bm\epsilon^{\bm f_{t+1}}_{t+1} \bm f_{t+1}^\top]=\bm 0.
\end{equation}
Moreover, $\bm\Phi^{\bm f_{t+1}}_t$ is the unique loading matrix satisfying \eqref{eq:ortho1al0} if and only if $\bm S_t$ and $\bm Q_{\bm f,t}$ are invertible.
\end{lemma}

\begin{proof}
By the support fact of Remark~\ref{rem:support}, $\E_t[\bm x_{t+1}\bm f_{t+1}^\top]=\E_t[\bm x_{t+1}\bm f_{t+1}^\top]\bm Q_{\bm f,t}^+\bm Q_{\bm f,t}$, so that
\[
\E_t[\bm\epsilon^{\bm f_{t+1}}_{t+1}\bm f_{t+1}^\top]
=\E_t[\bm x_{t+1}\bm f_{t+1}^\top]-\bm\Phi^{\bm f_{t+1}}_t\bm Q_{\bm f,t}
=\E_t[\bm x_{t+1}\bm f_{t+1}^\top](\bm I_{m_t}-\bm Q_{\bm f,t}^+\bm Q_{\bm f,t})=\bm 0,
\]
which is \eqref{eq:ortho3al0}. For any $\Fcal_t$-measurable $\bm\Phi_t$, decompose $\bm x_{t+1}-\bm\Phi_t\bm f_{t+1}=\bm\epsilon^{\bm f_{t+1}}_{t+1}+(\bm\Phi^{\bm f_{t+1}}_t-\bm\Phi_t)\bm f_{t+1}$. The cross term vanishes by \eqref{eq:ortho3al0}, and
\begin{equation}\label{eq:TSlossid}
\Ecal_t(\bm\Phi_t,\bm f_{t+1})=\Ecal_t(\bm\Phi^{\bm f_{t+1}}_t,\bm f_{t+1})+\tr\big[\bm S_t(\bm\Phi_t-\bm\Phi^{\bm f_{t+1}}_t)\bm Q_{\bm f,t}(\bm\Phi_t-\bm\Phi^{\bm f_{t+1}}_t)^\top\bm S_t^\top\big],
\end{equation}
where the second term is nonnegative whatever $\bm S_t$ is, which proves \eqref{eq:ortho1al0} simultaneously for every weight matrix. The second term vanishes if and only if $\bm S_t(\bm\Phi_t-\bm\Phi^{\bm f_{t+1}}_t)\bm Q_{\bm f,t}^{1/2}=\bm 0$. If $\bm S_t$ and $\bm Q_{\bm f,t}$ are invertible, this forces $\bm\Phi_t=\bm\Phi^{\bm f_{t+1}}_t$. If $\bm S_t$ is singular, $\bm\Phi_t=\bm\Phi^{\bm f_{t+1}}_t+\bm v_t\bm e_1^\top$ with $\bm 0\neq\bm v_t\in\ker\bm S_t$ is another minimizer. If $\bm Q_{\bm f,t}$ is singular, so is $\bm\Phi_t=\bm\Phi^{\bm f_{t+1}}_t+\bm v_t\bm w_t^\top$ with $\bm v_t\neq\bm 0$ and $\bm 0\neq\bm w_t\in\ker\bm Q_{\bm f,t}$.
\end{proof}

In particular, under \eqref{eq:coh1} the loadings minimize the weighted panel loss given the factors, simultaneously for every $\bm S_t$.

The following lemma records what factor--residual orthogonality \eqref{eq:ORTHO} means for the loadings and the factor component.

\begin{lemma}\label{lem:FOCf}
Given a factor representation \eqref{eq:linfacmodelPhi}, the following are equivalent:
\begin{enumerate}
\item Factors and residuals are orthogonal, \eqref{eq:ORTHO}.

\item Residuals are optimal, $\bm\epsilon_{t+1}=\bm\epsilon^{\bm f_{t+1}}_{t+1}$, or equivalently, factor components are optimal,
\begin{equation}\label{eq:FFopti}
\bm\Phi_t \bm f_{t+1} =\bm\Phi^{\bm f_{t+1}}_t \bm f_{t+1}.
\end{equation}

\item Projected loadings are regression loadings,
\begin{equation}\label{eq:PLTScoef}
\bm\Phi_t \bm Q_{\bm f,t} \bm Q_{\bm f,t}^+ = \bm\Phi^{\bm f_{t+1}}_t.
\end{equation}

\end{enumerate}
 If in addition the loadings are spanned by the factor second moments, $ \Ima\bm \Phi_t^\top\subseteq\Ima\bm Q_{\bm f,t}$, then any of the above is equivalent to
 \begin{equation}\label{eq:PLTScoefreg}
\bm\Phi_t = \bm\Phi^{\bm f_{t+1}}_t.
 \end{equation}
\end{lemma}

\begin{proof}
Assume \eqref{eq:ORTHO}. Then $\bm 0=\E_t[\bm\epsilon_{t+1}\bm f_{t+1}^\top]=\E_t[\bm x_{t+1}\bm f_{t+1}^\top]-\bm\Phi_t\bm Q_{\bm f,t}$, so that
\[
\bm\Phi^{\bm f_{t+1}}_t=\E_t[\bm x_{t+1}\bm f_{t+1}^\top]\bm Q_{\bm f,t}^+=\bm\Phi_t\bm Q_{\bm f,t}\bm Q_{\bm f,t}^+,
\]
which is \eqref{eq:PLTScoef}.

Assume \eqref{eq:PLTScoef}. By the support fact, $\bm Q_{\bm f,t}\bm Q_{\bm f,t}^+\bm f_{t+1}=\bm f_{t+1}$ almost surely. Therefore
\[
\bm\Phi_t\bm f_{t+1}=\bm\Phi_t\bm Q_{\bm f,t}\bm Q_{\bm f,t}^+\bm f_{t+1}=\bm\Phi^{\bm f_{t+1}}_t\bm f_{t+1},
\]
which is \eqref{eq:FFopti}, and equivalently $\bm\epsilon_{t+1}=\bm\epsilon^{\bm f_{t+1}}_{t+1}$.

Assume \eqref{eq:FFopti}. Then $\bm\epsilon_{t+1}=\bm\epsilon^{\bm f_{t+1}}_{t+1}$ satisfies \eqref{eq:ORTHO} by \eqref{eq:ortho3al0}. This closes the cycle and proves the equivalence of the three conditions.

For the last statement, if $\Ima\bm\Phi_t^\top\subseteq\Ima\bm Q_{\bm f,t}$ then $\bm\Phi_t\bm Q_{\bm f,t}\bm Q_{\bm f,t}^+=\bm\Phi_t$, so that \eqref{eq:PLTScoef} coincides with \eqref{eq:PLTScoefreg}.
\end{proof}

\subsection{Factors given loadings: the cross-sectional problem}\label{app:CSproblem}
Given $\bm\Phi_t$, we seek factors with
\begin{equation}\label{eq:optif}
 \Ecal_t(\bm\Phi_t,\bm f^{\bm\Phi_t}_{t+1})\le \Ecal_t(\bm\Phi_t, \bm f_{t+1}),\quad\text{for all random vectors $\bm f_{t+1}\in L^2_t$,}
\end{equation}
which is achieved by minimizing the integrand in \eqref{eq:Ecaldef} pointwise, that is, by solving for each realization of $\bm x_{t+1}$ the weighted least-squares problem
\begin{equation}\label{eq:GLSeq}
 \min_{\bm f_{t+1}\in\R^{m_t}} \| \bm S_t(\bm x_{t+1} - \bm\Phi_t\bm f_{t+1})\|_2^2.
\end{equation}
As \eqref{eq:GLSeq} is a convex quadratic problem, $\bm f_{t+1}$ solves it if and only if the normal equation
\begin{equation}\label{eq:normal}
(\bm\Phi_t^\top\bm S_t^\top\bm S_t\bm\Phi_t) \bm f_{t+1} = \bm\Phi_t^\top\bm S_t^\top\bm S_t \bm x_{t+1}
\end{equation}
holds, which in terms of the residuals reads as the \emph{cross-sectional orthogonality} condition
\begin{equation}\label{eq:CSortho}
\bm\Phi_t^\top\bm S_t^\top\bm S_t\bm\epsilon_{t+1}=\bm 0.
\end{equation}
The WLS factor portfolios \eqref{eq:SLSfact} are a particular solution. The three weightings of Section~\ref{sec:weightedconstructions} give three different answers: the unweighted case $\bm S_t=\bm I_{n_t}$ leads to OLS regressions and, in the joint problem of Appendix~\ref{app:jointproblem}, to uncentered principal components. The moment weighting $\bm S_t=\bm Q_t^{+1/2}$ returns the factor portfolios \eqref{eq:MWfactors}, by \eqref{eq:MWnosqrt}. The GLS weighting $\bm S_t=\bm\Sigma_t^{+1/2}$ returns the weight matrix $(\bm\Phi_t^\top\bm\Sigma_t^+\bm\Phi_t)^+\bm\Phi_t^\top\bm\Sigma_t^+$ by the same identity. Section~\ref{sec:computation} gives conditions under which the three coincide.

The factor component operator of the WLS factors is characterized as follows.

\begin{lemma}\label{lem:Pidefchar}
For the WLS factors \eqref{eq:SLSfact}, the factor component operator $\bm\Pi_t = \bm\Phi_t (\bm S_t\bm\Phi_t)^+\bm S_t $ is idempotent, $\bm\Pi_t^2=\bm\Pi_t$, with $\Ima\bm\Pi_t =\Ima(\bm\Phi_t\bm\Phi_t^\top\bm S_t^\top)\subseteq \Ima\bm\Phi_t$, with equality if and only if $\Ima\bm \Phi_t\cap\ker\bm S_t=\{\bm 0\}$, and $\ker\bm\Pi_t = \ker((\bm S_t\bm\Phi_t)^+\bm S_t)$. It is a projection along $\ker\bm\Pi_t$, which is in general not the orthogonal complement of $\Ima\bm\Pi_t$ in the Euclidean inner product. Moreover, WLS residual risk is unpriced, $\E_t[\bm\epsilon^{\bm\Phi_t}_{t+1}]=\bm 0$, if and only if $\bm\mu_t\in\Ima\bm\Pi_t$.
\end{lemma}

\begin{proof}
We claim that 
\begin{equation}\label{eq:prlemPidefchar1}
\Ima ((\bm S_t\bm\Phi_t)^+\bm S_t) = \Ima(\bm S_t\bm\Phi_t)^\top.
\end{equation}
Indeed, we have $\Ima(\bm S_t\bm\Phi_t)^\top=\Ima(\bm S_t\bm\Phi_t)^+ \supseteq \Ima((\bm S_t\bm\Phi_t)^+ \bm S_t)\supseteq \Ima((\bm S_t\bm\Phi_t)^+ \bm S_t\bm\Phi_t) = \Ima(\bm S_t\bm\Phi_t)^\top$, which proves \eqref{eq:prlemPidefchar1}. The lemma now follows.
\end{proof}

\begin{remark}[Fama--MacBeth Regressions]\label{rem:famamacbeth}
Condition \eqref{eq:CSortho} for $\bm S_t=\bm I_{n_t}$ is the population normal equation of the period-by-period cross-sectional regressions of \citet{famamacbeth73}, and general $\bm S_t$ covers their weighted variants. Cross-sectional orthogonality \eqref{eq:CSortho} and factor--residual orthogonality \eqref{eq:ORTHO} are first-order conditions of different problems, and neither implies the other. In particular, $\bm\Phi_t \E_t[ \bm f^{\bm\Phi_t}_{t+1}\bm\epsilon^{\bm\Phi_t\top}_{t+1} ] = \bm\Pi_t\bm Q_t (\bm I_{n_t}-\bm \Pi_t)^\top$ is nonzero in general, so the WLS factors satisfy \eqref{eq:CSortho} but can violate both \eqref{eq:ORTHO} and \eqref{eq:UNCORR}, see Example~\ref{ex:spanning2} in Appendix~\ref{app:proofs}. Corollary~\ref{cor:Phistar} below shows that both conditions hold jointly at a minimizer of the joint problem. 
\end{remark}

The minimal weighted panel loss attained in the cross-sectional problem is
\begin{equation}\label{eq:Ecalminalphabeta}
\begin{aligned}
\Ecal_t(\bm\Phi_t ,\bm f^{\bm\Phi_t}_{t+1}) &=\tr\E_t[\bm S_t (\bm I_{n_t}-\bm\Pi_t) \bm x_{t+1} \bm x_{t+1}^\top  (\bm I_{n_t}-\bm\Pi_t)^\top \bm S_t^\top]\\
&= \tr [\bm S_t (\bm I_{n_t}-\bm\Pi_t) \bm Q_t (\bm I_{n_t}-\bm\Pi_t)^\top \bm S_t^\top]\\
&=\tr [(\bm I_{n_t}-(\bm S_t \bm\Phi_t )(\bm S_t \bm\Phi_t)^+) (\bm S_t \bm Q_t \bm S_t^\top)],
\end{aligned}
\end{equation}
where the last equality uses the cyclic property of the trace and the fact that $(\bm S_t \bm\Phi_t )(\bm S_t \bm\Phi_t)^+$ and its complement are orthogonal projections, hence self-adjoint.

\subsection{The joint problem and the price of parsimony}\label{app:jointproblem}
Equation \eqref{eq:Ecalminalphabeta} is the minimal loss for a given loading matrix. Minimizing over the loading matrix as well requires a target rank $k_t\le n_t$, and gives the joint problem
\begin{equation}\label{eq:EcalminPhi}
\min_{\rank\bm\Phi_t\le k_t}\tr [(\bm I_{n_t}-(\bm S_t \bm\Phi_t )(\bm S_t \bm\Phi_t)^+) (\bm S_t \bm Q_t \bm S_t^\top)],
\end{equation}
where the minimum is over all $\Fcal_t$-measurable dimensions $m_t$ and $n_t\times m_t$ loading matrices with $\rank\bm\Phi_t\le k_t$. Note that $\Ima(\bm S_t\bm Q_t\bm S_t^\top)=\Ima(\bm S_t\bm Q_t)$, so $\rank(\bm S_t \bm Q_t \bm S_t^\top) = \rank\bm Q_t$ by \eqref{eq:NoNDS}. We recall that a linear space $V$ is called a $k$-dimensional principal subspace of a positive semidefinite matrix $\bm A$ if $V$ is spanned by $k$ eigenvectors corresponding to the largest eigenvalues of $\bm A$.

The degenerate regime is settled first.

\begin{lemma}[Zero Minimal Value]\label{lem:zeroloss}
The minimal value of \eqref{eq:EcalminPhi} is zero if and only if $k_t\ge\rank\bm Q_t$. In this case, a loading matrix $\bm\Phi_t$ with $\rank\bm\Phi_t\le k_t$ attains it, $\Ecal_t(\bm\Phi_t,\bm f^{\bm\Phi_t}_{t+1})=0$, if and only if
\begin{equation}\label{eq:zerolosschar}
\Ima(\bm S_t\bm\Phi_t)\supseteq\Ima(\bm S_t\bm Q_t).
\end{equation}
\end{lemma}

\begin{proof}
Write $\bm M_t\coloneqq\bm S_t\bm Q_t\bm S_t^\top$ and let $\bm P_V$ denote the orthogonal projection onto $V\coloneqq\Ima(\bm S_t\bm\Phi_t)$. By \eqref{eq:Ecalminalphabeta},
\[
\Ecal_t(\bm\Phi_t,\bm f^{\bm\Phi_t}_{t+1})=\tr[(\bm I_{n_t}-\bm P_V)\bm M_t]=\tr[(\bm I_{n_t}-\bm P_V)\bm M_t(\bm I_{n_t}-\bm P_V)],
\]
the trace of a symmetric positive semidefinite matrix, which vanishes if and only if $\bm M_t^{1/2}(\bm I_{n_t}-\bm P_V)=\bm 0$, that is, if and only if $\Ima\bm M_t\subseteq V$. Now $\Ima\bm M_t=\Ima(\bm S_t\bm Q_t^{1/2})=\Ima(\bm S_t\bm Q_t)$, which is \eqref{eq:zerolosschar}, and $\dim\Ima(\bm S_t\bm Q_t)=\rank\bm Q_t$ by \eqref{eq:NoNDS}, while $\dim V\le\rank\bm\Phi_t\le k_t$. Hence a zero value requires $k_t\ge\rank\bm Q_t$, and for $k_t\ge\rank\bm Q_t$ the choice $\bm\Phi_t=\bm Q_t$ is feasible and attains it.
\end{proof}

The parsimonious regime carries the content.

\begin{proposition}[Optimal Factors]\label{prop:minPhi}
Let $\lambda_{t,1}\ge \cdots \ge \lambda_{t,n_t}\ge 0$ denote the eigenvalues in non-increasing order of $\bm S_t \bm Q_t \bm S_t^\top$, and assume
\begin{equation}\label{eq:krankQ}
k_t\le\rank\bm Q_t.
\end{equation}
Then for every pair $(\bm\Phi_t,\bm f_{t+1})$ with $\rank\bm\Phi_t\le k_t$,
\[
\Ecal_t(\bm\Phi_t,\bm f_{t+1})\ \ge\ \sum_{i>k_t}\lambda_{t,i},
\]
with equality if and only if
\begin{equation}\label{eq:SPhieigen}
    \text{$\Ima(\bm S_t\bm\Phi_t)$ is a $k_t$-dimensional principal subspace of $\bm S_t \bm Q_t \bm S_t^\top$}
\end{equation}
and $\bm\Phi_t\bm f_{t+1}=\bm\Phi_t\bm f^{\bm\Phi_t}_{t+1}$ almost surely. Such $\bm\Phi_t$ always exist.
\end{proposition}

The proof uses the trace maximization of \cite[Corollary~4.3.18]{hor_joh_85}, going back to \citet{fan1949}: for a symmetric $n\times n$ matrix $\bm A$ with ordered eigenvalues $\lambda_1\ge\dots\ge\lambda_n$,
\begin{equation}\label{eq:tracemax}
\max_V \tr(\bm P_V\bm A)=\sum_{i=1}^k\lambda_i,
\end{equation}
where the maximum is taken over all $k$-dimensional subspaces $V\subseteq\R^n$ and $\bm P_V$ denotes the orthogonal projection onto $V$, and it is attained if $V$ is spanned by orthonormal eigenvectors corresponding to the $k$ largest eigenvalues of $\bm A$. The characterization of all maximizers is not part of the cited statement. Step~3 of the proof establishes it.

\begin{proof}
Write $\bm M_t\coloneqq\bm S_t \bm Q_t \bm S_t^\top$ and $V\coloneqq\Ima(\bm S_t\bm\Phi_t)$, and let $\bm P_V=(\bm S_t\bm\Phi_t)(\bm S_t\bm\Phi_t)^+$ denote the orthogonal projection onto $V$.

\emph{Step 1: reduction to the cross-sectional minimum.} Pointwise, $\bm S_t(\bm x_{t+1}-\bm\Phi_t\bm f_{t+1})=\bm S_t\bm\epsilon^{\bm\Phi_t}_{t+1}+\bm S_t\bm\Phi_t(\bm f^{\bm\Phi_t}_{t+1}-\bm f_{t+1})$, where $\bm S_t\bm\epsilon^{\bm\Phi_t}_{t+1}=(\bm I_{n_t}-\bm P_V)\bm S_t\bm x_{t+1}$ is orthogonal to the second summand, which lies in $V$. Hence, using \eqref{eq:Ecalminalphabeta},
\begin{equation}\label{eq:jointlossid}
\Ecal_t(\bm\Phi_t,\bm f_{t+1})=\tr[(\bm I_{n_t}-\bm P_V)\bm M_t]+\E_t[\|\bm S_t\bm\Phi_t(\bm f_{t+1}-\bm f^{\bm\Phi_t}_{t+1})\|_2^2].
\end{equation}

\emph{Step 2: the value.} We have $\tr[(\bm I_{n_t}-\bm P_V)\bm M_t]=\tr\bm M_t-\tr(\bm P_V\bm M_t)$ and $\dim V\le\rank\bm\Phi_t\le k_t$, so \eqref{eq:tracemax} gives $\Ecal_t(\bm\Phi_t,\bm f_{t+1})\ge\tr\bm M_t-\sum_{i\le k_t}\lambda_{t,i}=\sum_{i>k_t}\lambda_{t,i}$. The bound is attained: for any $\bm B$ whose columns are orthonormal eigenvectors of $\bm M_t$ corresponding to the $k_t$ largest eigenvalues, these lie in $\Ima\bm M_t\subseteq\Ima\bm S_t$, their eigenvalues being positive under \eqref{eq:krankQ}, the matrix $\bm\Phi_t\coloneqq\bm S_t^+\bm B$ satisfies $\Ima(\bm S_t\bm\Phi_t)=\Ima\bm B$ and $\rank\bm\Phi_t\le k_t$, and $\tr(\bm P_{\Ima\bm B}\bm M_t)=\sum_{i\le k_t}\lambda_{t,i}$ by \eqref{eq:tracemax}. This also proves the existence claim.

\emph{Step 3: the equality characterization.} Let $\bm q_1,\dots,\bm q_{n_t}$ be orthonormal eigenvectors of $\bm M_t$ for $\lambda_{t,1}\ge\dots\ge\lambda_{t,n_t}$ and set $c_i\coloneqq\|\bm P_V\bm q_i\|_2^2\in[0,1]$, so that $\tr(\bm P_V\bm M_t)=\sum_i\lambda_{t,i}c_i$ and $\sum_i c_i=\tr\bm P_V=\dim V\le k_t$. Then
\begin{align*}
\sum_i\lambda_{t,i}c_i&=\lambda_{t,k_t}\sum_i c_i+\sum_{i\le k_t}(\lambda_{t,i}-\lambda_{t,k_t})c_i+\sum_{i>k_t}(\lambda_{t,i}-\lambda_{t,k_t})c_i\\
&\le\lambda_{t,k_t}k_t+\sum_{i\le k_t}(\lambda_{t,i}-\lambda_{t,k_t})=\sum_{i\le k_t}\lambda_{t,i},
\end{align*}
where $\lambda_{t,k_t}>0$ by \eqref{eq:krankQ}, with equality if and only if $\sum_ic_i=k_t$, $c_i=1$ whenever $\lambda_{t,i}>\lambda_{t,k_t}$, and $c_i=0$ whenever $\lambda_{t,i}<\lambda_{t,k_t}$. Since $c_i=1$ means $\bm q_i\in V$ and $c_i=0$ means $\bm q_i\perp V$, equality holds if and only if $\dim V=k_t$, $V$ contains the span $V_>$ of the eigenspaces with eigenvalue exceeding $\lambda_{t,k_t}$, and $V\subseteq V_>\oplus E$ for the eigenspace $E$ of $\lambda_{t,k_t}$. In that case $V=V_>\oplus(V\cap E)$, and as every subspace of an eigenspace is spanned by eigenvectors, $V$ is spanned by $k_t$ eigenvectors corresponding to the $k_t$ largest eigenvalues of $\bm M_t$. The converse is the attainment statement of \eqref{eq:tracemax}. Equality in the trace bound is therefore \eqref{eq:SPhieigen}.

\emph{Step 4: the factors.} By \eqref{eq:jointlossid}, equality in the loss bound further requires $\bm S_t\bm\Phi_t\bm f_{t+1}=\bm S_t\bm\Phi_t\bm f^{\bm\Phi_t}_{t+1}$ almost surely. At equality, Step~3 gives $\dim V=k_t\ge\rank\bm\Phi_t\ge\rank(\bm S_t\bm\Phi_t)=\dim V$, so $\rank(\bm S_t\bm\Phi_t)=\rank\bm\Phi_t$, equivalently $\Ima\bm\Phi_t\cap\ker\bm S_t=\{\bm 0\}$, and therefore $\bm\Phi_t\bm f_{t+1}=\bm\Phi_t\bm f^{\bm\Phi_t}_{t+1}$ almost surely. Conversely, \eqref{eq:SPhieigen} and $\bm\Phi_t\bm f_{t+1}=\bm\Phi_t\bm f^{\bm\Phi_t}_{t+1}$ almost surely give equality in \eqref{eq:jointlossid}.
\end{proof}

The following invariance property is used below.

\begin{lemma}\label{lem:minPhireg}
Assume \eqref{eq:krankQ} and let $\bm\Phi_t$ be a minimizer of \eqref{eq:EcalminPhi}. Then
\begin{equation}\label{eq:assPhiinmuSigma}
\Ima(\bm S_t\bm\Phi_t)=(\bm S_t \bm Q_t \bm S_t^\top)\Ima(\bm S_t\bm\Phi_t) \subseteq\Ima(\bm S_t\bm Q_t),
\end{equation}
with equality in the inclusion if and only if equality holds in \eqref{eq:krankQ}.
\end{lemma}

\begin{proof}
By \eqref{eq:SPhieigen}, $\Ima(\bm S_t\bm\Phi_t)$ is spanned by eigenvectors of $\bm M_t\coloneqq\bm S_t\bm Q_t\bm S_t^\top$ whose eigenvalues are positive under \eqref{eq:krankQ}, so $\bm M_t$ acts bijectively on it, which is the equality in \eqref{eq:assPhiinmuSigma}. The inclusion holds as $\Ima\bm M_t=\Ima(\bm S_t\bm Q_t)$, and since $\dim\Ima(\bm S_t\bm\Phi_t)=k_t$ while $\dim\Ima(\bm S_t\bm Q_t)=\rank\bm Q_t$ by \eqref{eq:NoNDS}, it is an equality precisely when $k_t=\rank\bm Q_t$.
\end{proof}

The minimal value $\sum_{i>k_t}\lambda_{t,i}$ is the weighted panel loss that no factor representation of rank at most $k_t$ can undercut. It is measured in the units of $\bm S_t\bm x_{t+1}$ and scales with $\tr(\bm S_t\bm Q_t\bm S_t^\top)$, so only the relative quantity $\sum_{i>k_t}\lambda_{t,i}/\sum_i\lambda_{t,i}$ is meaningful, and only within a fixed $\bm S_t$.

In the unweighted case $\bm S_t=\bm I_{n_t}$ the joint problem is uncentered principal component analysis: if $m_t=k_t\le\rank\bm Q_t$ and the columns of $\bm\Phi_t$ are orthonormal eigenvectors corresponding to the $k_t$ largest eigenvalues of $\bm Q_t$, then $\bm\Phi_t$ is a minimizer of \eqref{eq:EcalminPhi} and the WLS factors $\bm f^{\bm\Phi_t}_{t+1}=\bm\Phi_t^\top\bm x_{t+1}$ are the first $k_t$ uncentered principal components of $\bm x_{t+1}$. For the moment weighting the joint problem degenerates.

\begin{corollary}[Moment Weighting]\label{cor:IMWexi}
Let $\bm S_t=\bm Q_t^{+1/2}$. Then any $\Fcal_t$-measurable $n_t\times m_t$ loading matrix $\bm\Phi_t$ satisfying either $\Ima\bm\Phi_t\supseteq\Ima\bm Q_t$ or the support condition \eqref{eq:Phistarass} is a minimizer of \eqref{eq:EcalminPhi} for $k_t=\rank\bm\Phi_t$, and the minimal value of \eqref{eq:EcalminPhi} is $\max\{\rank\bm Q_t-k_t,0\}$.
\end{corollary}

\begin{proof}
Here $\bm M_t=\bm Q_t^{+1/2}\bm Q_t\bm Q_t^{+1/2}=\bm P_{\bm Q,t}$ by \eqref{eq:sqrtproj}, the orthogonal projection onto $\Ima\bm Q_t$, whose eigenvalues are $1$ with multiplicity $\rank\bm Q_t$ and $0$ otherwise. The minimal value at rank $k_t$ is thus $\max\{\rank\bm Q_t-k_t,0\}$, and every $k_t$-dimensional subspace of $\Ima\bm Q_t$ with $k_t\le\rank\bm Q_t$ is a $k_t$-dimensional principal subspace of $\bm M_t$. Under the support condition, $k_t=\rank\bm\Phi_t\le\rank\bm Q_t$ and $\Ima(\bm Q_t^{+1/2}\bm\Phi_t)$ is a $k_t$-dimensional subspace of $\Ima\bm Q_t$, since $\bm Q_t^{+1/2}$ is injective on $\Ima\bm Q_t$ by \eqref{eq:sqrtproj}. Proposition~\ref{prop:minPhi} applies. If instead $\Ima\bm\Phi_t\supseteq\Ima\bm Q_t$, then $k_t=\rank\bm\Phi_t\ge\rank\bm Q_t$ and $\Ima(\bm Q_t^{+1/2}\bm\Phi_t)=\Ima\bm Q_t\supseteq\Ima(\bm Q_t^{+1/2}\bm Q_t)$. Lemma~\ref{lem:zeroloss} applies and the minimal value $(\rank\bm Q_t-k_t)^+=0$ is attained.
\end{proof}

At the moment weighting the joint problem therefore does not discriminate among loading matrices: every $\bm\Phi_t$ satisfying the support condition is a minimizer at its own rank, and the minimal value is an integer count of unspanned dimensions that carries no second-moment information. At a minimizing loading matrix, moreover, the choice of weighting stops mattering for the factors:

\begin{corollary}[Optimal Loading Matrices Are Consistent]\label{cor:Phistar}
Assume \eqref{eq:krankQ} and the support condition \eqref{eq:Phistarass}, and let $\bm\Phi_t$ be a minimizer of \eqref{eq:EcalminPhi} for a weight matrix $\bm S_t$ satisfying \eqref{eq:NoNDS}. Then $\bm\Phi_t= \bm\Phi^{\bm f_{t+1}^{\bm\Phi_t}}_t$, so that the pair $(\bm\Phi_t,\bm f^{\bm\Phi_t}_{t+1})$ satisfies \eqref{eq:coh1} and both first-order conditions \eqref{eq:ORTHO} and \eqref{eq:CSortho} hold. Consequently, by Theorem~\ref{thm:rigid}, the WLS factor portfolios coincide almost surely with the moment-weighted factor portfolios \eqref{eq:MWfactors}, whatever the weight matrix $\bm S_t$. The representation is coherent if and only if the risk-premium condition \eqref{eq:arpspanning} holds, and it then spans under the law of one price \eqref{eq:LoP}.
\end{corollary}

\begin{proof}
Denote by $\bm f_{t+1}= \bm f_{t+1}^{\bm\Phi_t}$ the WLS factors, and write $\bm M_t\coloneqq\bm S_t\bm Q_t\bm S_t^\top$ and $V\coloneqq\Ima(\bm S_t\bm\Phi_t)$. By Proposition~\ref{prop:minPhi}, \eqref{eq:SPhieigen} holds, so $V$ is invariant under $\bm M_t$ by Lemma~\ref{lem:minPhireg}. Invariance says that
 \begin{align*}
 \bm 0 &= (\bm S_t\bm\Phi_t )(\bm S_t \bm\Phi_t)^+  (\bm S_t \bm Q_t \bm S_t^\top) (\bm I_{n_t}-(\bm S_t \bm\Phi_t )(\bm S_t \bm\Phi_t)^+).
 \end{align*}
Since $\Ima (\bm S_t \bm\Phi_t)^+= \Ima (\bm S_t \bm\Phi_t)^\top$, after transposing therefore
  \begin{align*}
 \bm 0 &=  (\bm I_{n_t}-(\bm S_t \bm\Phi_t )(\bm S_t \bm\Phi_t)^+)  (\bm S_t \bm Q_t \bm S_t^\top) (\bm S_t \bm\Phi_t)^{+\top}\\
 &= \bm S_t ( \bm Q_t\bm S_t^\top (\bm S_t \bm\Phi_t)^{+\top}- \bm\Phi_t (\bm S_t \bm\Phi_t)^+  \bm S_t \bm Q_t \bm S_t^\top (\bm S_t \bm\Phi_t)^{+\top}).
 \end{align*}
In view of \eqref{eq:NoNDS} and \eqref{eq:Phistarass} therefore
  \begin{align*}
 \bm 0 &=    \bm Q_t\bm S_t^\top (\bm S_t \bm\Phi_t)^{+\top}- \bm\Phi_t (\bm S_t \bm\Phi_t)^+  \bm S_t \bm Q_t \bm S_t^\top (\bm S_t \bm\Phi_t)^{+\top}\\
 &=\E_t[ \bm x_{t+1}  \bm f_{t+1}^\top ] - \bm\Phi_t \bm Q_{\bm f,t}.
 \end{align*}
By the same token, we have $\Ima(\bm S_t \bm\Phi_t)^{+\top}=\Ima(\bm S_t \bm\Phi_t)\subseteq \Ima(\bm S_t \bm Q_t)=\Ima(\bm S_t \bm Q_t\bm S_t^\top)$ and hence $\Ima\bm Q_{\bm f,t} = \Ima((\bm S_t \bm\Phi_t)^+ \bm S_t \bm Q_t \bm S_t^\top) = \Ima(\bm S_t \bm\Phi_t)^+ = \Ima(\bm S_t \bm\Phi_t)^\top = \Ima \bm\Phi^\top_t$. Right multiplying the above equation by $\bm Q_{\bm f,t}^+$ thus gives
\[ \bm 0 = \E_t[ \bm x_{t+1} \bm f^{\top}_{t+1} ]\bm Q_{\bm f,t}^+ - \bm\Phi_t ,\]
which proves \eqref{eq:hatalphaeqal0}, that is, \eqref{eq:coh1} for the pair $(\bm\Phi_t,\bm f^{\bm\Phi_t}_{t+1})$. Factor--residual orthogonality \eqref{eq:ORTHO} follows from Lemma~\ref{lem:TSregral0}, and cross-sectional orthogonality \eqref{eq:CSortho} holds by construction of the WLS factors.

For the remaining statements, $\bm f_{t+1}$ is traded and satisfies $\bm\Phi_t=\bm\Phi^{\bm f_{t+1}}_t$ with \eqref{eq:Phistarass}, so Theorem~\ref{thm:rigid}~\ref{it:rigid2} identifies it almost surely with the moment-weighted factor portfolios \eqref{eq:MWfactors}, for every weight matrix $\bm S_t$. Corollary~\ref{cor:premium} then gives coherence if and only if \eqref{eq:arpspanning} holds, and the statement recorded after Corollary~\ref{cor:premium} gives spanning under \eqref{eq:LoP}.
\end{proof}
The rank bound \eqref{eq:krankQ} costs nothing here: under the support condition \eqref{eq:Phistarass} every loading matrix has $\rank\bm\Phi_t\le\rank\bm Q_t$, so a larger $k_t$ admits no additional loading matrices satisfying \eqref{eq:Phistarass} and does not reduce the minimal loss. For $\bm S_t=\bm I_{n_t}$, Corollary~\ref{cor:Phistar} together with \eqref{eq:SPhieigen} recovers Corollary~\ref{cor:OLScoh} on the principal subspaces of $\bm Q_t$, a strict subset of the $\bm Q_t$-invariant column spaces on which the OLS factor portfolios satisfy \eqref{eq:coh1}.

\subsection{The intercept regression and zero alpha}\label{app:intercept}
The representation \eqref{eq:linfacmodelPhi} contains no intercept, and \eqref{eq:coh2} is a restriction on the representation rather than a normalization. Empirical practice tests pricing instead through the zero-alpha restriction in the intercept representation \eqref{eq:linfacmodelN} of Section~\ref{sec:maintheorem}, with the regression coefficients as the canonical choice. In analogy to \eqref{eq:Ecaldef}, define the conditional weighted mean squared error
\begin{equation}\label{eq:Ecaldefalbe}
 \Ecal_t( \bm\alpha_t,\bm\beta_t,\bm f_{t+1}) \coloneqq \E_t[ \| \bm S_t(\bm x_{t+1} - \bm\alpha_t - \bm \beta_t\bm f_{t+1})\|_2^2].
\end{equation}

The following is the analogue of Lemma~\ref{lem:TSregral0} with an explicit intercept.

\begin{lemma}\label{lem:TSregr}
Given factors $\bm f_{t+1}$, the time-series regression coefficients
 \begin{equation}\label{eq:hatalphaeq}
  \bm\beta^{\bm f_{t+1}}_t \coloneqq \cov_t[\bm x_{t+1},\bm f_{t+1}]\bm \Sigma_{\bm f,t}^+ ,\quad \bm\alpha^{\bm f_{t+1}}_t \coloneqq \bm\mu_t -\bm\beta^{\bm f_{t+1}}_t\bm\mu_{\bm f,t}
\end{equation}
are optimal in the sense that
\begin{equation}\label{eq:ortho1}
 \Ecal_t( \bm\alpha^{\bm f_{t+1}}_t,\bm\beta^{\bm f_{t+1}}_t ,\bm f_{t+1})\le \Ecal_t(\bm\alpha_t,\bm\beta_t, \bm f_{t+1}),\quad\text{for all $\Fcal_t$-measurable $\bm\alpha_t$ and $\bm\beta_t$.}
\end{equation}
The implied residuals $\bm\eta^{\bm f_{t+1}}_{t+1}\coloneqq \bm x_{t+1}- \bm\alpha^{\bm f_{t+1}}_t -\bm\beta^{\bm f_{t+1}}_t\bm f_{t+1}$ satisfy the orthogonality conditions
\begin{equation}\label{eq:ortho3}
\E_t[\bm\eta^{\bm f_{t+1}}_{t+1}]=\bm 0\quad\text{and}\quad \cov_t[\bm f_{t+1},\bm\eta^{\bm f_{t+1}}_{t+1}]=\bm 0.
\end{equation}
Moreover, $\bm\alpha^{\bm f_{t+1}}_t$ and $\bm\beta^{\bm f_{t+1}}_t$ in \eqref{eq:hatalphaeq} are the unique coefficients satisfying \eqref{eq:ortho1} if and only if $\bm S_t$ and $\bm\Sigma_{\bm f,t}$ are invertible.
\end{lemma}

\begin{proof}
Denote by $\Scal_t(1,\bm f_{t+1})$ the linear space of conditional linear combinations $v_t + \bm w_t^\top\bm f_{t+1}$ for $\Fcal_t$-measurable $v_t$ and $\bm w_t$, and note that $\Scal_t(1,\bm f_{t+1}) = \Scal_t(\bm G_{t+1})$, where $\bm G_{t+1}^\top \coloneqq [1, (\bm f_{t+1}-\bm\mu_{\bm f, t})^\top]$, and the second moment matrix of $\bm G_{t+1}$ can be expressed in terms of the factor covariance matrix
\[ \E_t[\bm G_{t+1}\bm G_{t+1}^\top]=\begin{bmatrix}
    1 & \bm 0 \\ \bm 0 & \bm\Sigma_{\bm f, t}
\end{bmatrix}.\]
Hence in view of Proposition~\ref{prop:Sty} the conditional least-squares projection onto $\Scal_t(1,\bm f_{t+1})$ of any $X\in L^2_t$ is given by
\begin{align*}
    P_t[X\mid 1, \bm f_{t+1}] &= P_t[X\mid\bm G_{t+1}] = \big[\E_t[X],\cov_t[X,\bm f_{t+1}]\big]\begin{bmatrix}
    1 & \bm 0 \\ \bm 0 & \bm\Sigma_{\bm f, t}^+
\end{bmatrix} \begin{bmatrix}
    1 \\\bm f_{t+1}-\bm\mu_{\bm f, t}
\end{bmatrix}\\
&= \E_t[X] + \cov_t[X,\bm f_{t+1}]\bm\Sigma_{\bm f, t}^+ (\bm f_{t+1}-\bm\mu_{\bm f, t}).
\end{align*}
In particular, $P_t[V \mid 1, \bm f_{t+1}]= V$ for any $\Fcal_t$-measurable $V$.

Now apply this to $\bm S_t\bm x_{t+1}$. We find that $\bm S_t \bm x_{t+1} =\bm S_t \bm\alpha^{\bm f_{t+1}}_t + \bm S_t\bm\beta^{\bm f_{t+1}}_t\bm f_{t+1} + \bm S_t\bm\eta^{\bm f_{t+1}}_{t+1}$ is the desired decomposition, which by \eqref{eq:L2tmini} and \eqref{eq:L2tortho} satisfies \eqref{eq:ortho1} and
\begin{equation}\label{eq:ortho4}
    \bm S_t \E_t[\bm\eta^{\bm f_{t+1}}_{t+1} [1,\bm f_{t+1}^\top]] =\bm 0.
\end{equation}
As $\bm \mu_t\in\Ima\bm Q_t$ and the support of the conditional distribution of $\bm x_{t+1}$ is contained in $\Ima\bm Q_t$, see Remark~\ref{rem:support}, the columns of $ \bm\alpha^{\bm f_{t+1}}_t $ and $\bm\beta^{\bm f_{t+1}}_t$ and the support of $\bm\eta^{\bm f_{t+1}}_{t+1} $ lie in $\Ima\bm Q_t$ as well. Given \eqref{eq:NoNDS}, property~\eqref{eq:ortho4} thus implies $\E_t[\bm\eta^{\bm f_{t+1}}_{t+1} [1,\bm f_{t+1}^\top]] =\bm 0$, which again is equivalent to~\eqref{eq:ortho3}.

For the last statement, suppose first that $\bm S_t$ and $\bm\Sigma_{\bm f,t}$ are invertible, and let $(\bm\alpha_t,\bm\beta_t)$ also attain the minimum in \eqref{eq:ortho1}. The two minimizing fits coincide, $\bm S_t(\bm\alpha_t+\bm\beta_t\bm f_{t+1})=\bm S_t(\bm\alpha^{\bm f_{t+1}}_t+\bm\beta^{\bm f_{t+1}}_t\bm f_{t+1})$ almost surely, by the uniqueness part of Proposition~\ref{prop:Sty}~\ref{it:Sty3} applied componentwise, and invertibility of $\bm S_t$ removes $\bm S_t$. Taking $\E_t$ and $\cov_t[\cdot,\bm f_{t+1}]$ gives $(\bm\beta_t-\bm\beta^{\bm f_{t+1}}_t)\bm\Sigma_{\bm f,t}=\bm 0$, hence $\bm\beta_t=\bm\beta^{\bm f_{t+1}}_t$, and in turn $\bm\alpha_t=\bm\alpha^{\bm f_{t+1}}_t$. Conversely, if $\bm S_t\bm z_t=\bm 0$ for some $\bm z_t\neq\bm 0$, then $(\bm\alpha^{\bm f_{t+1}}_t+\bm z_t,\bm\beta^{\bm f_{t+1}}_t)$ also attains the minimum, and if $\bm\Sigma_{\bm f,t}\bm v_t=\bm 0$ for some $\bm v_t\neq\bm 0$, then $\bm v_t^\top(\bm f_{t+1}-\bm\mu_{\bm f,t})=0$ almost surely, so for any $\bm w_t\neq\bm 0$ the pair $(\bm\alpha^{\bm f_{t+1}}_t-\bm w_t\bm v_t^\top\bm\mu_{\bm f,t},\,\bm\beta^{\bm f_{t+1}}_t+\bm w_t\bm v_t^\top)$ leaves the fit unchanged almost surely and attains the minimum as well. This proves the lemma.
\end{proof}

The two sets of regression coefficients are related as follows.

\begin{lemma}\label{lem:Phialbe}
We have
\begin{equation}\label{eq:Phialbeeq1}
 \Ecal_t( \bm\alpha^{\bm f_{t+1}}_t,\bm\beta^{\bm f_{t+1}}_t ,\bm f_{t+1}) \le \Ecal_t(\bm \Phi^{\bm f_{t+1}}_t, \bm f_{t+1}),
 \end{equation}
 and the following conditions are equivalent:
 \begin{enumerate}
  \item\label{it:Phialbe1} Equality holds in \eqref{eq:Phialbeeq1}.
\item\label{it:Phialbe2} Residual risk is unpriced, \eqref{eq:coh2} for $\bm\epsilon_{t+1}=\bm\epsilon^{\bm f_{t+1}}_{t+1}$.

 \item\label{it:Phialbe3} Residuals coincide, $\bm\epsilon^{\bm f_{t+1}}_{t+1}=\bm\eta^{\bm f_{t+1}}_{t+1}$, or equivalently,
 \begin{equation}\label{eq:Phialbeeq3}
\bm\alpha^{\bm f_{t+1}}_t+\bm\beta^{\bm f_{t+1}}_t \bm f_{t+1} = \bm \Phi^{\bm f_{t+1}}_t \bm f_{t+1}.
 \end{equation}

 \item\label{it:Phialbe4} Asset risk premia are spanned by factors, $\bm\mu_t=\bm\Phi^{\bm f_{t+1}}_t\bm\mu_{\bm f,t}$, and projected loadings coincide,
\begin{equation}\label{eq:Phialbeeq4}
    \bm\beta^{\bm f_{t+1}}_t = \bm \Phi^{\bm f_{t+1}}_t \bm \Sigma_{\bm f,t}\bm \Sigma_{\bm f,t}^+.
 \end{equation}
 \end{enumerate}
 If in addition the regression loadings are spanned by the factor covariance,
 \begin{equation}\label{eq:ImaPhifin}
 \Ima\bm \Phi^{\bm f_{t+1}\top}_t\subseteq\Ima\bm\Sigma_{\bm f,t},
 \end{equation}
 then any of the above is equivalent to
 \begin{equation}\label{eq:Phialbeeq5}
\bm\alpha^{\bm f_{t+1}}_t=\bm 0\quad\text{and}\quad \bm\beta^{\bm f_{t+1}}_t = \bm \Phi^{\bm f_{t+1}}_t.
 \end{equation}
 Sufficient conditions for \eqref{eq:ImaPhifin} to hold are: (i) $\bm\Sigma_{\bm f,t}$ is invertible, or (ii) the law of one price holds, \eqref{eq:LoP}, and factors are portfolios, \eqref{eq:factorportfolios}, see Proposition~\ref{prop:tradeq}~\ref{it:tradeq1}.
  \end{lemma}

\begin{proof}
In view of their definitions \eqref{eq:hatalphaeqal0} and \eqref{eq:hatalphaeq}
and Proposition~\ref{prop:Sty}, we see that $\bm\alpha^{\bm f_{t+1}}_t+\bm\beta^{\bm f_{t+1}}_t \bm f_{t+1} = P_t[\bm x_{t+1}\mid 1,\bm f_{t+1}]$ and $\bm \Phi^{\bm f_{t+1}}_t \bm f_{t+1} =P_t[\bm x_{t+1}\mid \bm f_{t+1}]$ are the component-wise conditional least-squares projections of $\bm x_{t+1}$ onto $\Scal_t(1,\bm f_{t+1})$ and $\Scal_t(\bm f_{t+1})$, respectively. As the latter is a subspace, $\Scal_t(\bm f_{t+1})\subseteq\Scal_t(1,\bm f_{t+1}) $, the inequality \eqref{eq:Phialbeeq1} follows from \eqref{eq:L2tmini}.

\ref{it:Phialbe1}$\Leftrightarrow$\ref{it:Phialbe3}: This follows from \eqref{eq:L2tmini} and the uniqueness of the conditional least-squares projection in Proposition~\ref{prop:Sty}~\ref{it:Sty3}.

\ref{it:Phialbe2}$\Leftrightarrow$\ref{it:Phialbe3}: This follows from \eqref{eq:L2tortho}, using \eqref{eq:ortho3al0} and \eqref{eq:ortho3}, and the uniqueness of the conditional least-squares projection in Proposition~\ref{prop:Sty}~\ref{it:Sty3}.

\ref{it:Phialbe3}$\Leftrightarrow$\ref{it:Phialbe4}: The identity \eqref{eq:Phialbeeq3} can be equivalently rewritten as
\[ \bm\mu_t + \bm\beta^{\bm f_{t+1}}_t (\bm f_{t+1} -\bm\mu_{\bm f,t}) = \bm \Phi^{\bm f_{t+1}}_t \bm\mu_{\bm f,t} + \bm \Phi^{\bm f_{t+1}}_t(\bm f_{t+1} -\bm\mu_{\bm f,t}).\]
Taking conditional expectations on both sides, and using that the support of the conditional distribution of $\bm f_{t+1} -\bm\mu_{\bm f,t}$ is contained in $\Ima\bm\Sigma_{\bm f, t}$, see Remark~\ref{rem:support}, that $\Ima\bm\beta^{\bm f_{t+1}\top}_t \subseteq\Ima\bm\Sigma_{\bm f, t}$, and that $\bm \Sigma_{\bm f,t}^+\bm \Sigma_{\bm f,t}$ is the orthogonal projection onto $\Ima\bm \Sigma_{\bm f,t}$, shows that \eqref{eq:Phialbeeq3} and \eqref{eq:Phialbeeq4} are equivalent.

The last statement of the lemma follows by the same token, noting that $\Ima\bm \Phi^{\bm f_{t+1}\top}_t\subseteq\Ima \bm Q_{\bm f,t}=\Ima\bm\Sigma_{\bm f, t} + \Ima\bm\mu_{\bm f,t}$, by \eqref{eq:hatalphaeqal0}, and conditions (i) and (ii) both imply that $\bm\mu_{\bm f,t}\in\Ima\bm \Sigma_{\bm f,t}$.
\end{proof}

Assumption \eqref{eq:ImaPhifin} is tight, as \eqref{eq:Phialbeeq3} does not imply \eqref{eq:Phialbeeq5} in general.

\begin{example}[Spurious $\bm\alpha_t$]\label{ex:spuralpha}
Let $n_t=m_t=2$, and let $\bm x_{t+1}$ have conditional mean $\bm \mu_t = [0,1]^\top$ and covariance $\bm\Sigma_t=\bm I_2$. Assume factors are given by $\bm f_{t+1} = [\bm x_{t+1,1}, 1]^\top$. Elementary calculations give $\bm\mu_{\bm f, t}= [0,1]^\top$,
\[ \cov_t[\bm x_{t+1},\bm f_{t+1}] = \bm\Sigma_{\bm f,t}=\begin{bmatrix} 1 & 0 \\ 0 & 0\end{bmatrix},\quad  \E_t[\bm x_{t+1} \bm f_{t+1}^\top] =\bm Q_{\bm f,t} = \bm I_2, \]
so that
\[ \bm\alpha^{\bm f_{t+1}}_t = \begin{bmatrix} 0 \\ 1\end{bmatrix},\quad \bm\beta^{\bm f_{t+1}}_t = \begin{bmatrix} 1 & 0 \\ 0 & 0\end{bmatrix},\quad \bm \Phi^{\bm f_{t+1}}_t =\bm I_2,\quad\bm\epsilon^{\bm f_{t+1}}_{t+1}=\bm\eta^{\bm f_{t+1}}_{t+1} = \begin{bmatrix} 0 \\ \bm x_{t+1,2}-1\end{bmatrix}. \]
Hence residual risk is unpriced, \eqref{eq:coh2} for $\bm\epsilon_{t+1}=\bm\epsilon^{\bm f_{t+1}}_{t+1}$, while $\bm\alpha^{\bm f_{t+1}}_t\neq\bm 0$ and $\bm\beta^{\bm f_{t+1}}_t \neq \bm \Phi^{\bm f_{t+1}}_t $.
\end{example}

The zero-alpha restriction decomposes as follows. The statement holds for arbitrary factors and requires no pricing assumption.

\begin{lemma}[Decomposition of the Zero-Alpha Restriction]\label{lem:alphazero}
Given factors $\bm f_{t+1}$, the zero-alpha restriction holds, $\bm\alpha^{\bm f_{t+1}}_t=\bm 0$, if and only if both of the following hold:
\begin{enumerate}
\item\label{it:alphazero1} Residual risk is unpriced, \eqref{eq:coh2} for $\bm\epsilon_{t+1}=\bm\epsilon^{\bm f_{t+1}}_{t+1}$.
\item\label{it:alphazero2} The two regressions have the same factor component,
\begin{equation}\label{eq:fceq}
\bm\beta^{\bm f_{t+1}}_t\bm f_{t+1}=\bm\Phi^{\bm f_{t+1}}_t\bm f_{t+1}.
\end{equation}
\end{enumerate}
\end{lemma}

\begin{proof}
Suppose $\bm\alpha^{\bm f_{t+1}}_t=\bm 0$. As in the proof of Lemma~\ref{lem:Phialbe}, componentwise $\bm\alpha^{\bm f_{t+1}}_t+\bm\beta^{\bm f_{t+1}}_t\bm f_{t+1}=P_t[\bm x_{t+1}\mid 1,\bm f_{t+1}]$ and $\bm\Phi^{\bm f_{t+1}}_t\bm f_{t+1}=P_t[\bm x_{t+1}\mid\bm f_{t+1}]$. The former projection thus equals $\bm\beta^{\bm f_{t+1}}_t\bm f_{t+1}$, whose components lie in $\Scal_t(\bm f_{t+1})$, and each component of $\bm x_{t+1}-\bm\beta^{\bm f_{t+1}}_t\bm f_{t+1}$ is conditionally orthogonal to $\Scal_t(1,\bm f_{t+1})\supseteq\Scal_t(\bm f_{t+1})$ by \eqref{eq:L2tortho}. The uniqueness part of Proposition~\ref{prop:Sty}~\ref{it:Sty3} gives \eqref{eq:fceq}, which is \ref{it:alphazero2}, and hence condition~\ref{it:Phialbe3} of Lemma~\ref{lem:Phialbe}, and \ref{it:alphazero1} follows from \ref{it:Phialbe3}$\Leftrightarrow$\ref{it:Phialbe2}. Conversely, \ref{it:alphazero1} gives condition~\ref{it:Phialbe3} of Lemma~\ref{lem:Phialbe}, that is, $\bm\alpha^{\bm f_{t+1}}_t+\bm\beta^{\bm f_{t+1}}_t\bm f_{t+1}=\bm\Phi^{\bm f_{t+1}}_t\bm f_{t+1}$, and subtracting \eqref{eq:fceq} leaves $\bm\alpha^{\bm f_{t+1}}_t=\bm 0$.
\end{proof}

Lemma~\ref{lem:alphazero} isolates the wedge: condition~\ref{it:alphazero1} is \eqref{eq:coh2} for the regression residuals, and condition~\ref{it:alphazero2} is an extra requirement unrelated to pricing, so the zero-alpha restriction is strictly stronger than the pricing restriction. Example~\ref{ex:spuralpha} is a case where \ref{it:alphazero1} holds and \ref{it:alphazero2} fails: a degenerate factor covariance direction carries a nonzero risk premium, \eqref{eq:ImaPhifin} fails, and the premium is absorbed through the factor mean rather than through covariances. By Lemma~\ref{lem:Phialbe}, condition~\ref{it:alphazero2} follows from condition~\ref{it:alphazero1} under \eqref{eq:ImaPhifin}, which holds automatically for traded factors under the law of one price. The wedge then closes, which is Proposition~\ref{prop:tradeq} of Section~\ref{sec:OFM}. The law of one price cannot be dropped, and the mechanism is not confined to non-traded factors.

\begin{example}[A Coherent Traded Representation with Nonzero Alpha]\label{ex:sccounter}
Let $n_t=2$ and let $\bm x_{t+1}$ have $x_{t+1,1}=1$ almost surely and $x_{t+1,2}$ with zero mean and unit variance, so that $\bm\mu_t=[1,0]^\top$, $\bm\Sigma_t=\diag(0,1)$ and $\bm Q_t=\bm I_2$. The law of one price \eqref{eq:LoP} fails, as $\bm\mu_t\notin\Ima\bm\Sigma_t$, and indeed the first asset is an arbitrage. Take $\bm\Phi_t=[1,0]^\top$, so that $m_t=1$. Then \eqref{eq:MWfactors} gives $\bm W_t=[1,0]^\top$ and $f_{t+1}=x_{t+1,1}=1$, with $Q_{\bm f,t}=1$ and $\bm\Phi^{f_{t+1}}_t=\E_t[\bm x_{t+1}f_{t+1}]=\bm\mu_t=\bm\Phi_t$, so \eqref{eq:coh1} holds, while $\bm\epsilon_{t+1}=[0,x_{t+1,2}]^\top$ has zero mean, so \eqref{eq:coh2} holds, and $\text{SR}^\star_{\bm f,t}=\text{SR}^\star_t=\infty$, so \eqref{eq:factorportfolios} and \eqref{eq:maxSR} hold. The representation is coherent and spans. Yet $\Sigma_{\bm f,t}=0$, hence $\bm\beta^{f_{t+1}}_t=\bm 0$ and $\bm\alpha^{f_{t+1}}_t=\bm\mu_t\neq\bm 0$. Condition~\ref{it:alphazero1} of Lemma~\ref{lem:alphazero} holds and condition~\ref{it:alphazero2} fails.
\end{example}

Example~\ref{ex:sccounter} is the traded counterpart of Example~\ref{ex:spuralpha}: in both, a degenerate factor covariance direction carries a nonzero risk premium and \eqref{eq:ImaPhifin} fails. Under the law of one price this is impossible for traded factors, since a constant traded excess return with nonzero conditional mean has an unbounded conditional Sharpe ratio, excluded by Proposition~\ref{prop:LOPmain}~\ref{it:LOPmain2}.

For empirical pricing tests the delimitation is as follows. Alpha-based tests in the tradition of \citet{gib_ros_sha_89} test the correct hypothesis whenever the factors are traded and the law of one price holds: by Proposition~\ref{prop:tradeq}, the zero-alpha restriction is then the pricing restriction itself, in particular for the factor portfolios \eqref{eq:MWfactors} of Theorem~\ref{thm:rigid}. A wedge between the two, a \emph{spurious alpha}, requires the failure of condition~\ref{it:alphazero2} of Lemma~\ref{lem:alphazero}, hence non-traded factors or a failure of the law of one price, and since $\bm\alpha^{\bm f_{t+1}}_t=\bm 0$ implies the pricing restriction for arbitrary factors, by Lemma~\ref{lem:alphazero}, the caution concerns specifications with non-traded factors whose degenerate covariance directions carry premia. For those, the zero-alpha restriction is not a valid proxy for pricing, and the appropriate hypothesis is \eqref{eq:coh2} for the residuals of the regression without intercept.

\section{Proofs}\label{app:proofs}
This appendix proves the statements of the main text, together with the auxiliary results that are used only for these proofs. It builds on Appendices~\ref{app:matrix} to~\ref{app:optimalfactors}. Statements made in the other appendices are proved where they occur.

The proofs in this appendix repeatedly use the support fact of Remark~\ref{rem:support}: for any $\bm y_{t+1},\bm z_{t+1}\in L^2_t$, we have $\bm Q_{\bm y,t}\bm Q_{\bm y,t}^+\bm y_{t+1}=\bm y_{t+1}$ almost surely and $\Ima\E_t[\bm y_{t+1}\bm z_{t+1}^\top]\subseteq\Ima\bm Q_{\bm y,t}$.

\subsection{Proof of Proposition~\ref{prop:LOPmain}}
The proposition collects items \ref{it:LOP1}, \ref{it:LOP21}, \ref{it:LOP5} and \ref{it:LOP32} of Proposition~\ref{prop:LOP} in Appendix~\ref{app:CAP}. \qed

\subsection{Proof of Proposition~\ref{prop:span}}
\ref{it:span1}$\Leftrightarrow$\ref{it:span2}: Given \eqref{eq:LoP}, we can write $\bm\mu_t =\bm \Sigma_t^{1/2}\bm\nu_t$ for some $\Fcal_t$-measurable vector $\bm\nu_t\in\Ima\bm\Sigma_t$, and thus $\bm\mu_{\bm f,t} = \bm W_t^\top \bm\Sigma_t^{1/2}\bm\nu_t$. From \eqref{eq:SReq} we obtain $(\text{SR}_t^\star)^2 = \bm\nu_t^\top\bm\nu_t$ and, using the matrix identity $\bm A^\top (\bm A\bm A^\top)^+ = \bm A^+$ for any matrix $\bm A$, applying \eqref{eq:SReq} to the factor universe,
  \begin{align*}
 (\text{SR}_{\bm f,t}^\star)^2&=\bm\nu_t^\top (\bm W_t^\top \bm\Sigma_t^{1/2})^\top \big((\bm W_t^\top \bm\Sigma_t^{1/2}) (\bm W_t^\top \bm\Sigma_t^{1/2})^\top \big)^+ (\bm W_t^\top \bm\Sigma_t^{1/2})   \bm \nu_t \\
 &= \bm\nu_t^\top (\bm W_t^\top \bm\Sigma_t^{1/2})^+(\bm W_t^\top \bm\Sigma_t^{1/2})   \bm \nu_t = \bm\nu_t^\top \bm P_{\bm\Sigma_t^{1/2} \bm W_t } \bm\nu_t,
\end{align*}
where $\bm P_{\bm\Sigma_t^{1/2} \bm W_t }$ denotes the orthogonal projection onto $\Ima(\bm\Sigma_t^{1/2} \bm W_t )$. It follows that \eqref{eq:maxSR} holds if and only if $\bm\nu_t = \bm P_{\bm\Sigma_t^{1/2} \bm W_t } \bm\nu_t$ lies in the image of $\bm\Sigma_t^{1/2} \bm W_t $, which again is equivalent to \eqref{eq:spanNEWeq1}.

\ref{it:span2}$\Rightarrow$\ref{it:span3}: Given \eqref{eq:spanNEWeq1}, and because $\ker(\bm \Sigma_t^{1/2}\bm W_t)\subseteq \ker(\bm \Sigma_t\bm W_t)$, there exists a vector $\bm b_t\in\Ima(\bm W_t^\top\bm\Sigma_t)$ such that $\bm \mu_t=\bm\Sigma_t \bm W_t \bm b_t$ and therefore $\bm\Sigma_{\bm f,t}^+\bm \mu_{\bm f,t}=(\bm W_t^\top \bm\Sigma_t \bm W_t)^+ \bm W_t^\top\bm\Sigma_t\bm W_t \bm b_t =\bm b_t$. On the other hand, we have $\bm x_{t+1}=\bm\Sigma_t\bm \Sigma_t^+ \bm x_{t+1}$ with probability one, by Remark~\ref{rem:support} and $\Ima\bm Q_t=\Ima\bm\Sigma_t$ under \eqref{eq:LoP}. We obtain $\bm\mu_{\bm f,t}^\top \bm\Sigma_{\bm f,t}^+ \bm f_{t+1} = \bm b_t^\top \bm W_t^\top \bm x_{t+1} = \bm b_t^\top \bm W_t^\top \bm\Sigma_t\bm \Sigma_t^+ \bm x_{t+1}=\bm\mu_{t}^\top \bm\Sigma_{t}^+ \bm x_{t+1}$, with probability one. Using \eqref{eq:maxSR}, Proposition~\ref{prop:Xstar}~\ref{it:Xstar1} and Corollary~\ref{cor:SMQ} prove \eqref{eq:spanNEWeqX}.

\ref{it:span3}$\Rightarrow$\ref{it:span4}: This follows from Proposition~\ref{prop:Xstar}~\ref{it:Xstar4}.

\ref{it:span4}$\Rightarrow$\ref{it:span1}: This follows from Proposition~\ref{prop:Xstar}~\ref{it:Xstar5.1}.

This completes the proof of Proposition~\ref{prop:span}. \qed

\subsection{Proof of Proposition~\ref{prop:uncspan}}
By Proposition~\ref{prop:span}, factor spanning \eqref{eq:maxSR} is equivalent to \eqref{eq:spanNEWeq1}, so it suffices to relate \eqref{eq:spanNEWeq1} and unpriced residual risk.

Assume \eqref{eq:spanNEWeq1}. For factor portfolios, uncorrelatedness \eqref{eq:UNCORR} reads $\bm W_t^\top\bm\Sigma_t=\bm W_t^\top\bm\Sigma_t\bm W_t\bm\Phi_t^\top$, and transposing gives $\bm\Sigma_t\bm W_t=\bm\Pi_t\bm\Sigma_t\bm W_t$. Using the latter, \eqref{eq:spanNEWeq1} implies $\bm\Pi_t \bm\mu_t =\bm\Pi_t \bm\Sigma_t \bm W_t\bm c_t =\bm\Sigma_t \bm W_t\bm c_t=\bm\mu_t $, for some $\Fcal_t$-measurable vector $\bm c_t$. Hence $\E_t[\bm\epsilon_{t+1}]=(\bm I_{n_t}-\bm\Pi_t)\bm\mu_t=\bm 0$.

Conversely, assume $\E_t[\bm\epsilon_{t+1}]=\bm 0$. We only use the weaker uncorrelatedness of the factor component, $\cov_t[\bm\Phi_t\bm f_{t+1},\bm\epsilon_{t+1}]=\bm 0$, which for factor portfolios reads $\bm\Pi_t\bm\Sigma_t=\bm\Pi_t\bm\Sigma_t\bm\Pi_t^\top$. The right hand side is symmetric, so that $\bm\Pi_t\bm\Sigma_t=\bm\Sigma_t\bm\Pi_t^\top$. Combining $\E_t[\bm\epsilon_{t+1}]=\bm 0$, that is $\bm\mu_t=\bm\Pi_t\bm\mu_t$, with \eqref{eq:LoP}, $\bm\mu_t=\bm\Sigma_t\bm b_t$ for some $\Fcal_t$-measurable vector $\bm b_t$, we obtain $\bm\mu_t = \bm\Pi_t \bm\mu_t =\bm\Pi_t \bm\Sigma_t \bm b_t = \bm\Sigma_t\bm\Pi_t^\top \bm b_t = \bm\Sigma_t\bm W_t\bm\Phi_t^\top \bm b_t\in \Ima(\bm\Sigma_t\bm W_t)$, which is \eqref{eq:spanNEWeq1}. \qed

\subsection{Orthogonality and uncorrelatedness}\label{app:orthouncorr}
\begin{remark}\label{rem:propspan}
As the proof of Proposition~\ref{prop:uncspan} shows, the implication from unpriced residual risk \eqref{eq:coh2} to factor spanning \eqref{eq:maxSR} continues to hold under the weaker uncorrelatedness of the factor component, $\cov_t[\bm\Phi_t\bm f_{t+1},\bm\epsilon_{t+1}]=\bm 0$, in place of \eqref{eq:UNCORR}. In contrast, the converse implication does not hold without uncorrelatedness in general, as the trivial counterexample with $\bm\Phi_t=\bm 0$ and $\bm\mu_t\neq\bm 0$ shows.
\end{remark}

The following lemma records the pricing content of unpriced residual risk.

\begin{lemma}\label{lem:FOC1}
Given a factor representation \eqref{eq:linfacmodelPhi}, the following are equivalent:
\begin{enumerate}
\item Residual risk is unpriced, \eqref{eq:coh2}.

\item Asset risk premia are spanned by factors,
\begin{equation}\label{eq:ARPspf}
\bm\mu_t =\bm\Phi_t \bm\mu_{\bm f,t}.
\end{equation}
 \end{enumerate}
In either case the risk-premium condition \eqref{eq:arpspanning} holds. If in addition factors and residuals are orthogonal,
\eqref{eq:ORTHO},
then any of the above implies (and, if $\bm\mu_{\bm f,t}\neq \bm 0$ almost surely, follows from) uncorrelatedness
\eqref{eq:UNCORR}.
\end{lemma}

\begin{proof}
This follows from the elementary identities $\bm\mu_t =\bm\Phi_t \bm\mu_{\bm f,t}+ \bm\mu_{\bm \epsilon,t}$ and $\E_t[\bm\epsilon_{t+1} \bm f_{t+1}^\top]=\cov_t[\bm\epsilon_{t+1} ,\bm f_{t+1}]+\bm\mu_{\bm \epsilon,t}\bm\mu_{\bm f,t} ^\top$.
\end{proof}

The following example separates the orthogonality conditions for factor portfolios. Its second half shows in addition that the uncorrelatedness hypothesis of Proposition~\ref{prop:uncspan} is not implied by its conclusions: the pricing restriction \eqref{eq:coh2} and factor spanning \eqref{eq:maxSR} jointly do not deliver factor--residual uncorrelatedness \eqref{eq:UNCORR}. It exhibits OLS factor portfolios that satisfy cross-sectional orthogonality, $\bm\Phi_t^\top\bm\epsilon_{t+1}=\bm 0$, and matrix orthogonality \eqref{eq:matrixortho} while violating factor--residual orthogonality \eqref{eq:ORTHO} and uncorrelatedness \eqref{eq:UNCORR}, and it shows that the latter two conditions are not interchangeable.

\begin{example}\label{ex:spanning2}
Let $n_t=3$ and $m_t=2$, and let $\xi_1,\xi_2,\xi_3$ be conditionally uncorrelated random variables with conditional variances $\var_t[\xi_i]=a_i>0$, means $b_i\coloneqq\E_t[\xi_i]\in\R$, and second moments $q_i\coloneqq\E_t[\xi_i^2]=a_i+b_i^2$. Assume returns and loading matrix are given by
\[
\bm x_{t+1}=\begin{bmatrix} \xi_1\\ \xi_2\\ \frac{\rho}{a_1}\xi_1+\frac{\rho}{a_2}\xi_2+\xi_3\end{bmatrix},\quad
\bm\Phi_t=\begin{bmatrix} 1 & 0\\ 1 & 0\\ 0 & 1\end{bmatrix},
\]
for some $\rho\in\R$. This gives the following mean and covariance matrix of $\bm x_{t+1}$,
\[
\bm\mu_t=\begin{bmatrix} b_1\\ b_2\\ \frac{\rho}{a_1}b_1+\frac{\rho}{a_2}b_2+b_3\end{bmatrix},\quad
\bm\Sigma_t=\begin{bmatrix} a_1 & 0 & \rho\\ 0 & a_2 & \rho\\ \rho & \rho & \frac{\rho^2}{a_1}+\frac{\rho^2}{a_2}+a_3\end{bmatrix}.
\]
The loading matrix has full column rank $m_t=2$, with pseudoinverse
\[
\bm\Phi_t^+=\begin{bmatrix} \frac{1}{2} & \frac{1}{2} & 0\\ 0 & 0 & 1\end{bmatrix}.
\]
We consider the OLS factors \eqref{eq:SLSfact} for $\bm S_t=\bm I_3$, that is, the factor portfolios \eqref{eq:factorportfolios} for $\bm W_t^\top=\bm\Phi_t^+$,
\[
\bm f_{t+1}=\bm\Phi_t^+\bm x_{t+1}=\begin{bmatrix} \frac{1}{2}\bm x_{t+1,1}+\frac{1}{2}\bm x_{t+1,2}\\ \bm x_{t+1,3}\end{bmatrix}.
\]
The factor component operator $\bm\Pi_t=\bm\Phi_t\bm\Phi_t^+$ is the orthogonal projection onto $\Ima\bm\Phi_t$, so \eqref{eq:FCopProj} holds, and the residuals
\[
\bm\epsilon_{t+1}=(\bm I_3-\bm\Phi_t\bm\Phi_t^+)\bm x_{t+1}=\begin{bmatrix} \frac{\xi_1-\xi_2}{2}\\ \frac{\xi_2-\xi_1}{2}\\ 0\end{bmatrix}
\]
satisfy cross-sectional orthogonality, $\bm\Phi_t^\top\bm\epsilon_{t+1}=\bm 0$, by inspection, as well as matrix orthogonality \eqref{eq:matrixortho}, in line with Corollary~\ref{cor:rankdefWLS}. However,
\[
\cov_t[\bm\epsilon_{t+1},\bm f_{t+1}]=\frac{a_1-a_2}{4}\begin{bmatrix} 1 & 0\\ -1 & 0\\ 0 & 0\end{bmatrix},\quad
\E_t[\bm\epsilon_{t+1}\bm f_{t+1}^\top]=\cov_t[\bm\epsilon_{t+1},\bm f_{t+1}]+\bm\mu_{\bm\epsilon,t}\bm\mu_{\bm f,t}^\top,
\]
with $\bm\mu_{\bm\epsilon,t}=\frac{b_1-b_2}{2}[1,-1,0]^\top$ and $\bm\mu_{\bm f,t}=[\frac{b_1+b_2}{2},\,\frac{\rho}{a_1}b_1+\frac{\rho}{a_2}b_2+b_3]^\top$. In particular, the $(1,1)$ entry of $\E_t[\bm\epsilon_{t+1}\bm f_{t+1}^\top]$ equals $(q_1-q_2)/4$. Uncorrelatedness \eqref{eq:UNCORR} thus fails whenever $a_1\neq a_2$, and orthogonality \eqref{eq:ORTHO} fails whenever $q_1\neq q_2$. For $a_1=a_2$ and $b_1\neq b_2$ with $\bm\mu_{\bm f,t}\neq\bm 0$, uncorrelatedness \eqref{eq:UNCORR} holds while orthogonality \eqref{eq:ORTHO} fails, so the two conditions are not interchangeable, in line with Lemma~\ref{lem:FOC1}.

\medskip\noindent\emph{Spanning without uncorrelatedness.} Assume in addition that $b_1=b_2$ and $\rho\neq 0$. Then the pricing restriction \eqref{eq:coh2} holds by inspection, as $\bm\mu_{\bm\epsilon,t}=\frac{b_1-b_2}{2}[1,-1,0]^\top=\bm 0$. Moreover, straightforward algebra shows that
\[
\bm\Sigma_t \bm W_t
\begin{bmatrix}0\\ b_1/\rho\end{bmatrix}
=\bm\mu_t,
\]
provided that
$b_3=a_3 b_1/\rho$,
which establishes \eqref{eq:spanNEWeq1}, and hence \eqref{eq:maxSR}. However, for $a_1\neq a_2$, uncorrelatedness \eqref{eq:UNCORR} fails, and so does the weaker uncorrelatedness of the factor component, as $\cov_t[\bm\Phi_t\bm f_{t+1},\bm\epsilon_{t+1}]=\bm\Phi_t\cov_t[\bm f_{t+1},\bm\epsilon_{t+1}]\neq\bm 0$.
\end{example}

The following lemma clarifies the mechanism underlying this failure and places Example~\ref{ex:spanning2} in a broader context.\footnote{In Example~\ref{ex:spanning2}, the vector $\bm c_t$ equals $[0, b_1/\rho]^\top$.}

\begin{lemma}\label{lem:CEX}
Assume the law of one price \eqref{eq:LoP} holds. For factor portfolios \eqref{eq:factorportfolios}, the pricing restriction \eqref{eq:coh2} together with factor spanning \eqref{eq:maxSR} imply that there exists an $\Fcal_t$-measurable vector $\bm c_t\in\R^{m_t}$ such that
\[
\bm\Sigma_t \bm W_t \bm c_t
= \bm\Pi_t\bm\Sigma_t \bm W_t \bm c_t.
\]
However, the corresponding matrix identity $\bm\Sigma_t\bm W_t=\bm\Pi_t\bm\Sigma_t\bm W_t$, and hence factor--residual uncorrelatedness \eqref{eq:UNCORR}, need not hold in general.
\end{lemma}

\begin{proof}
Factor spanning~\eqref{eq:spanNEWeq1} implies that there exists an $\Fcal_t$-measurable vector $\bm c_t\in\R^{m_t} $ such that $\bm\mu_t=\bm\Sigma_t\bm W_t\bm c_t$. The pricing restriction \eqref{eq:coh2} is equivalent to $\bm\mu_t=\bm\Phi_t\bm\mu_{\bm f,t}=\bm\Pi_t\bm\mu_t$, by Lemma~\ref{lem:FOC1}. The claimed vector equality follows. The last statement follows by means of the counterexample given in Example~\ref{ex:spanning2}.
\end{proof}

\subsection{The bijection between weight and loading matrices}\label{app:bijection}
This subsection develops the machinery behind Theorem~\ref{thm:rigid}. The proof follows in the next subsection, where existence and uniqueness are a two-line corollary of Theorem~\ref{thm:rigidgen}. No rank or invertibility assumption is used anywhere in this subsection. The pseudoinverse identities \eqref{eq:pinvid}--\eqref{eq:sqrtproj} of Appendix~\ref{app:matrix} are used throughout.

Both weight matrices and loading matrices live in one linear space.

Recall from Section~\ref{sec:cost} the set $\Vcal_t=\{\bm A\in\R^{n_t\times m_t}:\Ima\bm A\subseteq\Ima\bm Q_t\}$, for a fixed number of columns $m_t$. It is a linear subspace of $\R^{n_t\times m_t}$ of dimension $m_t\rank\bm Q_t$, and the orthogonal projection onto it is $\bm A\mapsto\bm P_{\bm Q,t}\bm A$.

For a symmetric positive semidefinite $n_t\times n_t$ matrix $\bm A$ define
\begin{equation}\label{eq:Psidef}
\Psi_{\bm A}(\bm X)\coloneqq\bm A\bm X(\bm X^\top\bm A\bm X)^+.
\end{equation}
By \eqref{eq:TSreg}, the regression loading matrix of the traded factors $\bm f_{t+1}=\bm W_t^\top\bm x_{t+1}$ is exactly
\begin{equation}\label{eq:PsiQ}
\bm\Phi_t^{\bm f_{t+1}}=\bm Q_t\bm W_t(\bm W_t^\top\bm Q_t\bm W_t)^+=\Psi_{\bm Q_t}(\bm W_t).
\end{equation}

\begin{theorem}[Existence and Uniqueness, General Form]\label{thm:rigidgen}
The map $\bm W_t\mapsto\bm W_t^\top\bm x_{t+1}$ is a bijection from $\Vcal_t$ onto the set of traded factors, taken modulo almost sure equality. Moreover $\Psi_{\bm Q_t}$ maps $\Vcal_t$ bijectively onto $\Vcal_t$, and its inverse is $\Psi_{\bm Q_t^+}$. Explicitly, for $\bm W_t,\bm\Phi_t\in\Vcal_t$,
\begin{equation}\label{eq:duality}
\bm\Phi_t=\bm Q_t\bm W_t(\bm W_t^\top\bm Q_t\bm W_t)^+
\qquad\Longleftrightarrow\qquad
\bm W_t=\bm Q_t^+\bm\Phi_t(\bm\Phi_t^\top\bm Q_t^+\bm\Phi_t)^+.
\end{equation}
\end{theorem}

\begin{proof}
\emph{Step 1: the parameterization.} By the support fact of Remark~\ref{rem:support}, $\Ima\bm Q_t$ is the smallest subspace containing $\bm x_{t+1}$ almost surely. Hence $\bm N_t^\top\bm x_{t+1}=\bm 0$ almost surely if and only if the columns of $\bm N_t$ are orthogonal to $\Ima\bm Q_t$, that is, if and only if $\bm P_{\bm Q,t}\bm N_t=\bm 0$. Both claims follow.

\emph{Step 2: two bijections of $\Vcal_t$.} The maps $\bm M_{\bm Q^{1/2}}:\bm A\mapsto\bm Q_t^{1/2}\bm A$, with inverse $\bm M_{\bm Q^{+1/2}}:\bm A\mapsto\bm Q_t^{+1/2}\bm A$, and $\bm J:\bm A\mapsto(\bm A^+)^\top$, an involution, are bijections of $\Vcal_t$ into itself. Both maps send $\Vcal_t$ into $\Vcal_t$ by \eqref{eq:sqrtproj}, and $\bm Q_t^{+1/2}\bm Q_t^{1/2}\bm A=\bm P_{\bm Q,t}\bm A=\bm A$ for $\bm A\in\Vcal_t$, and symmetrically.
$\bm J$ maps $\Vcal_t$ into $\Vcal_t$ by \eqref{eq:pinvim}. By \eqref{eq:pinvid}, $\big((\bm A^+)^\top\big)^+=\big((\bm A^+)^+\big)^\top=\bm A^\top$, hence $\bm J(\bm J(\bm A))=\bm A$.

\emph{Step 3: the duality.} Let $\bm X\in\Vcal_t$ and put $\bm V\coloneqq\bm Q_t^{1/2}\bm X\in\Vcal_t$. By \eqref{eq:pinvkey},
\[
\Psi_{\bm Q_t}(\bm X)=\bm Q_t^{1/2}\bm V(\bm V^\top\bm V)^+=\bm Q_t^{1/2}(\bm V^+)^\top
=\bm Q_t^{1/2}\bm J\big(\bm Q_t^{1/2}\bm X\big),
\]
that is, $\Psi_{\bm Q_t}=\bm M_{\bm Q^{1/2}}\circ\bm J\circ\bm M_{\bm Q^{1/2}}$ on $\Vcal_t$. The same computation with $\bm Q_t$ replaced by $\bm Q_t^+$, whose square root is $\bm Q_t^{+1/2}$ and whose image is again $\Ima\bm Q_t$, gives $\Psi_{\bm Q_t^+}=\bm M_{\bm Q^{+1/2}}\circ\bm J\circ\bm M_{\bm Q^{+1/2}}$ on $\Vcal_t$. By Step 2 both are compositions of bijections of $\Vcal_t$, and since $\bm J$ is an involution and $\bm M_{\bm Q^{+1/2}}$ inverts $\bm M_{\bm Q^{1/2}}$,
\[
\Psi_{\bm Q_t^+}\circ\Psi_{\bm Q_t}
=\bm M_{\bm Q^{+1/2}}\bm J\,\bm M_{\bm Q^{+1/2}}\bm M_{\bm Q^{1/2}}\bm J\,\bm M_{\bm Q^{1/2}}
=\bm M_{\bm Q^{+1/2}}\bm J\bm J\,\bm M_{\bm Q^{1/2}}=\id,
\]
and symmetrically $\Psi_{\bm Q_t}\circ\Psi_{\bm Q_t^+}=\id$.
\end{proof}

\subsection{Proof of Theorem~\ref{thm:rigid} and Corollary~\ref{cor:premium}}
\ref{it:rigid1} Necessity of \eqref{eq:Phistarass}: by the support fact of Remark~\ref{rem:support}, $\Ima\E_t[\bm x_{t+1}\bm f_{t+1}^\top]\subseteq\Ima\bm Q_t$, hence $\Ima\bm\Phi_t^{\bm f_{t+1}}\subseteq\Ima\bm Q_t$ by \eqref{eq:TSreg}. This holds for arbitrary, not necessarily traded, factors. Sufficiency follows from~\ref{it:rigid2}.

\ref{it:rigid2} By Theorem~\ref{thm:rigidgen} we may take $\bm W_t\in\Vcal_t$ without loss of generality, and two traded factors agree almost surely if and only if their weight matrices in $\Vcal_t$ agree. Given \eqref{eq:Phistarass} we have $\bm\Phi_t\in\Vcal_t$, so by \eqref{eq:PsiQ} and Theorem~\ref{thm:rigidgen} the equation $\bm\Phi_t^{\bm f_{t+1}}=\bm\Phi_t$ has the unique solution $\bm W_t=\Psi_{\bm Q_t^+}(\bm\Phi_t)=\bm Q_t^+\bm\Phi_t(\bm\Phi_t^\top\bm Q_t^+\bm\Phi_t)^+$ in $\Vcal_t$. Transposing and using the symmetry of $(\bm\Phi_t^\top\bm Q_t^+\bm\Phi_t)^+$ and of $\bm Q_t^+$ gives \eqref{eq:MWfactors}. The identity \eqref{eq:MWnosqrt} is \eqref{eq:pinvid} applied to $\bm A=\bm Q_t^{+1/2}\bm\Phi_t$, using $\bm Q_t^{+1/2}\bm Q_t^{+1/2}=\bm Q_t^+$.

\ref{it:rigid3} With $\bm G_t\coloneqq\bm\Phi_t^\top\bm Q_t^+\bm\Phi_t$ we have $\bm\Pi_t=\bm\Phi_t\bm G_t^+\bm\Phi_t^\top\bm Q_t^+$, which is the standard formula for the orthogonal projection onto $\Ima\bm\Phi_t$ in the semi-inner product \eqref{eq:Qplusinnercrit} on $\Ima\bm Q_t$. Explicitly, $\bm\Pi_t^2=\bm\Phi_t\bm G_t^+\bm G_t\bm G_t^+\bm\Phi_t^\top\bm Q_t^+=\bm\Pi_t$, and $\Ima\bm\Pi_t\subseteq\Ima\bm\Phi_t$ with equality because $\bm\Pi_t\bm\Phi_t=\bm\Phi_t\bm G_t^+\bm G_t=\bm\Phi_t$, the last step because $\bm G_t^+\bm G_t$ is the orthogonal projection onto $\Ima\bm G_t=\Ima(\bm Q_t^{+1/2}\bm\Phi_t)^\top=\Ima\bm\Phi_t^\top$, the last equality since $\bm Q_t^{+1/2}$ is injective on $\Ima\bm Q_t\supseteq\Ima\bm\Phi_t$ by \eqref{eq:sqrtproj} and \eqref{eq:Phistarass}, and $\bm Q_t^+\bm\Pi_t=\bm Q_t^+\bm\Phi_t\bm G_t^+\bm\Phi_t^\top\bm Q_t^+$ is symmetric, which is $\bm Q_t^+$-self-adjointness, so $\langle\bm\Pi_t\bm x_{t+1},(\bm I_{n_t}-\bm\Pi_t)\bm x_{t+1}\rangle_{\bm Q^+}=0$.

The operator satisfying \ref{it:rigid3} is moreover unique in the class considered: any candidate $\bm\Phi_t\bm W_t^\top$ with $\Ima\bm W_t\subseteq\Ima\bm Q_t$ vanishes on $\ker\bm Q_t=(\Ima\bm Q_t)^\perp\subseteq(\Ima\bm W_t)^\perp=\ker\bm W_t^\top$, and on $\Ima\bm Q_t$ the orthogonal projection onto $\Ima\bm\Phi_t$ for the inner product \eqref{eq:Qplusinnercrit} is unique. Hence \ref{it:rigid3} forces $\bm\Pi_t=\bm\Phi_t\bm G_t^+\bm\Phi_t^\top\bm Q_t^+$. If $\rank\bm\Phi_t=m_t$, then $\bm\Phi_t$ has a left inverse, so $\bm\Phi_t\bm W_t^\top=\bm\Phi_t\bm G_t^+\bm\Phi_t^\top\bm Q_t^+$ implies $\bm W_t^\top=\bm G_t^+\bm\Phi_t^\top\bm Q_t^+$, which is \eqref{eq:MWfactors}: condition \ref{it:rigid3} implies \ref{it:rigid2}, and all three conditions are equivalent.

\eqref{eq:rankinherit} From \eqref{eq:MWfactors}, $\bm Q_{\bm f,t}=\bm W_t^\top\bm Q_t\bm W_t=\bm G_t^+\bm G_t\bm G_t^+=\bm G_t^+$, using $\bm\Phi_t^\top\bm Q_t^+\bm Q_t\bm Q_t^+\bm\Phi_t=\bm G_t$, which is the Moore--Penrose identity $\bm Q_t^+\bm Q_t\bm Q_t^+=\bm Q_t^+$ and needs no assumption. Hence $\rank\bm Q_{\bm f,t}=\rank\bm G_t=\rank(\bm Q_t^{+1/2}\bm\Phi_t)=\rank\bm\Phi_t$, the last equality because $\bm Q_t^{+1/2}$ is injective on $\Ima\bm Q_t\supseteq\Ima\bm\Phi_t$ by \eqref{eq:sqrtproj}, and $\rank\bm W_t=\rank(\bm Q_t^+\bm\Phi_t\bm G_t^+)=\rank\bm G_t$ by the same injectivity. Finally $\bm W_t^\top\bm\Phi_t=\bm G_t^+\bm G_t$ is the orthogonal projection onto $\Ima\bm G_t=\Ima\bm Q_{\bm f,t}$, which equals $\bm I_{m_t}$ if and only if $\rank\bm G_t=m_t$.

\emph{Corollary~\ref{cor:premium}.} \eqref{eq:coh1} holds by construction. By~\ref{it:rigid3}, $\Ima\bm\Pi_t=\Ima\bm\Phi_t$, so the last statement of Lemma~\ref{lem:Pidefchar} gives $\E_t[\bm\epsilon_{t+1}]=\bm 0$ if and only if $\bm\mu_t\in\Ima\bm\Phi_t$, which is \eqref{eq:coh2}. \qed

\begin{remark}[Why the Loading Condition Cannot Be Weakened]\label{rem:twoinverse}
Assume \eqref{eq:Phistarass}. The traded solutions of the orthogonality condition $\E_t[\bm\epsilon_{t+1}\bm f_{t+1}^\top]=\bm 0$, that is, of \eqref{eq:ORTHO}, written in terms of their canonical weight matrices $\bm W_t\in\Vcal_t$ of Theorem~\ref{thm:rigidgen}, are exactly
\[
\bm W_t=\bm Q_t^+\bm\Phi_t\bm M,\qquad
\bm M=\bm M^\top \text{ with } \bm M\bm G_t\bm M=\bm M,\qquad
\bm G_t=\bm\Phi_t^\top\bm Q_t^+\bm\Phi_t,
\]
that is, the symmetric $\{2\}$-inverses of $\bm G_t$. This family includes $\bm M=\bm 0$, giving the null factors $\bm f_{t+1}=\bm 0$, for which orthogonality holds trivially. \eqref{eq:coh1} selects the single member $\bm M=\bm G_t^+$, by Theorem~\ref{thm:rigid}. Indeed, orthogonality reads $(\bm I_{n_t}-\bm\Phi_t\bm W_t^\top)\bm Q_t\bm W_t=\bm 0$, that is, $\bm Q_t\bm W_t=\bm\Phi_t\bm M$ with $\bm M\coloneqq\bm W_t^\top\bm Q_t\bm W_t$ symmetric. For $\bm W_t\in\Vcal_t$ this gives $\bm W_t=\bm Q_t^+\bm\Phi_t\bm M$, and substituting back, $\bm M=\bm M\bm G_t\bm M$. Conversely, for symmetric $\bm M$ with $\bm M\bm G_t\bm M=\bm M$, the matrix $\bm W_t=\bm Q_t^+\bm\Phi_t\bm M$ satisfies $\bm W_t^\top\bm Q_t\bm W_t=\bm M\bm G_t\bm M=\bm M$ and $\bm Q_t\bm W_t=\bm P_{\bm Q,t}\bm\Phi_t\bm M=\bm\Phi_t\bm M$, using \eqref{eq:Phistarass}. Orthogonality follows.
\end{remark}

\subsection{The characterization of admissible loading matrices}
We verify that a loading matrix admits a coherent factor representation if and only if it satisfies \eqref{eq:admissible}.
\emph{Necessity.} \eqref{eq:coh1} and Theorem~\ref{thm:rigid}~\ref{it:rigid1} give \eqref{eq:Phistarass}, the argument there applying to arbitrary factors. \eqref{eq:coh2} and Lemma~\ref{lem:FOC1} give $\bm\mu_t=\bm\Phi_t\bm\mu_{\bm f,t}\in\Ima\bm\Phi_t$, which is \eqref{eq:arpspanning}. Combining \eqref{eq:coh1} with \eqref{eq:ortho3al0} and Lemma~\ref{lem:FOC1} gives uncorrelatedness \eqref{eq:UNCORR}.

\emph{Sufficiency.} Take the traded factors \eqref{eq:MWfactors}, which are well defined by \eqref{eq:Phistarass}. By Corollary~\ref{cor:premium} and \eqref{eq:arpspanning} the representation is coherent.

For the last statement, let $(\bm\Phi_t,\bm f_{t+1})$ be coherent and traded. By the uncorrelatedness implied by coherence, recorded in Section~\ref{sec:spanning}, factors and residuals are uncorrelated, \eqref{eq:UNCORR}, and residual risk is unpriced, \eqref{eq:coh2}. Assuming \eqref{eq:LoP}, Proposition~\ref{prop:uncspan} gives factor spanning \eqref{eq:maxSR}, so \eqref{eq:factorportfolios} and \eqref{eq:maxSR} hold. \qed

\subsection{Proof of Proposition~\ref{prop:tradeq}}
We prove part~\ref{it:tradeq1} first. The proof of part~\ref{it:tradeq2} uses it.

\ref{it:tradeq1}: By \eqref{eq:LoP} there exists an $\Fcal_t$-measurable vector $\bm\nu_t$ with $\bm\mu_t=\bm\Sigma_t\bm\nu_t$. Hence
\[
\bm\mu_{\bm f,t}=\bm W_t^\top\bm\Sigma_t\bm\nu_t=(\bm W_t^\top\bm\Sigma_t^{1/2})(\bm\Sigma_t^{1/2}\bm\nu_t)\in\Ima(\bm W_t^\top\bm\Sigma_t^{1/2})=\Ima\bm\Sigma_{\bm f,t},
\]
where the last identity follows from $\bm\Sigma_{\bm f,t}=(\bm W_t^\top\bm\Sigma_t^{1/2})(\bm W_t^\top\bm\Sigma_t^{1/2})^\top$. By \eqref{eq:hatalphaeqal0}, $\bm\Phi^{\bm f_{t+1}\top}_t=\bm Q_{\bm f,t}^+\E_t[\bm f_{t+1}\bm x_{t+1}^\top]$, so that $\Ima\bm\Phi^{\bm f_{t+1}\top}_t\subseteq\Ima\bm Q_{\bm f,t}^+=\Ima\bm Q_{\bm f,t}$. Since $\bm Q_{\bm f,t}=\bm\Sigma_{\bm f,t}+\bm\mu_{\bm f,t}\bm\mu_{\bm f,t}^\top$, we obtain
\[
\Ima\bm\Phi^{\bm f_{t+1}\top}_t\subseteq\Ima\bm\Sigma_{\bm f,t}+\Ima\bm\mu_{\bm f,t}=\Ima\bm\Sigma_{\bm f,t},
\]
which is \eqref{eq:ImaPhifin}.

Since $\bm\mu_{\bm f,t}\in\Ima\bm\Sigma_{\bm f,t}$, the extended Sherman--Morrison formula \eqref{eq:SMformula} of Proposition~\ref{prop:LOP}, applied to $(\bm\Sigma_{\bm f,t},\bm\mu_{\bm f,t})$ in place of $(\bm\Sigma_t,\bm\mu_t)$, gives
\[
\bm Q_{\bm f,t}^+=\bm\Sigma_{\bm f,t}^+-c_t^{-1}\bm\Sigma_{\bm f,t}^+\bm\mu_{\bm f,t}\bm\mu_{\bm f,t}^\top\bm\Sigma_{\bm f,t}^+,
\qquad
\bm\mu_{\bm f,t}^\top\bm Q_{\bm f,t}^+=c_t^{-1}\bm\mu_{\bm f,t}^\top\bm\Sigma_{\bm f,t}^+,
\]
for $c_t\coloneqq 1+\bm\mu_{\bm f,t}^\top\bm\Sigma_{\bm f,t}^+\bm\mu_{\bm f,t}>0$. Substituting $\E_t[\bm x_{t+1}\bm f_{t+1}^\top]=\cov_t[\bm x_{t+1},\bm f_{t+1}]+\bm\mu_t\bm\mu_{\bm f,t}^\top$ into \eqref{eq:TSreg} and collecting terms,
\[
\bm\Phi^{\bm f_{t+1}}_t
=\bm\beta^{\bm f_{t+1}}_t+\bigl(\bm\mu_t-\bm\beta^{\bm f_{t+1}}_t\bm\mu_{\bm f,t}\bigr)c_t^{-1}\bm\mu_{\bm f,t}^\top\bm\Sigma_{\bm f,t}^+
=\bm\beta^{\bm f_{t+1}}_t+\bm\alpha^{\bm f_{t+1}}_t\bm\mu_{\bm f,t}^\top\bm Q_{\bm f,t}^+,
\]
using \eqref{eq:hatalphaeq}, which is \eqref{eq:betaPhi}. The two loading matrices therefore coincide if and only if $\bm\alpha^{\bm f_{t+1}}_t\bm\mu_{\bm f,t}^\top\bm Q_{\bm f,t}^+=\bm 0$, equivalently $\bm\alpha^{\bm f_{t+1}}_t\bm\mu_{\bm f,t}^\top=\bm 0$, as $\bm\mu_{\bm f,t}^\top\bm Q_{\bm f,t}^+=\bm 0$ forces $\bm\mu_{\bm f,t}=\bm 0$. This proves \ref{it:tradeq1}.

\ref{it:tradeq2}: Given \eqref{eq:ImaPhifin}, Lemma~\ref{lem:Phialbe} shows that unpriced residual risk, its condition~\ref{it:Phialbe2}, is equivalent to \eqref{eq:Phialbeeq5} and hence implies $\bm\alpha^{\bm f_{t+1}}_t=\bm 0$. The converse implication is condition~\ref{it:alphazero1} of Lemma~\ref{lem:alphazero}. Together these are the equivalence \eqref{eq:tradeq}, and under either side \eqref{eq:betaPhi} gives $\bm\beta^{\bm f_{t+1}}_t=\bm\Phi^{\bm f_{t+1}}_t$. Coincidence of residuals, $\bm\eta^{\bm f_{t+1}}_{t+1}=\bm\epsilon^{\bm f_{t+1}}_{t+1}$, is condition~\ref{it:Phialbe3} of Lemma~\ref{lem:Phialbe}.

For the statement following the proposition: given \eqref{eq:coh1}, the remaining property of $(\bm\Phi_t,\bm f_{t+1})$ is \eqref{eq:coh2}, which by \eqref{eq:tradeq} is the zero-alpha restriction. Spanning then follows from the statement recorded after Corollary~\ref{cor:premium}. In that case \eqref{eq:betaPhi} gives $\bm\beta^{\bm f_{t+1}}_t=\bm\Phi^{\bm f_{t+1}}_t$, which equals $\bm\Phi_t$ by \eqref{eq:coh1}.

In particular \ref{it:tradeq1} and \ref{it:tradeq2} apply to the moment-weighted factor portfolios \eqref{eq:MWfactors} of Theorem~\ref{thm:rigid}. \qed

\subsection{Proof of Proposition~\ref{prop:rankdef}}
Recall the factor component operator $\bm\Pi_t=\bm\Phi_t\bm W_t^\top$ and write $\bm M_t\coloneqq\bm I_{n_t}-\bm\Pi_t$. The residuals $\bm\epsilon_{t+1}=\bm M_t\bm x_{t+1}$ have second moment matrix $\bm Q_{\bm\epsilon,t}=\bm M_t\bm Q_t\bm M_t^\top$.

Assume \eqref{eq:ORTHO}. Then $\bm 0=\E_t[\bm\epsilon_{t+1}\bm f_{t+1}^\top]=\bm M_t\bm Q_t\bm W_t$, and transposing, $\bm W_t^\top\bm Q_t\bm M_t^\top=\bm 0$. Hence both $\bm\Pi_t\bm Q_t\bm M_t^\top=\bm\Phi_t(\bm W_t^\top\bm Q_t\bm M_t^\top)=\bm 0$ and $\bm\Pi_t^2\bm Q_t\bm M_t^\top=\bm\Pi_t\bm\Phi_t(\bm W_t^\top\bm Q_t\bm M_t^\top)=\bm 0$, and thus
\[
\bm\Pi_t\bm Q_{\bm\epsilon,t}=\bm\Pi_t\bm M_t\bm Q_t\bm M_t^\top=\bm\Pi_t\bm Q_t\bm M_t^\top-\bm\Pi_t^2\bm Q_t\bm M_t^\top=\bm 0.
\]
Assume instead \eqref{eq:FCopProj}. Then $\bm\Pi_t\bm M_t=\bm\Pi_t-\bm\Pi_t^2=\bm 0$, so again $\bm\Pi_t\bm Q_{\bm\epsilon,t}=\bm\Pi_t\bm M_t\bm Q_t\bm M_t^\top=\bm 0$. In either case \eqref{eq:matrixortho} holds. Consequently, $\Ima\bm Q_{\bm\epsilon,t}\subseteq\ker\bm\Pi_t$ and $\rank\bm Q_{\bm\epsilon,t}\le n_t-\rank\bm\Pi_t<n_t$, where $\bm\Pi_t\neq\bm 0$ follows from non-degeneracy. Moreover, $\bm Q_{\bm\epsilon,t}=\bm\Sigma_{\bm\epsilon,t}+\bm\mu_{\bm\epsilon,t}\bm\mu_{\bm\epsilon,t}^\top\succeq\bm\Sigma_{\bm\epsilon,t}$ implies $\ker\bm Q_{\bm\epsilon,t}\subseteq\ker\bm\Sigma_{\bm\epsilon,t}$ and thus $\rank\bm\Sigma_{\bm\epsilon,t}\le\rank\bm Q_{\bm\epsilon,t}$. This proves \eqref{eq:RDambient}. \qed

\subsection{Rank deficiency relative to the support of returns}
The bound \eqref{eq:RDambient} counts the directions the residuals lose in the ambient space. The following lemma, invoked in Section~\ref{sec:rankdeficiency}, places the loss inside the support of returns. The two bounds are not comparable. For invertible $\bm Q_t$ the lemma follows from \eqref{eq:RDambient}, while for singular $\bm Q_t$ it is the informative statement, since $n_t-\rank\bm\Pi_t$ may exceed $\rank\bm Q_t$.

\begin{lemma}[Rank Deficiency Relative to the Support of Returns]\label{lem:rankdefsupp}
Under the hypotheses of Proposition~\ref{prop:rankdef}, assume in addition that $\Ima\bm\Pi_t\cap\Ima\bm Q_t\neq\{\bm 0\}$, which holds under \eqref{eq:ORTHO}, and under \eqref{eq:FCopProj} whenever $\bm Q_t$ is invertible. Then
\begin{equation}\label{eq:RD}
\rank \bm \Sigma_{\bm\epsilon,t}\le \rank \bm Q_{\bm\epsilon,t}<\rank\bm Q_t.
\end{equation}
For the WLS factor portfolios \eqref{eq:SLSfact}, the intersection condition and thus \eqref{eq:RD} hold whenever $\bm\Phi_t\neq\bm 0$, the support condition \eqref{eq:Phistarass} holds, and the weighting satisfies \eqref{eq:NoNDS}.
\end{lemma}

\begin{proof}
In the notation of the proof of Proposition~\ref{prop:rankdef}, $\bm Q_{\bm\epsilon,t}=(\bm M_t\bm Q_t^{1/2})(\bm M_t\bm Q_t^{1/2})^\top$, so that, by rank--nullity applied to the restriction of $\bm M_t$ to $\Ima\bm Q_t^{1/2}=\Ima\bm Q_t$,
\[
\rank\bm Q_{\bm\epsilon,t}=\rank(\bm M_t\bm Q_t^{1/2})=\rank\bm Q_t-\dim(\ker\bm M_t\cap\Ima\bm Q_t).
\]
It thus suffices to exhibit a nonzero $\bm v_t\in\ker\bm M_t\cap\Ima\bm Q_t$. Under \eqref{eq:ORTHO}, the identity $\bm M_t\bm Q_t\bm W_t=\bm 0$ gives $\bm Q_t\bm W_t=\bm\Pi_t\bm Q_t\bm W_t$, and $\bm Q_t\bm W_t\neq\bm 0$, as $\bm\Pi_t\bm Q_t\neq\bm 0$ implies $\bm W_t^\top\bm Q_t\neq\bm 0$. Any nonzero column combination $\bm v_t=\bm Q_t\bm W_t\bm a_t$ then satisfies $\bm M_t\bm v_t=\bm 0$ and lies in $\Ima\bm\Pi_t\cap\Ima\bm Q_t$, so the intersection condition holds under \eqref{eq:ORTHO}. Under \eqref{eq:FCopProj}, the idempotency of $\bm\Pi_t$ gives $\ker\bm M_t=\Ima\bm\Pi_t$, so any nonzero $\bm v_t\in\Ima\bm\Pi_t\cap\Ima\bm Q_t$ works, and if $\bm Q_t$ is invertible the intersection condition reduces to $\bm\Pi_t\neq\bm 0$.

For the WLS factor portfolios, \eqref{eq:FCopProj} holds by Lemma~\ref{lem:Pidefchar}. Assume $\bm\Phi_t\neq\bm 0$ and $\Ima\bm\Phi_t\subseteq\Ima\bm Q_t$. If $\bm S_t\bm\Phi_t=\bm 0$ then the columns of $\bm\Phi_t$ lie in $\Ima\bm Q_t\cap\ker\bm S_t=\{\bm 0\}$ by \eqref{eq:NoNDS}, contradicting $\bm\Phi_t\neq\bm 0$. Hence $\bm S_t\bm\Phi_t\neq\bm 0$, and $\bm\Pi_t\neq\bm 0$ follows from $\bm S_t\bm\Pi_t\bm\Phi_t=(\bm S_t\bm\Phi_t)(\bm S_t\bm\Phi_t)^+(\bm S_t\bm\Phi_t)=\bm S_t\bm\Phi_t\neq\bm 0$. By Lemma~\ref{lem:Pidefchar}, $\Ima\bm\Pi_t\subseteq\Ima\bm\Phi_t\subseteq\Ima\bm Q_t$, so any nonzero $\bm v_t\in\Ima\bm\Pi_t$ lies in $\Ima\bm\Pi_t\cap\Ima\bm Q_t$. Non-degeneracy holds as well, as writing $\bm v_t=\bm Q_t\bm u_t$ gives $\bm\Pi_t\bm Q_t\bm u_t=\bm\Pi_t\bm v_t=\bm v_t\neq\bm 0$.
\end{proof}

The intersection condition cannot be dropped under \eqref{eq:FCopProj} when $\ker\bm Q_t\neq\{\bm 0\}$. For $n_t=2$, $\bm x_{t+1}=\xi_{t+1}[1,1]^\top$ with $\E_t[\xi_{t+1}^2]>0$, $\bm\Phi_t=[1,0]^\top$ and OLS weights $\bm W_t^\top=\bm\Phi_t^+$, we have $\bm\Pi_t\bm Q_t\neq\bm 0$ and $\rank\bm Q_{\bm\epsilon,t}=\rank\bm Q_t=1$, so \eqref{eq:RD} fails. Here $\Ima\bm\Phi_t\not\subseteq\Ima\bm Q_t$.

\subsection{Proof of Corollary~\ref{cor:rankdefWLS}}
By Lemma~\ref{lem:Pidefchar}, the factor component operator $\bm\Pi_t=\bm\Phi_t(\bm S_t\bm\Phi_t)^+\bm S_t$ of the factor portfolios \eqref{eq:SLSfact} is a projection, so \eqref{eq:FCopProj} holds, and $\bm\Pi_t\bm\epsilon^{\bm\Phi_t}_{t+1}=\bm\Pi_t(\bm I_{n_t}-\bm\Pi_t)\bm x_{t+1}=\bm 0$ pointwise, whence \eqref{eq:matrixortho}. Consequently, $\Ima\bm Q_{\bm\epsilon,t}\subseteq\ker\bm\Pi_t$ and
\[
\rank\bm Q_{\bm\epsilon,t}\le n_t-\rank\bm\Pi_t=n_t-\rank(\bm S_t\bm\Phi_t),
\]
where we use that the idempotent $\bm\Pi_t$ satisfies $\rank\bm\Pi_t=\tr\bm\Pi_t=\tr((\bm S_t\bm\Phi_t)^+\bm S_t\bm\Phi_t)=\rank(\bm S_t\bm\Phi_t)$. The bound $\rank\bm\Sigma_{\bm\epsilon,t}\le\rank\bm Q_{\bm\epsilon,t}$ follows as in the proof of Proposition~\ref{prop:rankdef}. \qed

\subsection{Proof of Corollary~\ref{cor:OLScoh}}
Assume \eqref{eq:Phistarass} and let $\bm P_t\coloneqq\bm\Phi_t\bm\Phi_t^+$, the Euclidean orthogonal projection onto $\Ima\bm\Phi_t$. The OLS factor portfolios have weight matrix $\bm W_t=(\bm\Phi_t^+)^\top\in\Vcal_t$, by \eqref{eq:pinvim} and \eqref{eq:Phistarass}, and factor component operator $\bm\Phi_t\bm\Phi_t^+=\bm P_t$.

We claim that $\bm\Phi_t=\bm\Phi_t^{\bm f_{t+1}}$ holds if and only if $\bm Q_t^+\bm P_t$ is symmetric, at any rank of $\bm\Phi_t$. If $\bm\Phi_t=\bm\Phi_t^{\bm f_{t+1}}$, then $\bm P_t$ is the $\bm Q_t^+$-orthogonal projection onto $\Ima\bm\Phi_t$ by Theorem~\ref{thm:rigid}~\ref{it:rigid3}, so $\bm Q_t^+\bm P_t$ is symmetric. Conversely, suppose $\bm Q_t^+\bm P_t$ is symmetric. With $\bm G_t\coloneqq\bm\Phi_t^\top\bm Q_t^+\bm\Phi_t$, both $\bm P_t$ and $\bm\Phi_t\bm G_t^+\bm\Phi_t^\top\bm Q_t^+$ are then idempotent with image $\Ima\bm\Phi_t$, $\bm Q_t^+$-self-adjoint, and zero on $\ker\bm Q_t$, for $\bm P_t$ because $\ker\bm Q_t=(\Ima\bm Q_t)^\perp\subseteq(\Ima\bm\Phi_t)^\perp=\ker\bm P_t$, so both equal the $\bm Q_t^+$-orthogonal projection onto $\Ima\bm\Phi_t$ of the proof of Theorem~\ref{thm:rigid}~\ref{it:rigid3}, which is unique in that class. Transposing $\bm P_t=\bm\Phi_t\bm G_t^+\bm\Phi_t^\top\bm Q_t^+$, using the symmetry of $\bm P_t$, $\bm G_t^+$ and $\bm Q_t^+$, and multiplying from the right by $\bm\Phi_t^{+\top}$ gives
\[
\bm\Phi_t^{+\top}=\bm P_t\bm\Phi_t^{+\top}=\bm Q_t^+\bm\Phi_t\bm G_t^+\bm\Phi_t^\top\bm\Phi_t^{+\top}=\bm Q_t^+\bm\Phi_t\bm G_t^+,
\]
where the first equality uses $\Ima\bm\Phi_t^{+\top}=\Ima\bm\Phi_t$, by \eqref{eq:pinvim}, and the last uses $\bm G_t^+\bm\Phi_t^\top\bm\Phi_t^{+\top}=\bm G_t^+\bm\Phi_t^+\bm\Phi_t=\bm G_t^+$, from $\Ima\bm G_t^+\subseteq\Ima\bm\Phi_t^\top$. The OLS weight matrix is therefore \eqref{eq:MWfactors}, and $\bm\Phi_t=\bm\Phi_t^{\bm f_{t+1}}$ follows by Theorem~\ref{thm:rigid}~\ref{it:rigid2}. This proves the claim. It remains to identify the criterion: multiplying $\bm Q_t^+\bm P_t$ on both sides by $\bm Q_t$, and back by $\bm Q_t^+$ for the converse using $\bm P_t\bm Q_t\bm Q_t^+=\bm P_t$ under \eqref{eq:Phistarass}, symmetry of $\bm Q_t^+\bm P_t$ is equivalent to symmetry of $\bm P_t\bm Q_t$, equivalently to $(\bm I_{n_t}-\bm P_t)\bm Q_t\bm P_t=\bm 0$, which is \eqref{eq:Qinvariant}. A subspace is invariant under a symmetric matrix if and only if it is spanned by eigenvectors of that matrix, by the spectral theorem. It need not be a sum of eigenspaces: for $\bm Q_t=\diag(2,2,1)$ every line in $\spn\{\bm e_1,\bm e_2\}$ is invariant, while the eigenspaces are $\spn\{\bm e_1,\bm e_2\}$ and $\spn\{\bm e_3\}$. \qed

\subsection{Proof of Corollary~\ref{cor:cohgap}}
By Theorem~\ref{thm:rigidgen} and \eqref{eq:PsiQ}, the set of loading matrices attainable as $\bm\Phi_t^{\bm f_{t+1}}$ for traded $\bm f_{t+1}$ is exactly $\Vcal_t$. Fix $\bm A\in\Vcal_t$ and decompose
\[
\bm\Phi_t-\bm A=\bm B+\bm C,\qquad \bm B\coloneqq\bm P_{\bm Q,t}\bm\Phi_t-\bm A,\qquad \bm C\coloneqq(\bm I_{n_t}-\bm P_{\bm Q,t})\bm\Phi_t.
\]
The columns of $\bm B$ lie in $\Ima\bm Q_t$ and those of $\bm C$ in its orthogonal complement, so $\bm B^\top\bm C=\bm 0$ and
$(\bm\Phi_t-\bm A)^\top(\bm\Phi_t-\bm A)=\bm B^\top\bm B+\bm C^\top\bm C\succeq\bm C^\top\bm C$.
By Weyl monotonicity the singular values of $\bm\Phi_t-\bm A$ dominate those of $\bm C$ termwise, hence $\|\bm\Phi_t-\bm A\|\ge\|\bm C\|$ for every unitarily invariant norm, with equality at $\bm A=\bm P_{\bm Q,t}\bm\Phi_t$, which lies in $\Vcal_t$. By Theorem~\ref{thm:rigidgen} the minimum is attained by the traded factors with weight matrix $\Psi_{\bm Q_t^+}(\bm P_{\bm Q,t}\bm\Phi_t)$, that is, the moment-weighted factor portfolios of $\bm P_{\bm Q,t}\bm\Phi_t$. The minimum $\|\bm C\|$ vanishes if and only if $\bm P_{\bm Q,t}\bm\Phi_t=\bm\Phi_t$, which is \eqref{eq:Phistarass}. \qed

\subsection{Weighting invariance}
The following lemma, invoked throughout Section~\ref{sec:computation}, shows that the WLS weight matrix is invariant to stripping a factor structure aligned with $\bm\Phi_t$ from the weighting matrix. It generalizes \citet[Theorem~2]{lu_sch_12} to singular matrices, of which the classical GLS--OLS equivalences are their Theorem~1.

\begin{lemma}[Weighting Invariance]\label{lem:GLSinv}
Let $\bm A_t$ be a symmetric positive semidefinite $\Fcal_t$-measurable $n_t\times n_t$ matrix and $(\bm C_t,\bm D_t)$ a regular decomposition of $\bm A_t$ with respect to $\bm\Phi_t$, defined by \eqref{eq:regulardec} and \eqref{eq:regularim} with $\bm A_t$ in place of $\bm Q_t$. Then
\begin{equation}\label{eq:GLSinv}
(\bm\Phi_t^\top\bm A_t^+\bm\Phi_t)^+\bm\Phi_t^\top\bm A_t^+=(\bm\Phi_t^\top\bm D_t^+\bm\Phi_t)^+\bm\Phi_t^\top\bm D_t^+.
\end{equation}
\end{lemma}

\begin{proof}
We drop the subscript $t$ throughout the proof and set $\bm S\coloneqq\bm\Phi^\top\bm D^+\bm\Phi$, so that the right-hand side of \eqref{eq:GLSinv} reads $\bm S^+\bm\Phi^\top\bm D^+$. Since $\bm D\bm D^+$ is the orthogonal projection onto $\Ima\bm D\supseteq\Ima\bm\Phi$, we have $\bm D\bm D^+\bm\Phi=\bm\Phi$. For $\bm w\in\ker\bm S$, moreover, $0=\bm w^\top\bm S\bm w=\|\bm D^{+1/2}\bm\Phi\bm w\|_2^2$ implies $\bm\Phi\bm w=\bm D^{1/2}\bm D^{+1/2}\bm\Phi\bm w=\bm 0$. Hence $\ker\bm S\subseteq\ker\bm\Phi$, and therefore $\Ima\bm S=\Ima\bm\Phi^\top$.

Both summands of $\bm A=\bm\Phi\bm C\bm\Phi^\top+\bm D$ are positive semidefinite, so that $\bm v^\top\bm A\bm v=0$ forces $\bm D\bm v=\bm 0$, whence $\ker\bm A\subseteq\ker\bm D$, while $\Ima\bm A\subseteq\Ima\bm\Phi+\Ima\bm D=\Ima\bm D$. Together, $\Ima\bm A=\Ima\bm D$. Moreover,
\[
\bm A\bm D^+\bm\Phi=\bm\Phi\bm C\bm S+\bm D\bm D^+\bm\Phi=\bm\Phi(\bm I_{m_t}+\bm C\bm S),
\]
where $\bm I_{m_t}+\bm C\bm S$ is invertible, as the nonzero spectrum of $\bm C\bm S$ coincides with that of the positive semidefinite matrix $\bm S^{1/2}\bm C\bm S^{1/2}$ and is therefore nonnegative. Multiplying by $\bm A^+$ from the left and by $(\bm I_{m_t}+\bm C\bm S)^{-1}$ from the right gives
\[
\bm A^+\bm\Phi=\bm D^+\bm\Phi(\bm I_{m_t}+\bm C\bm S)^{-1},
\]
using that $\bm A^+\bm A$ is the orthogonal projection onto $\Ima\bm A=\Ima\bm D$, which contains the columns of $\bm D^+\bm\Phi$. Transposing, and setting $\bm T\coloneqq(\bm I_{m_t}+\bm S\bm C)^{-1}$, yields
\[
\bm\Phi^\top\bm A^+=\bm T\bm\Phi^\top\bm D^+,\qquad \bm\Phi^\top\bm A^+\bm\Phi=\bm T\bm S.
\]
Since $\bm T$ is invertible, $\Ima(\bm T\bm S)^\top=\Ima(\bm S\bm T^\top)=\Ima\bm S$, so $(\bm T\bm S)^+(\bm T\bm S)$ is the orthogonal projection onto $\Ima\bm S$, which equals $\bm S\bm S^+$ by symmetry of $\bm S$. For any $\bm v\in\Ima\bm S$ therefore
\[
(\bm T\bm S)^+\bm T\bm v=(\bm T\bm S)^+\bm T\bm S\bm S^+\bm v=\bm S\bm S^+\bm S^+\bm v=\bm S^+\bm v.
\]
The columns of $\bm\Phi^\top\bm D^+$ lie in $\Ima\bm\Phi^\top=\Ima\bm S$. Hence
\[
(\bm\Phi^\top\bm A^+\bm\Phi)^+\bm\Phi^\top\bm A^+=(\bm T\bm S)^+\bm T\bm\Phi^\top\bm D^+=\bm S^+\bm\Phi^\top\bm D^+,
\]
which is \eqref{eq:GLSinv}.
\end{proof}

\subsection{Proof of Proposition~\ref{prop:COMO}}
The proof builds on the following lemma.

\begin{lemma}[Decomposition Lemma]\label{lem:regdec}
A regular decomposition of $\bm Q_t$ with respect to $\bm\Phi_t$ exists if and only if \eqref{eq:Phistarass} holds. In that case, one choice is
\begin{equation}\label{eq:regdecchoice}
\bm C_t=\big(1+\tr(\bm\Phi_t^\top\bm Q_t^+\bm\Phi_t)\big)^{-1}\bm I_{m_t},\qquad \bm D_t=\bm Q_t-\bm\Phi_t\bm C_t\bm\Phi_t^\top,
\end{equation}
for which moreover $\Ima\bm D_t=\Ima\bm Q_t$.
\end{lemma}

\begin{proof}
Let $(\bm C_t,\bm D_t)$ be a regular decomposition. From \eqref{eq:regulardec} we obtain $\bm 0\preceq\bm D_t\preceq\bm Q_t$. For any $\bm v\in\ker\bm Q_t$ therefore $0\le\bm v^\top\bm D_t\bm v\le\bm v^\top\bm Q_t\bm v=0$, hence $\bm D_t\bm v=\bm 0$. Thus $\ker\bm Q_t\subseteq\ker\bm D_t$, and taking orthogonal complements gives $\Ima\bm D_t\subseteq\Ima\bm Q_t$. Combined with \eqref{eq:regularim}, condition \eqref{eq:Phistarass} follows.

Conversely, assume \eqref{eq:Phistarass} and define $(\bm C_t,\bm D_t)$ by \eqref{eq:regdecchoice}, which are $\Fcal_t$-measurable, see Appendix~\ref{app:matrix}. Set $\bm B_t\coloneqq\bm Q_t^{+1/2}\bm\Phi_t$ and $c_t\coloneqq(1+\tr(\bm B_t^\top\bm B_t))^{-1}$, so that $\bm C_t=c_t\bm I_{m_t}$, using $\bm Q_t^{+1/2}\bm Q_t^{+1/2}=\bm Q_t^+$. As $\bm Q_t\bm Q_t^+$ is the orthogonal projection onto $\Ima\bm Q_t$, assumption \eqref{eq:Phistarass} gives $\bm\Phi_t=\bm Q_t\bm Q_t^+\bm\Phi_t=\bm Q_t^{1/2}\bm B_t$. Hence
\[
\bm D_t=\bm Q_t-c_t\bm\Phi_t\bm\Phi_t^\top=\bm Q_t^{1/2}\bm M_t\bm Q_t^{1/2},\qquad \bm M_t\coloneqq\bm Q_t\bm Q_t^+-c_t\bm B_t\bm B_t^\top.
\]
The symmetric matrix $\bm M_t$ vanishes on $\ker\bm Q_t$, as $\bm B_t^\top\bm v=\bm\Phi_t^\top\bm Q_t^{+1/2}\bm v=\bm 0$ for $\bm v\in\ker\bm Q_t$. For any nonzero $\bm v\in\Ima\bm Q_t$, using $\lambda_{\max}(\bm B_t\bm B_t^\top)\le\tr(\bm B_t^\top\bm B_t)$,
\[
\bm v^\top\bm M_t\bm v\ge\big(1-c_t\tr(\bm B_t^\top\bm B_t)\big)\|\bm v\|_2^2=c_t\|\bm v\|_2^2>0.
\]
Hence $\bm M_t\succeq\bm 0$ and $\ker\bm M_t=\ker\bm Q_t$, so that $\Ima\bm M_t=\Ima\bm Q_t$. Since $\bm Q_t^{1/2}$ maps $\Ima\bm Q_t$ bijectively onto itself, we conclude that $\bm D_t\succeq\bm 0$ and $\Ima\bm D_t=\Ima\bm Q_t\supseteq\Ima\bm\Phi_t$. Thus $(\bm C_t,\bm D_t)$ is a regular decomposition, and the last statement holds.
\end{proof}

\begin{proof}[Proof of Proposition~\ref{prop:COMO}]
Existence of a regular decomposition is Lemma~\ref{lem:regdec}. The first expression in \eqref{eq:MWD} is \eqref{eq:MWfactors}, and the second equality is Lemma~\ref{lem:GLSinv} applied with $\bm A_t=\bm Q_t$, whose hypotheses are \eqref{eq:regulardec} and \eqref{eq:regularim}.
\end{proof}

\subsection{Closed moments}
The following lemma, referenced in Section~\ref{sec:computation}, gives the moments of the moment-weighted factor representation in terms of a regular decomposition.

\begin{lemma}[Closed Moments]\label{lem:closedmoments}
Let $(\bm C_t,\bm D_t)$ be a regular decomposition of $\bm Q_t$ with respect to $\bm\Phi_t$. The moments of the factor representation \eqref{eq:linfacmodelPhi} implied by the moment-weighted factor portfolios of Proposition~\ref{prop:COMO} are
\begin{equation}\label{eq:closedmoments}
\begin{aligned}
\bm\mu_{\bm f,t}&=(\bm\Phi_t^\top\bm D_t^+\bm\Phi_t)^+\bm\Phi_t^\top\bm D_t^+\bm\mu_t,\\
\bm Q_{\bm f,t}&=\bm\Phi_t^+\bm\Phi_t\bm C_t\bm\Phi_t^+\bm\Phi_t+(\bm\Phi_t^\top\bm D_t^+\bm\Phi_t)^+,\\
\bm Q_{\bm\epsilon,t}&=\bm D_t-\bm\Phi_t(\bm\Phi_t^\top\bm D_t^+\bm\Phi_t)^+\bm\Phi_t^\top.
\end{aligned}
\end{equation}
\end{lemma}

\begin{proof}
Write $\bm W_t^\top$ for \eqref{eq:MWD}, $\bm\Pi_t=\bm\Phi_t\bm W_t^\top$ for the corresponding factor component operator, and $\bm G_t\coloneqq\bm D_t^{+1/2}\bm\Phi_t$, so that $\bm\Phi_t=\bm D_t\bm D_t^+\bm\Phi_t=\bm D_t^{1/2}\bm G_t$ by \eqref{eq:regularim}. We collect four facts.
\begin{enumerate}
\item\label{it:prGLSreg1} $\bm W_t^\top\bm\Phi_t=(\bm G_t^\top\bm G_t)^+(\bm G_t^\top\bm G_t)=\bm G_t^+\bm G_t$ is the orthogonal projection onto $\Ima\bm G_t^\top=\Ima\bm\Phi_t^\top$, hence $\bm W_t^\top\bm\Phi_t=\bm\Phi_t^+\bm\Phi_t$. Here $\Ima\bm G_t^\top=\Ima\bm\Phi_t^\top$ follows from $\bm\Phi_t^\top=\bm G_t^\top\bm D_t^{1/2}$ and $\bm G_t^\top=\bm\Phi_t^\top\bm D_t^{+1/2}$.
\item\label{it:prGLSreg2} $\bm\Pi_t^2=\bm\Phi_t\bm\Phi_t^+\bm\Phi_t\bm W_t^\top=\bm\Pi_t$ is a projection, with $\Ima\bm\Pi_t=\Ima\bm\Phi_t$, as $\bm\Pi_t\bm\Phi_t=\bm\Phi_t\bm\Phi_t^+\bm\Phi_t=\bm\Phi_t$. This proves \eqref{eq:FCopProj}.
\item\label{it:prGLSreg3} $\bm D_t\bm W_t=\bm D_t\bm D_t^+\bm\Phi_t(\bm\Phi_t^\top\bm D_t^+\bm\Phi_t)^+=\bm\Phi_t(\bm\Phi_t^\top\bm D_t^+\bm\Phi_t)^+$, hence $(\bm I_{n_t}-\bm\Pi_t)\bm D_t\bm W_t=\bm 0$ by \ref{it:prGLSreg2}.
\item\label{it:prGLSreg4} $\bm\Pi_t\bm D_t=\bm\Phi_t(\bm\Phi_t^\top\bm D_t^+\bm\Phi_t)^+\bm\Phi_t^\top\bm D_t^+\bm D_t=\bm\Phi_t(\bm\Phi_t^\top\bm D_t^+\bm\Phi_t)^+\bm\Phi_t^\top$ is symmetric, using $\bm\Phi_t^\top\bm D_t^+\bm D_t=(\bm D_t\bm D_t^+\bm\Phi_t)^\top=\bm\Phi_t^\top$.
\end{enumerate}
For orthogonality \eqref{eq:ORTHO}, using \eqref{eq:regulardec} and $(\bm I_{n_t}-\bm\Pi_t)\bm\Phi_t=\bm 0$,
\[
\E_t[\bm\epsilon_{t+1}\bm f_{t+1}^\top]=(\bm I_{n_t}-\bm\Pi_t)\bm Q_t\bm W_t=(\bm I_{n_t}-\bm\Pi_t)\bm\Phi_t\bm C_t\bm\Phi_t^\top\bm W_t+(\bm I_{n_t}-\bm\Pi_t)\bm D_t\bm W_t=\bm 0,
\]
by \ref{it:prGLSreg2} and \ref{it:prGLSreg3}. The identity $\bm\mu_{\bm f,t}=\bm W_t^\top\bm\mu_t$ holds by definition. For $\bm Q_{\bm f,t}=\bm W_t^\top\bm Q_t\bm W_t$, fact~\ref{it:prGLSreg1} and its transpose give $\bm W_t^\top\bm\Phi_t\bm C_t\bm\Phi_t^\top\bm W_t=\bm\Phi_t^+\bm\Phi_t\bm C_t\bm\Phi_t^+\bm\Phi_t$, and
\[
\bm W_t^\top\bm D_t\bm W_t=(\bm\Phi_t^\top\bm D_t^+\bm\Phi_t)^+(\bm\Phi_t^\top\bm D_t^+\bm D_t\bm D_t^+\bm\Phi_t)(\bm\Phi_t^\top\bm D_t^+\bm\Phi_t)^+=(\bm\Phi_t^\top\bm D_t^+\bm\Phi_t)^+,
\]
using $\bm D_t^+\bm D_t\bm D_t^+=\bm D_t^+$ and $\bm X^+\bm X\bm X^+=\bm X^+$. For the residual second moment matrix, $(\bm I_{n_t}-\bm\Pi_t)\bm\Phi_t=\bm 0$ gives
\[
\bm Q_{\bm\epsilon,t}=(\bm I_{n_t}-\bm\Pi_t)\bm Q_t(\bm I_{n_t}-\bm\Pi_t)^\top=(\bm I_{n_t}-\bm\Pi_t)\bm D_t(\bm I_{n_t}-\bm\Pi_t)^\top=\bm D_t-\bm\Pi_t\bm D_t,
\]
where the last equality uses \ref{it:prGLSreg4}, symmetry of $\bm\Pi_t\bm D_t$, and $\bm\Pi_t\bm D_t\bm\Pi_t^\top=\bm\Phi_t(\bm\Phi_t^\top\bm D_t^+\bm\Phi_t)^+(\bm\Pi_t\bm\Phi_t)^\top=\bm\Pi_t\bm D_t$. This is \eqref{eq:closedmoments}. Finally, $\bm\mu_{\bm\epsilon,t}=(\bm I_{n_t}-\bm\Pi_t)\bm\mu_t$ vanishes if and only if $\bm\mu_t=\bm\Pi_t\bm\mu_t$, which by \ref{it:prGLSreg2} holds if and only if $\bm\mu_t\in\Ima\bm\Phi_t$.
\end{proof}

\subsection{Proof of Corollary~\ref{prop:MWGLS}}
\begin{proof}
Under the law of one price \eqref{eq:LoP}, $\Ima\bm Q_t=\Ima\bm\Sigma_t+\Ima\bm\mu_t=\Ima\bm\Sigma_t$. Admissibility \eqref{eq:admissible} therefore gives an $\Fcal_t$-measurable vector $\bm b_t$ with $\bm\mu_t=\bm\Phi_t\bm b_t$ and $\Ima\bm\Phi_t\subseteq\Ima\bm Q_t=\Ima\bm\Sigma_t$. Hence $\bm Q_t=\bm\Sigma_t+\bm\mu_t\bm\mu_t^\top=\bm\Phi_t\bm b_t\bm b_t^\top\bm\Phi_t^\top+\bm\Sigma_t$, and $(\bm b_t\bm b_t^\top,\bm\Sigma_t)$ is a regular decomposition of $\bm Q_t$ with respect to $\bm\Phi_t$. Lemma~\ref{lem:GLSinv} applied with $\bm A_t=\bm Q_t$ and $\bm D_t=\bm\Sigma_t$ gives \eqref{eq:MWGLS}.
\end{proof}

\subsection{Proof of Proposition~\ref{lem:propGLSsimple}}
\begin{proof}
\emph{Case $\sigma_t>0$.} Set $\bm M_t\coloneqq\bm R_t\bm L_t\bm R_t^\top+\sigma_t^2\bm I_{n_t}$, which is symmetric positive definite. Then $(\bm C_t,\bm M_t)$ is a regular decomposition of $\bm Q_t$ with respect to $\bm\Phi_t$, and Lemma~\ref{lem:GLSinv} gives $(\bm\Phi_t^\top\bm Q_t^+\bm\Phi_t)^+\bm\Phi_t^\top\bm Q_t^+=(\bm\Phi_t^\top\bm M_t^{-1}\bm\Phi_t)^+\bm\Phi_t^\top\bm M_t^{-1}$. From $\bm R_t^\top\bm\Phi_t=\bm 0$ we obtain $\bm M_t\bm\Phi_t=\sigma_t^2\bm\Phi_t$, hence $\bm M_t^{-1}\bm\Phi_t=\sigma_t^{-2}\bm\Phi_t$ and, by symmetry, $\bm\Phi_t^\top\bm M_t^{-1}=\sigma_t^{-2}\bm\Phi_t^\top$. Therefore
\[
(\bm\Phi_t^\top\bm M_t^{-1}\bm\Phi_t)^+\bm\Phi_t^\top\bm M_t^{-1}=(\sigma_t^{-2}\bm\Phi_t^\top\bm\Phi_t)^+\sigma_t^{-2}\bm\Phi_t^\top=(\bm\Phi_t^\top\bm\Phi_t)^+\bm\Phi_t^\top=\bm\Phi_t^+.
\]

\emph{Case $\sigma_t=0$.} Set $\bm A_t\coloneqq\bm\Phi_t\bm C_t\bm\Phi_t^\top$ and $\bm B_t\coloneqq\bm R_t\bm L_t\bm R_t^\top$, both symmetric positive semidefinite with $\bm A_t\bm B_t=\bm B_t\bm A_t=\bm 0$, by $\bm R_t^\top\bm\Phi_t=\bm 0$. Then $\bm Q_t^+=\bm A_t^++\bm B_t^+$, as is checked by verifying the four Moore--Penrose conditions using $\bm A_t\bm B_t^+=\bm B_t\bm A_t^+=\bm 0$. Since the columns of $\bm B_t^+$ lie in $\Ima\bm B_t\perp\Ima\bm\Phi_t$, we get $\bm\Phi_t^\top\bm Q_t^+=\bm\Phi_t^\top\bm A_t^+$, so the left-hand side of \eqref{eq:condOLSid} equals $\bm G_t^+\bm\Phi_t^\top\bm A_t^+$ with $\bm G_t\coloneqq\bm\Phi_t^\top\bm A_t^+\bm\Phi_t$.

The hypothesis $\Ima(\bm\Phi_t\bm C_t)=\Ima\bm\Phi_t$ gives $\Ima\bm A_t=\Ima(\bm\Phi_t\bm C_t^{1/2})=\Ima\bm\Phi_t$, using $\Ima(\bm\Phi_t\bm C_t)\subseteq\Ima(\bm\Phi_t\bm C_t^{1/2})$ and $\rank(\bm\Phi_t\bm C_t)=\rank(\bm\Phi_t\bm C_t^{1/2})$, the latter because $\bm C_t\bm\Phi_t^\top\bm v=\bm 0$ if and only if $\bm C_t^{1/2}\bm\Phi_t^\top\bm v=\bm 0$. Hence $\Ima\bm A_t^+=\Ima\bm A_t=\Ima\bm\Phi_t$, so $\bm A_t^+\bm\Phi_t\bm\Phi_t^+=\bm A_t^+$ and therefore $\bm G_t\bm\Phi_t^+=\bm\Phi_t^\top\bm A_t^+\bm\Phi_t\bm\Phi_t^+=\bm\Phi_t^\top\bm A_t^+$. For $\bm w\in\ker\bm G_t$, moreover, $0=\bm w^\top\bm G_t\bm w=\|\bm A_t^{+1/2}\bm\Phi_t\bm w\|_2^2$ implies $\bm\Phi_t\bm w=\bm A_t^{1/2}\bm A_t^{+1/2}\bm\Phi_t\bm w=\bm 0$, since $\bm A_t^{1/2}\bm A_t^{+1/2}$ is the orthogonal projection onto $\Ima\bm A_t=\Ima\bm\Phi_t$. Hence $\ker\bm G_t\subseteq\ker\bm\Phi_t$, so $\Ima\bm G_t=\Ima\bm\Phi_t^\top$ and $\bm G_t^+\bm G_t$ is the orthogonal projection onto $\Ima\bm\Phi_t^\top$, which contains the columns of $\bm\Phi_t^+$. Combining,
\[
\bm G_t^+\bm\Phi_t^\top\bm A_t^+=\bm G_t^+\bm G_t\bm\Phi_t^+=\bm\Phi_t^+,
\]
which is \eqref{eq:condOLSid}.
\end{proof}

\subsection{Remarks on computation}
\begin{remark}[Relation to Kozak--Nagel]\label{rem:KNhedged}
At $\sigma_t>0$ the structure \eqref{eq:condOLS} has the shape of \citet[Equation~(19)]{kozaknagel23}, and at $\sigma_t=0$ exactly the shape of the decomposition in their Proposition~2, $\bm\Sigma_t=\bm X_t\bm\Psi_t\bm X_t^\top+\bm U_t\bm\Omega_t\bm U_t^\top$ with $\bm U_t^\top\bm X_t=\bm 0$, with the difference that theirs decomposes the return covariance matrix and ours the second moment matrix. Section~\ref{sec:rulesout} places their Proposition~2 itself, whose condition is the $\bm\Sigma_t$-invariance of $\Ima\bm X_t$. Proposition~\ref{lem:propGLSsimple} adds the concrete structural form under which the invariance obtains, and its case $\sigma_t=0$ is the any-rank counterpart of their decomposition, with no invertibility required. Without invertibility of $\bm\Sigma_t$ and of the factor covariance matrix, two steps of their spanning lemma have no counterpart: the support condition \eqref{eq:Phistarass} is no longer automatic, so traded factors carrying the given loadings need not exist at all, by Theorem~\ref{thm:rigid}, and the decomposition their argument rests on requires the conditions of \citet[Theorem~1]{lu_sch_12}, which positive definiteness supplies. Corollary~\ref{cor:OLScoh} is the invariance characterization that survives at any rank.

\end{remark}

\begin{remark}[Singular $\bm Q_t$ Gives a Short Panel]\label{rem:shortpanel}
Let $\rank\bm Q_t=r_t<n_t$ and let $\bm B_t$ be an $n_t\times r_t$ matrix with orthonormal columns spanning $\Ima\bm Q_t$. By Remark~\ref{rem:support}, $\bm x_{t+1}=\bm B_t\bm z_{t+1}$ almost surely for $\bm z_{t+1}\coloneqq\bm B_t^\top\bm x_{t+1}$, and $\bm Q_{\bm z,t}=\bm B_t^\top\bm Q_t\bm B_t$ is nonsingular. Under the support condition \eqref{eq:Phistarass} write $\bm\Phi_t=\bm B_t\bm\Phi_{\bm z,t}$. Since $(\bm B_t\bm A)^+=\bm A^+\bm B_t^\top$ by orthonormality, $\bm Q_t^+=\bm B_t\bm Q_{\bm z,t}^{-1}\bm B_t^\top$ and $\bm W_t^\top=\bm W_{\bm z,t}^\top\bm B_t^\top$: the factors, residuals, the interval \eqref{eq:admissible} and the risk-premium gap coincide with those of the $r_t$-asset economy in $\bm z_{t+1}$ with loadings $\bm\Phi_{\bm z,t}$. A singular second moment matrix thus pertains to an economy whose effective cross-section has dimension $r_t$. Without the support condition the reduction discards the part of $\bm\Phi_t$ outside $\Ima\bm Q_t$, whose size is the loading gap of Corollary~\ref{cor:cohgap}.

In a $\Twin$-month window the reduction applies to $\hat{\bm Q}_t\coloneqq\bm Y_t^\top\bm Y_t/\Twin$, with $\bm Y_t$ the $\Twin\times n_t$ matrix of window returns, and two qualifications. First, $r_t=\rank\bm Y_t\le\min(n_t,\Twin)$, at equality in every window of the panels of Section~\ref{sec:empirical}, and every inversion runs at $\Twin\times\Twin$ size, since $\hat{\bm Q}_t^+=\bm Y_t^\top\hat{\bm G}_t^{+2}\bm Y_t/\Twin$ for the Gram matrix $\hat{\bm G}_t=\bm Y_t\bm Y_t^\top/\Twin$. Second, the reduction is exact only within the window: $\bm x_{t+1}$ generally lies outside $\Ima\hat{\bm Q}_t$, which leaves the factors unchanged, as the weights annihilate the orthogonal complement, but gives the one-step-ahead residuals a component the reduced economy discards. This fixed-$\Twin$, larger-$n_t$ configuration is that of the short-panel literature, which infers conditionally on the factor realizations, targeting the ex-post risk premia of \citet[Equation~(5)]{raponirobottizaffaroni19}, with consistency from the growing cross-section \citep{fortin2025eigenvalue,fortingagliardiniscaillet24}. The Gram matrix is the sample counterpart, up to normalization by $n_t$ in place of $\Twin$, of the cross-sectional second moment matrix of returns in \citet[Equation~(6)]{fortin2025eigenvalue}. The design of Section~\ref{sec:empirical} shares the shape but not the statistics: its cross-section is a portfolio universe, not a growing sample, and it attempts no within-window inference, for which that fixed-$\Twin$ theory is the natural candidate, see Section~\ref{sec:guidelines}. The source of the deficiency also differs: in the window design it is the number of observations, as the closing paragraph of Section~\ref{sec:admissibilityspanning} discusses, while here it is the return distribution itself, so the reduction is a statement about the population.
\end{remark}

\section{Existence of Factor Representations}\label{app:existence}
This appendix answers the existence question for factor representations. Section~\ref{sec:impossible} shows that any representation \eqref{eq:linfacmodelPhi} whose factors and residuals are orthogonal, \eqref{eq:ORTHO}, induces the decomposition \eqref{eq:Qdecomp} of the second moment matrix, $\bm Q_t=\bm\Phi_t\bm Q_{\bm f,t}\bm\Phi_t^\top+\bm Q_{\bm\epsilon,t}$. If residual risk is in addition unpriced, \eqref{eq:coh2}, then also $\bm\mu_t=\bm\Phi_t\bm\mu_{\bm f,t}$, with $\bm\mu_{\bm f,t}\bm\mu_{\bm f,t}^\top\preceq\bm Q_{\bm f,t}$ since $\bm Q_{\bm f,t}-\bm\mu_{\bm f,t}\bm\mu_{\bm f,t}^\top=\bm\Sigma_{\bm f,t}\succeq\bm 0$. Proposition~\ref{prop:REPR} establishes the converse: for any $\bm\Phi_t$, $\bm C_t$, $\bm D_t$ decomposing $\bm Q_t$ there exists, on a possibly enlarged probability space, a factor representation with orthogonal factors and residuals and moments $\bm Q_{\bm f,t}=\bm C_t$ and $\bm Q_{\bm\epsilon,t}=\bm D_t$, and analogously for the mean under the matching condition on $\bm b_t$.

Proposition~\ref{prop:REPR} generalizes \cite[Exercise~9.2.2]{mar_ken_bib_79} by allowing $\bm C_t$, $\bm D_t$, and $\bm\Phi_t$ to be singular. Papers working with approximate factor structures, such as \citet{chamberlainrothschild83}, typically characterize the moment structure directly without constructing an explicit factor representation. Proposition~\ref{prop:REPR} shows that such a representation always exists, so that factor--residual orthogonality entails no loss of generality.

\begin{proposition}[Factor Representation of Second Moment and Mean Decompositions]\label{prop:REPR}
Assume that the return second moment matrix admits a decomposition
\begin{equation}\label{eq:REPReq1}
  \bm Q_t = \bm\Phi_t\bm C_t\bm\Phi_t^\top + \bm D_t,
\end{equation}
for some $\Fcal_t$-measurable, symmetric positive semidefinite $m_t\times m_t$ and
$n_t\times n_t$ matrices $\bm C_t$ and $\bm D_t$. Then there exist
$\R^{m_t}$- and $\R^{n_t}$-valued random vectors $\bm f_{t+1}$ and
$\bm\epsilon_{t+1}$, defined on some possibly enlarged probability space, such that the return
vector admits the representation \eqref{eq:linfacmodelPhi}, factors and residuals are orthogonal, \eqref{eq:ORTHO}, and the second moments match,
\begin{equation}\label{eq:cfCceD}
    \bm Q_{\bm f,t}=\bm C_t, \quad \bm Q_{\bm\epsilon,t}=\bm D_t.
\end{equation}
If in addition the return mean vector admits a decomposition
\begin{equation}\label{eq:meandec}
    \bm\mu_t = \bm\Phi_t\bm b_t,
\end{equation}
for some $\Fcal_t$-measurable $m_t$-vector $\bm b_t$ satisfying
\begin{equation}\label{eq:bbC}
    \bm b_t\bm b_t^\top \preceq \bm C_t,
\end{equation}
then $\bm f_{t+1}$ and $\bm\epsilon_{t+1}$ can be chosen such that, in addition, the means match,
\begin{equation}\label{eq:meanmatch}
    \E_t[\bm f_{t+1}]=\bm b_t,\quad \E_t[\bm\epsilon_{t+1}]=\bm 0,
\end{equation}
so that residual risk is unpriced, \eqref{eq:coh2}, and factors and residuals are uncorrelated, \eqref{eq:UNCORR}. Condition \eqref{eq:bbC} cannot be dropped: \eqref{eq:cfCceD} and \eqref{eq:meanmatch} force $\cov_t[\bm f_{t+1}]=\bm C_t-\bm b_t\bm b_t^\top\succeq\bm 0$.
\end{proposition}

The proof is constructive, providing explicit expressions for $\bm f_{t+1}$ and $\bm\epsilon_{t+1}$ in terms of the matrices in \eqref{eq:REPReq1}, returns $\bm x_{t+1}$, and auxiliary noise augmentations that cancel out in the representation \eqref{eq:linfacmodelPhi}.

\begin{proof}
We define $\bm\Psi_t \coloneqq \bm\Phi_t\bm C_t^{1/2}$, so that $\bm\Phi_t\bm C_t\bm\Phi_t^\top=\bm\Psi_t\bm\Psi_t^\top$ and, by \eqref{eq:REPReq1}, $\Ima\bm\Psi_t=\Ima(\bm\Psi_t\bm\Psi_t^\top)\subseteq\Ima\bm Q_t$, using the fact that $\Ima(\bm A+\bm B)=\Ima\bm A+\Ima\bm B$ for symmetric positive semidefinite matrices $\bm A,\bm B$.

Note that $\bm K_t\coloneqq \bm\Psi_t\bm\Psi_t^\top - \bm\Psi_t\bm\Psi_t^\top\bm Q_t^+\bm\Psi_t\bm\Psi_t^\top$ is symmetric positive semidefinite.\footnote{$\bm K_t$ is the parallel sum $\bm\Psi_t\bm\Psi_t^\top:\bm D_t$, where the parallel sum of symmetric positive semidefinite matrices is $\bm A:\bm B\coloneqq\bm A(\bm A+\bm B)^+\bm B$, with $\Ima(\bm A:\bm B)=\Ima\bm A\cap\Ima\bm B$, see \cite{and_duf_69}. In particular, $\bm K_t = \bm\Psi_t\bm\Psi_t^\top (\bm\Psi_t\bm\Psi_t^\top+\bm D_t)^+ \bm D_t$ and $\Ima\bm K_t=\Ima\bm\Psi_t\cap\Ima\bm D_t$.\label{fn:ps}} This property follows from \cite[Theorem 1]{alb_69}, as $\bm K_t$ is equal to the generalized Schur complement of the upper-left block of the symmetric matrix
 \[
\begin{bmatrix}
  \bm Q_t &\bm\Psi_t\bm\Psi_t^\top\\
 \bm\Psi_t\bm\Psi_t^\top & \bm\Psi_t\bm\Psi_t^\top
\end{bmatrix} =
\begin{bmatrix}
 \bm\Psi_t  \\
  \bm\Psi_t
\end{bmatrix} \begin{bmatrix}
 \bm\Psi_t  \\
  \bm\Psi_t
\end{bmatrix} ^\top
+ \begin{bmatrix}
 \bm D_t  & \bm 0 \\
  \bm 0 & \bm 0
\end{bmatrix},
\]
which is positive semidefinite by assumption.

Now define
\begin{equation}\label{eq:proofpropREPReq1}
 \begin{bmatrix} \bm g_{t+1}\\ \bm \epsilon_{t+1}  \end{bmatrix} \coloneqq    \begin{bmatrix} \bm\Psi_t^\top \bm Q_t^+ & (\bm I_{m_t}-\bm\Psi_t^+\bm\Psi_t) & \bm \Psi_t^+ \\  \bm I_{n_t}-\bm\Psi_t\bm\Psi_t^\top\bm Q_t^+ & \bm 0 & -\bm\Psi_t\bm\Psi_t^+ \end{bmatrix} \begin{bmatrix} \bm x_{t+1}  \\ \bm y_{t+1} \\ \bm z_{t+1}\end{bmatrix},
\end{equation}
where $\bm y_{t+1}$ and $\bm z_{t+1}$ are auxiliary $\R^{m_t}$- and $\R^{n_t}$-valued random vectors with zero conditional means, $\E_t[\bm y_{t+1}]=\bm 0$ and $\E_t[\bm z_{t+1}]=\bm 0$, orthogonal to all returns and mutually, $\E_t[\bm y_{t+1}\bm x_{t+1}^\top]=\bm 0$, $\E_t[\bm z_{t+1}\bm x_{t+1}^\top]=\bm 0$, and $\E_t[\bm y_{t+1}\bm z_{t+1}^\top]=\bm 0$, and with second moment matrices $\E_t[\bm y_{t+1}\bm y_{t+1}^\top]=\bm I_{m_t}$ and $\E_t[\bm z_{t+1}\bm z_{t+1}^\top]=\bm K_t$.

These serve as noise augmentations such that the resulting second moment matrix of $\bm g_{t+1}$, $\bm \epsilon_{t+1}$ equals
\begin{equation}\label{eq:covgeps}
     \E_t\left[ \begin{bmatrix} \bm g_{t+1}\\ \bm \epsilon_{t+1}  \end{bmatrix}\begin{bmatrix} \bm g_{t+1}\\ \bm \epsilon_{t+1}  \end{bmatrix}^\top \right] = \begin{bmatrix} \bm I_{m_t} & \bm 0 \\ \bm 0 & \bm D_t   \end{bmatrix}.
\end{equation}
Indeed, \eqref{eq:covgeps} follows directly from \eqref{eq:proofpropREPReq1} by tedious but elementary calculations, using $\Ima\bm\Psi_t\subseteq \Ima\bm Q_t$ and that $\bm Q_t^+\bm Q_t$ is the orthogonal projection onto $\Ima\bm Q_t$.\footnote{In particular, if $\rank\bm\Psi_t=m_t$ then $\bm I_{m_t}-\bm\Psi_t^+\bm\Psi_t=\bm 0$, so that we could omit $\bm y_{t+1}$. Similarly, $\bm z_{t+1}=\bm 0$ if $\bm K_t=\bm 0$, which is equivalent to $\Ima\bm\Psi_t\cap\Ima\bm D_t=\{\bm 0\}$, see Footnote~\ref{fn:ps}.}

On the other hand, left-matrix multiplication of \eqref{eq:proofpropREPReq1} by $[ \bm\Psi_t , \bm I_{n_t} ]$ shows that the noise augmentations cancel out and we obtain $\bm\Psi_t\bm g_{t+1}+\bm \epsilon_{t+1} = \bm x_{t+1}$. Setting $\bm f_{t+1}\coloneqq \bm C_t^{1/2}\bm g_{t+1}$ gives \eqref{eq:linfacmodelPhi} with, by \eqref{eq:covgeps}, $\E_t[\bm\epsilon_{t+1}\bm f_{t+1}^\top]=\E_t[\bm\epsilon_{t+1}\bm g_{t+1}^\top]\bm C_t^{1/2}=\bm 0$ and $\bm Q_{\bm f,t}=\bm C_t^{1/2}\E_t[\bm g_{t+1}\bm g_{t+1}^\top]\bm C_t^{1/2}=\bm C_t$, which proves the first part of the proposition.

For the second part, assume \eqref{eq:meandec} and \eqref{eq:bbC}, and set $\widetilde{\bm C}_t\coloneqq\bm C_t-\bm b_t\bm b_t^\top$, which is symmetric positive semidefinite by \eqref{eq:bbC}. The centered returns $\bm x^0_{t+1}=\bm x_{t+1}-\bm\mu_t$ have second moment matrix
\[
\bm\Sigma_t=\bm Q_t-\bm\mu_t\bm\mu_t^\top=\bm\Phi_t\widetilde{\bm C}_t\bm\Phi_t^\top+\bm D_t,
\]
by \eqref{eq:REPReq1} and \eqref{eq:meandec}. Apply the construction above to $\bm x^0_{t+1}$ with the decomposition $(\widetilde{\bm C}_t,\bm D_t)$: it yields $\bm f^0_{t+1}$ and $\bm\epsilon^0_{t+1}$ with $\bm x^0_{t+1}=\bm\Phi_t\bm f^0_{t+1}+\bm\epsilon^0_{t+1}$, $\E_t[\bm\epsilon^0_{t+1}\bm f^{0\top}_{t+1}]=\bm 0$, $\bm Q_{\bm f^0,t}=\widetilde{\bm C}_t$ and $\bm Q_{\bm\epsilon^0,t}=\bm D_t$. Moreover, $\E_t[\bm f^0_{t+1}]=\bm 0$ and $\E_t[\bm\epsilon^0_{t+1}]=\bm 0$, since \eqref{eq:proofpropREPReq1} is linear in $(\bm x^0_{t+1},\bm y_{t+1},\bm z_{t+1})$, all three of which have zero conditional mean. Then $\bm f_{t+1}\coloneqq\bm f^0_{t+1}+\bm b_t$ and $\bm\epsilon_{t+1}\coloneqq\bm\epsilon^0_{t+1}$ satisfy \eqref{eq:linfacmodelPhi} and \eqref{eq:meanmatch}, along with
\begin{gather*}
\bm Q_{\bm f,t}=\widetilde{\bm C}_t+\bm b_t\bm b_t^\top=\bm C_t,\qquad
\bm Q_{\bm\epsilon,t}=\bm D_t,\\
\E_t[\bm\epsilon_{t+1}\bm f_{t+1}^\top]=\E_t[\bm\epsilon^0_{t+1}\bm f^{0\top}_{t+1}]+\E_t[\bm\epsilon^0_{t+1}]\bm b_t^\top=\bm 0,
\end{gather*}
and uncorrelatedness \eqref{eq:UNCORR} follows as $\cov_t[\bm\epsilon_{t+1},\bm f_{t+1}]=\E_t[\bm\epsilon_{t+1}\bm f_{t+1}^\top]-\E_t[\bm\epsilon_{t+1}]\E_t[\bm f_{t+1}]^\top=\bm 0$.

For the necessity of \eqref{eq:bbC}, conditions \eqref{eq:cfCceD} and \eqref{eq:meanmatch} give $\cov_t[\bm f_{t+1}]=\bm Q_{\bm f,t}-\E_t[\bm f_{t+1}]\E_t[\bm f_{t+1}]^\top=\bm C_t-\bm b_t\bm b_t^\top$, which is positive semidefinite. This completes the proof of Proposition~\ref{prop:REPR}.
\end{proof}

\begin{remark}[Covariance Counterpart]\label{rem:REPRcov}
Read on the centered returns $\bm x^0_{t+1}=\bm x_{t+1}-\bm\mu_t$, whose second moment matrix is $\bm\Sigma_t$, the construction yields the classical covariance statement: a decomposition $\bm\Sigma_t=\bm\Phi_t\bm C_t\bm\Phi_t^\top+\bm D_t$ with symmetric positive semidefinite $\bm C_t$ and $\bm D_t$ admits a representation \eqref{eq:linfacmodelPhi} with $\cov_t[\bm f_{t+1}]=\bm C_t$, $\cov_t[\bm\epsilon_{t+1}]=\bm D_t$ and uncorrelated factors and residuals, and under \eqref{eq:meandec} the means can be matched with no condition on $\bm b_t$: apply the construction to $\bm x^0_{t+1}$ with $(\bm C_t,\bm D_t)$ and set $\bm f_{t+1}=\bm f^0_{t+1}+\bm b_t$, $\bm\epsilon_{t+1}=\bm\epsilon^0_{t+1}$, which leaves all covariances unchanged. No analogue of \eqref{eq:bbC} appears because the covariance version does not pin down the factor second moment matrix.
\end{remark}

The auxiliary vectors $\bm y_{t+1}$ and $\bm z_{t+1}$ are a technical device without economic content. They cancel out in the representation \eqref{eq:linfacmodelPhi}, as the left multiplication in the proof shows, and serve only to match the prescribed, possibly singular, factor and residual moment matrices. No pricing implications attach to them.

\end{appendix}

\end{document}